\documentclass[a4paper,11pt]{article}
\usepackage{jheppub} 

\usepackage[T1]{fontenc} 
\usepackage{tikz}
\usepackage{graphicx}
\usepackage{float} 
\usepackage{amsmath}
\usepackage{amssymb}
\usepackage{subcaption}
\usetikzlibrary{decorations.pathreplacing}
\usetikzlibrary{quotes,angles}
\usepackage[normalem]{ulem}
\def\beq{\begin{equation}}
\def\eeq{\end{equation}}
\def\bea{\begin{eqnarray}}
\def\eea{\end{eqnarray}}

\allowdisplaybreaks

\title{Energy-Energy Correlator at Hadron Colliders: Celestial Blocks and Singularities}

\author[1]{Hao Chen,}
\author[2]{Hongyi Ruan,}
\author[2,3]{and Hua Xing Zhu}

\affiliation[1]{Center for Theoretical Physics, Massachusetts Institute of Technology, 77 Massachusetts Avenue, Cambridge, MA 02139, U.S.A}
\affiliation[2]{School of Physics, Peking University, Beijing 100871, China}
\affiliation[3]{Center for High Energy Physics, Peking University, Beijing 100871, China}

\emailAdd{hao\_chen@mit.edu}
\emailAdd{ruanhongyi@stu.pku.edu.cn}
\emailAdd{zhuhx@pku.edu.cn}

\abstract{
Energy-energy correlator (EEC) is an event shape observable that characterizes the distribution of energy flux in collision events. We initiate the study of full-range EEC at hadron colliders, generalizing the extensively studied EEC in $e^+e^-$ collision as well as the transverse EEC in hadron collisions. We derive celestial blocks from Lorentz symmetry to perform partial wave decomposition of the EEC at hadron colliders. These celestial blocks are essentially conformal blocks on the {\it 2d} celestial sphere, which have additional dependence on the collinear spin of ``light-ray transition matrix'' along the collision axis. In this work, we perform the leading-order (LO) analytic calculation of this observable in pure Yang-Mills theory and use it as an example to illustrate the block decomposition. Numerically, the block expansion demonstrates superior accuracy in the collinear limit compared to conventional power series expansion. Analytically, we observe in this example that the block coefficients exhibit analyticity in both collinear and transverse spin. In addition, we analyze several kinematic limits at LO --- collinear, back-to-back, opposite coplanar and Regge limit. While the first three limits naturally generalize their $e^+e^-$ collision counterparts or transverse EEC and are governed by soft-collinear dynamics, the Regge limit requires complete angular dependence and reveals BFKL physics. Phenomenologically, we propose a realistic experimental setup and briefly discuss how the convolution of parton distribution function modifies the perturbative EEC result. Our work suggests that the full-range EEC at hadron colliders is an elegant observable which probes a broader kinematic space and connects various regimes of different QCD dynamics through a single measurement.
}

\notoc

\begin{document} 
\maketitle

\newpage
\tableofcontents

\section{Introduction}
\label{sec:intro}

The energy-energy correlator (EEC) was first introduced over forty years ago in~\cite{Basham:1978bw} as an infrared and collinear (IRC) safe event shape observable in $e^+e^-$ colliders, specifically for precision tests of quantum chromodynamics (QCD)~\cite{Basham:1978zq}. This observable reflects the geometric characteristics of final-state energy flow distribution in particle collisions, providing profound insights into the underlying theory.

Recently, EEC has garnered renewed interest due to our improved understanding of its properties. The operator definition of EEC involves the energy flow operator~\cite{Sveshnikov:1995vi,Korchemsky:1999kt,Bauer:2008dt,Hofman:2008ar}
\begin{equation}
    \label{energyflow}
    \mathcal{E}(\vec n)=\lim\limits_{r\to\infty} \int_{0}^{\infty}dt\, r^2 \bar{n}^i T_{0i}(t,r\vec n),
\end{equation}
where $T_{\mu\nu}$ is the stress tensor of the theory, and $\vec n$ is a unit vector that specifies the direction of the calorimeter. The EEC is defined as the Fourier transform of a four-point Wightman function~\cite{Hofman:2008ar,Belitsky:2013xxa,Belitsky:2013ofa,Belitsky:2013bja},
\begin{equation}
    \label{EEC_formula}
    \frac{1}{\sigma_{\text{tot}}}\frac{d\sigma}{dz}=\frac{
        \int d^4x\, e^{iq\cdot x}\langle\mathcal{O}(x)\mathcal{E}(\vec n_1)\mathcal{E}(\vec n_2)\mathcal{O}^\dagger(0)\rangle
    }{
        \int d^4x\, e^{iq\cdot x} \langle\mathcal{O}(x)\mathcal{O}^\dagger(0)\rangle
    },
\end{equation}
where $z\equiv (1-\vec n_1\cdot\vec n_2)/2$. With this definition, sophisticated position-space calculation techniques for EEC and related correlators have been developed in~\cite{Belitsky:2013xxa,Henn:2019gkr,Chicherin:2020azt}. These techniques are extremely efficient in conformal field theories (CFT), which have achieved the next-to-next-to-leading order (NNLO) analytic result for EEC in $\mathcal{N}=4$ super Yang-Mills theory~\cite{Henn:2019gkr}.
In QCD, EEC has been calculated analytically at NLO~\cite{Dixon:2018qgp,Luo:2019nig,Gao:2020vyx} and numerically at NNLO~\cite{DelDuca:2016csb,DelDuca:2016ily} using momentum-space loop integration techniques. Recently, multi-point energy correlators $\langle \mathcal{E}(\vec{n}_1) \cdots \mathcal{E}(\vec{n}_k) \rangle$ in $\mathcal{N}=4$ SYM and QCD have been extensively studied in~\cite{Chen:2019bpb,Chen:2020vvp,Chen:2020adz,Chen:2021gdk,Chang:2022ryc,Chen:2022jhb,Yan:2022cye,Yang:2022tgm,Yang:2024gcn,Gao:2024wcg,Chicherin:2024ifn,He:2024hbb}. 

EEC is a collider observable that has deep connection to many fundamental field-theoretical concepts. The positivity of EEC, as a consequence of ANEC positivity~\cite{Hartman:2016lgu,Faulkner:2016mzt}, provides collider bound for some parameters in CFT~\cite{Hofman:2008ar,Hofman:2016awc,Cordova:2017zej}. 
ANEC positivity is also closely associated with properties under the renormalization group flow~\cite{Hartman:2023qdn,Hartman:2023ccw,Hartman:2024xkw,Zamolodchikov:1986gt,Komargodski:2011vj}. EEC is one of the most important examples of matrix element of light-ray operators~\cite{Hofman:2008ar,Kravchuk:2018htv,Caron-Huot:2022eqs,Balitsky:1987bk}, whose collinear limit can be formulated as the operator product expansion (OPE) in terms of light-ray operators~\cite{Hofman:2008ar,Kologlu:2019mfz,Chang:2020qpj}. EEC also probes the large spin physics~\cite{Alday:2007mf,Fitzpatrick:2012yx,Komargodski:2012ek,Alday:2013cwa,Alday:2015eya,Alday:2015ota} through its back-to-back limit~\cite{Korchemsky:2019nzm,Chen:2023wah}. On the other hand, we have alternative QCD factorization descriptions for both collinear~\cite{Dixon:2019uzg,Chen:2023zzh} and back-to-back limit~\cite{Moult:2018jzp,Ebert:2020sfi,Duhr:2022yyp}. 
These observations highlight how insights from formal theory can strengthen our understanding of collider phenomenology, bridging the gap between the two areas.
Further discussions of EEC in formal theory can be found in~\cite{Kologlu:2019bco,Firat:2023lbp,Chicherin:2023gxt,Chen:2024iuv,Gonzo:2020xza,Herrmann:2024yai,Cuomo:2025pjp}.

Over past few years, EEC has found many interesting applications in collider phenomenology. Phenomenological studies of EEC has motivated several experimental groups to do corresponding measurement in various colliders~\cite{CMS:2024mlf,ALICE:2024dfl,CMS:2024ovv,Tamis:2023guc,ALICE:2025igw,CMS:2025ydi}.
For example, based on the collinear limit of projected energy correlators~\cite{Chen:2020vvp,Komiske:2022enw,Chen:2023zlx}, CMS collaboration extracted the most precise $\alpha_s$ value obtained using jet substructure observables~\cite{CMS:2024mlf}.
Other phenomenological generalizations and applications of EECs can be found in~\cite{Li:2021zcf,Jaarsma:2022kdd,Jaarsma:2023ell,Lee:2023tkr,Lee:2023npz,
Liu:2022wop,Liu:2023aqb,Cao:2023oef,Li:2023gkh,Liu:2024kqt,Liu:2024lxy,Chen:2024bpj,Guo:2024jch,
Holguin:2022epo,Xiao:2024rol,Holguin:2024tkz,Holguin:2023bjf,
Andres:2022ovj,Andres:2023xwr,Andres:2023ymw,Andres:2024ksi,Andres:2024hdd,Andres:2024xvk,Yang:2023dwc,Barata:2023bhh,Barata:2024ieg,Barata:2024wsu,Barata:2024bmx,Singh:2024vwb,Bossi:2024qho,
Lee:2022ige,Craft:2022kdo,
Alipour-fard:2024szj,Budhraja:2024tev,Budhraja:2024xiq,
Li:2021txc,Li:2020bub,Guo:2024vpe,Schindler:2023cww,Lee:2024esz,Kang:2023gvg,Kang:2023big,Csaki:2024zig,Csaki:2024joe,Riembau:2024tom,Chen:2024nfl,Cao:2023qat,Devereaux:2023vjz,Barata:2023zqg,Ricci:2022htc,Chen:2022swd,
Lee:2024jnt,Barata:2024apg,Lin:2024lsj,Alipour-fard:2025dvp,Bhattacharya:2025bqa,Mantysaari:2025mht,Budhraja:2025ulx,Barata:2025fzd,Apolinario:2025vtx}.

Compared with the in-depth understanding of EEC in electron-positron collisions, our knowledge of EEC at hadron colliders remains limited, particularly regarding the theoretical framework and analytic computations. A natural extension of the EEC to hadron colliders is the transverse energy-energy correlator (TEEC)~\cite{Ali1984,ATLAS:2015yaa,ATLAS:2017qir,ATLAS:2023tgo,Ali:2012rn,Alvarez:2023fhi,Gao:2019ojf,Gao:2023ivm,Kang:2023oqj,Kang:2024otf}, which is defined as 
\begin{equation}
    \label{TEEC}
    \frac{d\sigma}{d\cos\phi}=\sum\limits_{a,b}\int d\sigma_{pp\to a+b+X}\frac{2E_{T,a}E_{T,b}}{{|\sum_i E_{T,i}|}^2}\delta(\cos\phi_{ab}-\cos\phi).
\end{equation}
Here, $\phi$ is the azimuthal angle separation of the two detectors, i.e., the angular separation in the plane transverse to the scattering axis. The variable $\phi_{ab}$ represents the azimuthal angle difference between particles $a$ and $b$, which are summed over all particle pairs in a given event. 
$E_{T,i}$ is the transverse energy of $i$-th particle in the final state, measured relative to the collision beam. Similar to EEC, TEEC exhibits singularities at the two endpoints and must be resummed to yield predictions that are comparable with experimental data. This resummation has been performed in the back-to-back limit at next-to-next-to-leading logarithm (NNLL) accuracy a few years ago~\cite{Gao:2019ojf}, and more recently at \(\text{N}^3\text{LL}\) accuracy~\cite{Gao:2023ivm}.

However, as a function of single kinematic variable $\phi$, TEEC integrates out the rapidity information of calorimeters, which may hinder its ability to capture other interesting physics properties. 
By contrast, EEC at hadron colliders with full solid angle dependence, defined as
\begin{equation}
    \label{HEEC}
    \frac{d^2\Sigma}{d\Omega_a d\Omega_b}=\sum\limits_{i,j}\int d\sigma_{pp\to i+j+X}\,E_i E_j\,\delta^{(2)}(\Omega_a-\Omega_{p_i})\delta^{(2)}(\Omega_b-\Omega_{p_j})\,,
\end{equation}
has richer kinematic space to explore.
For instance, it is challenging to directly use TEEC to probe the Regge limit/forward scattering limit, where the center-of-mass energy \( s \) is much larger than other scales. The Regge limit is an important concept in both formal quantum field theory and phenomenology~\cite{Gribov:2003nw,Forshaw:1997dc,Fadin:1975cb,Balitsky:1978ic}, e.g., revealing the asymptotic high-energy behavior of scattering amplitudes ~\cite{Falcioni:2021dgr,Caola:2021izf,Gao:2024fyz}, providing insights into the AdS/CFT correspondence~\cite{Cornalba:2006xk,Cornalba:2006xm,Cornalba:2007zb,Kulaxizi:2017ixa,Li:2017lmh}, understanding the small Bjorken-$x$ behavior of parton distribution functions (PDFs)~\cite{Catani:1994sq,Lipatov:1996ts,Ball:2017otu,Rothstein:2016bsq,Neill:2023jcd}, describing diffractive processes in high-energy collisions~\cite{Berera:1995fj,Collins:1997sr}.
Using EEC at hadron colliders~\eqref{HEEC}, the Regge limit can be accessible when the rapidity difference between two calorimeters is very large, much like the proposal of Mueller-Navelet jets~\cite{Mueller:1986ey}. 

Compared to the EEC in $e^+e^-$ colliders and the TEEC, the EEC at hadron colliders~\eqref{HEEC} remains far from complete in terms of theoretical development. 
First, $e^+e^-$ colliders offer a cleaner experimental environment, where the EEC can be formulated as a correlation function~\eqref{EEC_formula} involving a local operator $\mathcal{O}$, typically representing the electromagnetic current. In contrast, hadron colliders feature initial-state hadrons undergoing complicated scattering processes, significantly hampering first-principle calculations of the EEC. In particular, the energy flow distribution at hadron colliders is strongly affected by initial-state QCD radiation.
Second, unlike the TEEC which depends solely on the azimuthal angle separation between the two detectors, the EEC at hadron colliders also exhibits a nontrivial dependence on their rapidities. As a result, it is generally a function of three independent kinematic variables, leading to a more intricate parametrization of the observable. 
Third, as in the case of the TEEC, the EEC at hadron colliders receives contributions from multiple partonic scattering channels and involves convolutions with parton distribution functions (PDFs). 
Despite these complications, the EEC at hadron colliders offers valuable opportunities to probe the finer structure of collision events and to study the interplay between initial-state radiation and final-state measurements.

In this work, we take a first step toward addressing all three challenges outlined above, with a particular focus on the pure gluon scattering channel. To the best of our knowledge, no analytic result exists for the EEC at hadron colliders, even at the leading nontrivial order. We present the first leading-order (LO) analytic result for pure gluon scattering; calculation of other partonic channels in QCD should follow straightforwardly, and the corresponding results are included in the supplementary file \texttt{EEC\_result\_all\_channel.m}. In addition, we provide supporting numerical simulations that include the convolutions with PDFs. These results allow us to systematically explore various kinematic limits, including the collinear limit, the opposite coplanar limit, the back-to-back limit, and the Regge limit. For each of these regimes, we use fixed-order factorization approximation that successfully reproduces the leading-power (LP) behavior. Although we do not propose a complete factorization formula, our analysis offers a new perspective on the EEC at hadron colliders as a tool for probing diverse kinematic regimes—particularly the Regge limit—and opens a pathway for future studies of related topics such as BFKL dynamics.

In addition, we introduce the celestial blocks for the EEC at hadron colliders by following a procedure analogous to that developed for the $e^+e^-$ colliders~\cite{Kologlu:2019mfz}. The celestial block decomposition, which can be regarded as a partial wave decomposition at cross section level that respects Lorentz symmetry, offers an alternative representation of our perturbative results. It also serves as a nontrivial consistency check for the intricate pure gluon scattering contribution to the EEC at hadron colliders, as one can, in principle, isolate the contribution of each light-ray operator in the collinear limit. This, in turn, provides insight into higher-twist effects in QCD. Through these studies, we aim to offer new perspectives and research directions for future explorations of the EEC at hadron colliders.

The structure of this paper is as follows. In Sec.~\ref{sec:ccb}, we review the celestial blocks in \( e^+e^- \) annihilation and introduce their generalization to the EEC at hadron colliders. This provides a kinematic expansion basis for the EEC at hadron colliders. In Sec.~\ref{sec:gluon}, we analytically compute the LO contribution to the EEC at hadron colliders from pure gluon scattering, based on five-point scattering amplitudes. We also apply the celestial blocks derived in the previous section to decompose our result, which not only serves as a nontrivial cross-check but also offers a compact parametrization. In addition, we discuss the light-ray operators that contribute in the OPE limit. We observe that our OPE data exhibit analyticity in transverse spin, and we derive the relevant expressions using the Lorentzian inversion formula. In Sec.~\ref{sec:factorization}, we present fixed-order factorization analysis in various kinematic limits, highlighting the Regge limit as particularly intriguing yet largely unexplored in the context of the EEC. In Sec.~\ref{sec:plot}, we visualize the pure gluon scattering contribution to the EEC at hadron colliders, both with and without convolution over PDFs. Finally, we conclude with future directions in Sec.~\ref{sec:conclusion}.

\section{Collider celestial block}
\label{sec:ccb}

As correlation functions of light-ray operators, EECs inherit nice group-theoretic structures from the Lorentz symmetry. In particular, their partial wave decomposition in the collinear limit has the physical interpretation as light-ray OPE~\cite{Hofman:2008ar,Kologlu:2019mfz}. The decomposition basis, known as celestial blocks~\cite{Kologlu:2019mfz}, organizes the descendant contributions associated with their corresponding light-ray operator. 

The explicit form of celestial blocks depends on the initial-state collision setup (e.g., the collision axis or the dimension and spin of local operator $\mathcal{O}$ in Eq.~\eqref{EEC_formula}), as well as the number of detectors used for the final-state measurement. The first celestial block was obtained for EEC inside scalar source~\cite{Kologlu:2019mfz}, as we will briefly review in Section~\ref{sec:ccb_intro}. It was later generalized to spinning source in~\cite{Chang:2020qpj} and to three-point energy correlator in~\cite{Chang:2022ryc,Chen:2022jhb}. 

In this section, we are going to consider EEC in the hadron collider setup and derive the corresponding celestial blocks. Compared with previous examples of celestial blocks, the hadron collider celestial blocks contain the information of collision axis of initial-state particles.

\subsection{Review of the celestial blocks for $e^+e^-$ annihilation} 
\label{sec:ccb_intro}

The celestial blocks for EEC inside the scalar source can be calculated by the Casimir differential equation~\cite{Kologlu:2019mfz}, which is essentially the same as computing conformal blocks in conformal field theories~\cite{Dolan:2003hv}. Celestial block basis nicely organizes the collinear limit of EEC, where the following light-ray OPE applies
\begin{equation}\label{eq:schematic_OPE}
    \mathcal{E}(n_1)\mathcal{E}(n_2) = \sum_{\delta,j} c_{\delta,j}\left[ \theta^{\delta-6}\mathbb{O}^{[J=3]}_{\delta,j}(n_2) + \text{celestial descendants of $\mathbb{O}^{[J=3]}_{\delta,j}$}
    \right]\,.
\end{equation}
Here $\theta$ is the opening angle between $\mathcal{E}(n_1)$ and $\mathcal{E}(n_2)$, $\delta$ and $j$ denote the celestial dimension and transverse spin of the (primary) light-ray operator $\mathbb{O}^{[J=3]}_{\delta,j}$, and $c_{\delta,j}$ is the OPE coefficient. $J$ is the spin of the corresponding local operator, which is fixed to be $3$ by conformal symmetry and will be modified if scaling invariance is broken~\cite{Chen:2023zzh,Chen:2024nyc}. Roughly speaking, the celestial descendants can be viewed as derivatives of the corresponding primary operator on the celestial sphere and they are the subleading contributions in the collinear limit $\theta\to 0$. 

Inserting light-ray OPE \eqref{eq:schematic_OPE} into the correlation function gives the celestial block decomposition
\begin{equation}
\frac{\langle \tilde{\mathcal{O}}(-q) \mathcal{E}(n_1) \mathcal{E}(n_2) \tilde{\mathcal{O}}^\dagger(q)\rangle}{\langle \tilde{\mathcal{O}}(-q)\tilde{\mathcal{O}}^\dagger(q)\rangle}
= \sum_{\delta,j} c_{\delta,j} f_{\delta,j}(n_1,n_2;q)\,,
\end{equation}
where $f_{\delta,j}$ is the sum of 1-point correlation function of $\mathbb{O}^{[J=3]}_{\delta,j}$ and its descendants
\begin{equation}
 f_{\delta,j}(n_1,n_2;q) =  \theta^{\delta-6} 
 \frac{\langle \tilde{\mathcal{O}}(-q) \mathbb{O}^{[J=3]}_{\delta,j}(n_2) \tilde{\mathcal{O}}^\dagger(q)\rangle}{\langle \tilde{\mathcal{O}}(-q)\tilde{\mathcal{O}}^\dagger(q)\rangle} + \text{celestial descendants}\,.
\end{equation}
In the center of mass frame, scalar source is invariant under rotation and hence its corresponding EEC receives no contribution from non-vanishing transverse spin light-ray operators.

On the other hand, based on dimensional analysis and boost property $\mathcal{E}(\lambda n)=\lambda^{-3} \mathcal{E}(n)$, EEC inside a scalar source has the functional form
\begin{equation}\label{eq:scalar_EEC_functional_form}
    \frac{\langle \tilde{\mathcal{O}}(-q) \mathcal{E}(n_1) \mathcal{E}(n_2) \tilde{\mathcal{O}}^\dagger(q)\rangle}{\langle \tilde{\mathcal{O}}(-q)\tilde{\mathcal{O}}^\dagger(q)\rangle}
    = \frac{(q^2)^{4}}{(n_1\cdot q)^3 (n_2\cdot q)^3} f(\zeta)\,,
\end{equation}
where $\zeta=\frac{(n_1\cdot n_2) q^2}{2(n_1\cdot q)(n_2\cdot q)}$. 
Note that we work in 4d spacetime here and the generic $d$-dimensional spacetime expressions can be found in~\cite{Kologlu:2019mfz}.
We rewrite $f_{\delta,j=0}(n_1,n_2;q)$ in the same way as
\begin{equation}
    f_{\delta,0}(n_1,n_2;q) = \frac{(q^2)^{4}}{(n_1\cdot q)^3 (n_2\cdot q)^3} f_{\delta,0}(\zeta)\,.
\end{equation}

The light-ray operator $\mathbb{O}^{[J=3]}_{\delta,j}$ and its descendants belong to the same Lorentz representation and obey the equation
\begin{equation}
    [C_2,\mathbb{O}^{[J=3]}_{\delta,j}(n)]=\frac{1}{2}[M_{\mu\nu},[M^{\mu\nu},\mathbb{O}^{[J=3]}_{\delta,j}(n)]] =\lambda_{\delta,j}\mathbb{O}^{[J=3]}_{\delta,j}(n)
\end{equation}
with $\lambda_{\delta,j}=\delta(\delta-2)+j^2$. The same equation holds if $\mathbb{O}^{[J=3]}_{\delta,j}$ is replaced by its descendants. $C_2=\frac{1}{2}M_{\mu\nu}M^{\mu\nu}$ is the quadratic Casimir operator for Lorentz group. The measurement of EEC transforms covariantly under Lorentz transformation and the action of $C_2$ on EEC is 
\begin{equation}
    \langle \tilde{\mathcal{O}}(-q) \left[C_2,\mathcal{E}(n_1) \mathcal{E}(n_2)\right]\tilde{\mathcal{O}}^\dagger(q)\rangle
    =-\frac{1}{2}\left(\mathcal{L}_{\mu\nu}(n_1)+\mathcal{L}_{\mu\nu}(n_2)\right)^2
    \langle \tilde{\mathcal{O}}(-q) \mathcal{E}(n_1) \mathcal{E}(n_2) \tilde{\mathcal{O}}^\dagger(q)\rangle\,,
\end{equation}
where we define $\mathcal{L}_{\mu\nu}(x)$ as the differential operator
\begin{equation}
    \mathcal{L}_{\mu\nu}(x)\equiv x_\mu \frac{\partial}{\partial x^\nu} - x_\nu \frac{\partial}{\partial x^\mu}\,.
\end{equation}
Therefore, the celestial block associated with $\mathbb{O}^{[J=3]}_{\delta,0}(n)$ satisfies the celestial Casimir equation
\begin{equation}
    -\frac{1}{2}\left(\mathcal{L}_{\mu\nu}(n_1)+\mathcal{L}_{\mu\nu}(n_2)\right)^2 f_{\delta,0}(n_1,n_2;q) = \lambda_{\delta,0} f_{\delta,0}(n_1,n_2;q)\,,
\end{equation}
or equivalently, an ordinary differential equation for $f_{\delta,0}(\zeta)$, 
\begin{equation}\label{eq:scalar_block}
    \left[4\zeta^2 (1-\zeta)\frac{d^2}{d\zeta^2} +4\zeta(6-7\zeta)\frac{d}{d\zeta}+12(2-3\zeta)\right]f_{\delta,0}(\zeta) = \delta(\delta-2) f_{\delta,0}(\zeta)\,.
\end{equation}
The solution of $f_{\delta,0}(\zeta)$ can be expressed in terms of hypergeometric function
\begin{equation}
    f_{\delta,0}(\zeta) = \zeta^{\frac{\delta}{2}-3} \,_2F_1(\delta/2, \delta/2,\delta;\zeta)\,,
    \label{eq:ee_celestial_block}
\end{equation}
Note that we have imposed the physical boundary condition $f_{\delta,0}\sim \zeta^{\frac{\delta}{2}-3}\sim \theta^{\delta-6}$ to remove another solution to Eq. \eqref{eq:scalar_block}.

Light-ray operators with transverse spin do not have corresponding scalar source celestial block due to mismatch of symmetry. Considering spinning source or higher point energy correlators is helpful to probe non-zero transverse spin~\cite{Chang:2020qpj,Chang:2022ryc,Chen:2022jhb}. As we will see soon, in the hadron collider configuration, the corresponding celestial block can also have transverse spin.

\subsection{Celestial blocks for hadron colliders}
\label{sec:ccb_hadron_collider}

EEC at hadron colliders can be expressed as the matrix element
\begin{equation}
   \frac{ \langle P_1 P_2 \left| \mathcal{E}(n_a) \mathcal{E}(n_b)\right| P_1 P_2\rangle }{\langle P_1 P_2 | P_1 P_2\rangle}\,.
\end{equation}
For proton-proton collisions, $P_1, P_2$ are the momenta of the incoming protons. In the hadron center of mass frame, we can choose the collision axis to be $z$-axis and hence the proton momenta are
\begin{equation}\label{eq:hadron_com}
    P_1^\mu = P n_1^\mu,\quad P_2^\mu = P n_2^\mu\,.
\end{equation}
where $n_1^\mu =(1,0,0,1),\,n_2^\mu=(1,0,0,-1)$ if the proton mass is neglected in high-energy collisions.

For simplicity, we assume these protons are unpolarized and the Lorentz invariance constrains EEC to be the function of 6 scalar products
\begin{equation}
   n_a\cdot n_b,\quad P_1\cdot P_2,\quad n_a\cdot P_1,\quad n_a\cdot P_2,\quad n_b\cdot P_1,\quad n_b\cdot P_2. 
\end{equation}
Applying the same dimensional analysis and Lorentz symmetry as used for Eq.~\eqref{eq:scalar_EEC_functional_form}, the general functional form of hadron collider EEC is
\begin{equation}
    \frac{ \langle P_1 P_2 \left| \mathcal{E}(n_a) \mathcal{E}(n_b)\right| P_1 P_2\rangle }{\langle P_1 P_2 | P_1 P_2\rangle} =
    \frac{P_1\cdot P_2}{(n_a \cdot n_b)^3} F(u,v,w)\,,
\end{equation}
where $u,v,w$ are the dimensionless combinations of Lorentz invariants that ensures the invariance under scaling $n_a\to \lambda_a n_a$ and $n_b\to \lambda_b n_b$:
\beq
\label{eq: uvw}
u=\frac{(n_a\cdot n_b)(n_1\cdot n_2)}{(n_1\cdot n_a)(n_2\cdot n_b)},
\quad v=\frac{(n_2\cdot n_a)(n_1\cdot n_b)}{(n_1\cdot n_a)(n_2\cdot n_b)},
\quad w=\frac{P_1 \cdot n_a}{P_2\cdot n_a}.
\eeq
$u,v$ can be regarded as the cross ratios formed by 4 null vectors $n_a,n_b,n_1,n_2$ on the celestial sphere. $w$ is related to the rapidity $\displaystyle Y=\frac{1}{2} \ln \frac{E+p_z}{E-p_z}$ through $\displaystyle w = e^{-2 Y_a}$ in the hadron center of mass frame \eqref{eq:hadron_com}. Expressing in terms of hadron collider variables, $v$ is the exponential of rapidity difference between $n_a$ and $n_b$
\begin{equation}
    v=e^{2 (Y_a-Y_b)}\,,
\end{equation}
and $u$ is a more complicated combination of rapidity and azimuthal angle differences
\begin{equation}
    u = 1-2 e^{Y_a-Y_b} \cos(\phi_a-\phi_b) +e^{2(Y_a-Y_b)}.
\end{equation}

Following the case of scalar source, light-ray OPE \eqref{eq:schematic_OPE} provides a nice decomposition of $F(u,v,w)$ in the collinear limit
\begin{equation}
    F(u,v,w) = \sum_{h,\bar{h}} F_{h,\bar{h}}(z,\bar{z},w)\,,
\end{equation}
where the labels $h, \bar{h}$ are related to celestial dimension $\delta$ and transverse spin $j$ through $h=\frac{\delta - j}{2},\bar{h}=\frac{\delta+j}{2}$. As is common in conformal theory, we also change the variable from $(u,v)$ to $(z,\bar{z})$
\begin{equation}
    u=z\bar{z}\,,\qquad v=(1-z)(1-\bar{z})\,,
\end{equation}
which will result in a compact form of celestial block.
In particular, $F_{h,\bar{h}}(z,\bar{z},w)$ satisfies the Casimir differential equation in a factorized way
\begin{equation}
   \left[ (1-z)z^2\partial_z^2+(w\partial_{w}-1)z^2\partial_z   
   +(z\leftrightarrow \bar{z})
   \right] F_{h,\bar{h}}(z,\bar{z},w)=\left( h(h-1) +\bar{h}(\bar{h}-1)\right) F_{h,\bar{h}}(z,\bar{z},w)\,.
   \label{eq:casimir_eq}
\end{equation}
Roughly speaking, we can think of $(\delta,j)$ or $(h,\bar{h})$ as the labels conjugate to variables $(z,\bar{z})$. The natural step then is to diagonalize the third variable $w$, with a label $\gamma$ that is independent of Casimir eigenvalue $\lambda_{\delta,j}=2(h(h-1)+\bar{h}(\bar{h}-1))$. We notice that the Casimir differential operator is homogeneous in $w$ due to the combination $w \partial_w$. Therefore the simplest way to diagonalize is through Mellin transformation:
\begin{equation}
   F_{h,\bar{h}}(z,\bar{z},w) = \int \frac{d\gamma}{2\pi i} c_{\delta, j, \gamma} w^{\gamma} G_{\delta, j}^{(\gamma)} (z,\bar{z})
\end{equation}
Based on this ansatz, the solution to the Casimir equation \eqref{eq:casimir_eq} is given by 
\beq
G^{(\gamma)}_{\delta,j}(z,\bar{z})= 
\left[z^{h} \, _2F_1(h,h-\gamma ;2 h;z)\right] 
\left[ \bar{z}^{\bar{h}} \, _2F_1(\bar{h},\bar{h}-\gamma ;2 \bar{h};\bar{z})\right] + (z\leftrightarrow \bar{z})\,.
\label{eq:celestial_blocks_def}
\eeq
This coincides with the conformal block in $d=2$~\cite{Belavin:1984vu}.
 
In this paper, we define the celestial blocks for hadron colliders as
\begin{equation}
    F_{\delta,j,\gamma}(z,\bar{z},w) = w^\gamma G^{(\gamma)}_{\delta,j}(z,\bar{z})\,,
    \label{eq:celestial_blocks_def1}
\end{equation}
and dynamical information is included in the coefficients $c_{\delta,j,\gamma}$ in the celestial block decomposition
\begin{equation} \label{eq:celestial_block_decomposition}
    F(z,\bar{z},w) = \sum_{\delta,j} \int \frac{d\gamma}{2\pi i} c_{\delta, j, \gamma} F_{\delta,j,\gamma}(z,\bar{z},w)\,.
\end{equation}

\begin{figure}[ht]
    \centering
    \includegraphics[width=0.4\textwidth]{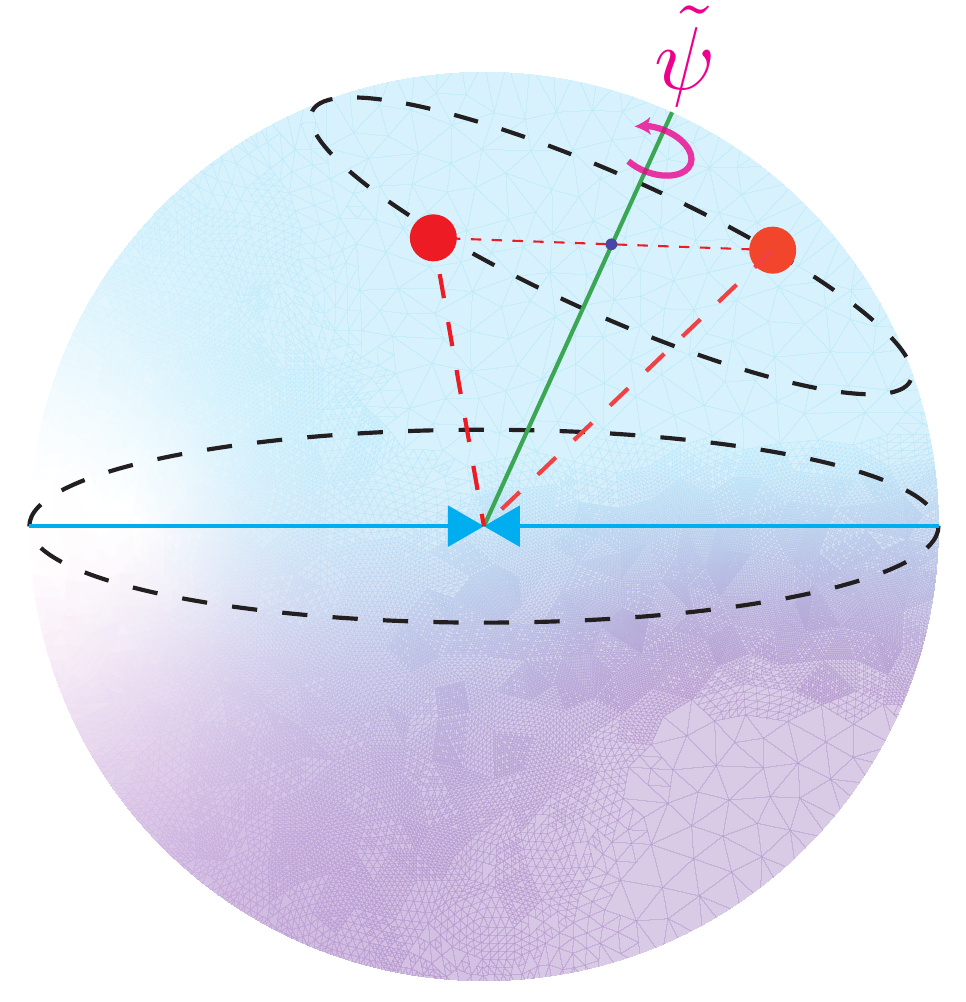}
    \caption{The celestial block decomposition is most naturally understood in the collinear limit, where the two detectors (red dots) are close to each other. In this limit, the opening angle between the two detectors is controlled by \( |z| \), while its phase $\arg z$, denoted by \( \psi \) in Sec.~\ref{sec:block_expansion}, can be intuitively viewed as the azimuthal angle \( \tilde\psi \) around the green axis. Although \( w \) is defined to be the rapidity of one detector, it effectively represents the rapidity of the green line in the collinear limit. Thus, the decomposition can be viewed as replacing the detector pair with a tower of operators of various celestial twists and spins along the green line. Away from the collinear limit, the geometric meaning of variables $z,\bar{z}, w$ becomes obscure.}
    \label{fig:block decomposition}
\end{figure}

To gain an intuition about the variables in the celestial blocks, we illustrate the collision and collinear measurement configuration in Fig.~\ref{fig:block decomposition}. In the collinear limit, the leading behaviors of $|z|, \arg z, w$ reveal three key geometric relations: (1) $|z|$ is proportional to the detector opening angle, (2) $\arg z$ corresponds to the azimuthal angle $\tilde{\psi}$ in Fig.~\ref{fig:block decomposition}, and (3) the variable $w$ encodes the orientation of the detector pair relative to the collision axis. 

In hadron colliders, the dominant contributions in high-energy hard scatterings often come from partonic collisions. Their probability distributions in momentum fraction $x$ are described by PDF $f_a(x;\mu)$. In this section, we suppress the scale dependence in PDF for simplicity (see, e.g.~\cite{Collins:1989gx,Collins:2011zzd,Ellis:1996mzs} for details). EEC schematically factorized as
\begin{align}
    \label{eq:hadron EEC}
    &\langle P_1 P_2 \left| \mathcal{E}(n_a) \mathcal{E}(n_b)\right| P_1 P_2\rangle = \nonumber\\
    &\qquad \sum_{\alpha,\beta}\int_0^1 dx_1 dx_2 \; x_1 x_2 f_{\alpha}(x_1) f_{\beta}(x_2) \langle \alpha(p_1) \beta(p_2) \left| \mathcal{E}(n_a) \mathcal{E}(n_b)\right| \alpha(p_1) \beta(p_2)\rangle\Big|_{p_1=x_1 P_1, p_2=x_2 P_2} \,.
\end{align}
Compared with hadron center of mass frame, the parton center of mass frame is relatively boosted along $z$-axis due the unequal momentum fractions. We notice that $(u,v)$ or $(z,\bar{z})$ is invariant under the $z$-axis boost, while $w$ in these two frames are related by
\begin{equation}
    w_{\text{Parton}} = \frac{x_1}{x_2} w_{\text{Hadron}}\,.
\end{equation}
For parton collision matrix element $\langle \alpha(p_1) \beta(p_2) \left| \mathcal{E}(n_a) \mathcal{E}(n_b)\right| \alpha(p_1) \beta(p_2)\rangle$, we choose the celestial block decomposition convention
\begin{equation}
    \label{eq:parton block}
    \langle \alpha(p_1) \beta(p_2) \left| \mathcal{E}(n_a) \mathcal{E}(n_b)\right| \alpha(p_1) \beta(p_2)\rangle
    = \frac{1}{(n_a\cdot n_b)^3} \sum_{\delta,j} \int \frac{d\gamma}{2\pi i} \tilde{c}^{\alpha\beta}_{\delta, j, \gamma} F_{\delta,j,\gamma}(z,\bar{z},w_{\text{Parton}})\,.
\end{equation}
The hadron block coefficients $c_{\delta,j,\gamma}$ can be obtained from 
the parton block coefficients $\tilde{c}^{\alpha\beta}_{\delta,j,\gamma}$ after convolution with parton distribution functions
\begin{equation}
    c_{\delta,j,\gamma} = \sum_{\alpha,\beta}\int_0^1 dx_1 dx_2 \;\frac{x_1^{1+\gamma} x_2^{1-\gamma} f_{\alpha}(x_1) f_{\beta}(x_2) \tilde{c}^{\alpha\beta}_{\delta,j,\gamma}}{(P_1\cdot P_2)\,\langle P_1 P_2 | P_1 P_2\rangle}
    = \sum_{\alpha,\beta} \frac{\tilde{f}_\alpha(1+\gamma) \tilde{f}_\beta(1-\gamma) \tilde{c}^{\alpha \beta}_{\delta,j,\gamma}}{(P_1\cdot P_2)\,\langle P_1 P_2 | P_1 P_2\rangle}\,,
\end{equation}
where $\tilde{f}_\alpha(N)$ is the moment of $f_\alpha(x)$
\begin{equation}
    \tilde{f}_\alpha(N)=\int_0^1 dx\, x^{N} f_\alpha(x)\,.
\end{equation}
We notice the simplicity of block coefficient $c_{\delta,j,\gamma}$ for only involving two moments of PDFs $\tilde{f}_\alpha(1+\gamma), \tilde{f}_\beta(1-\gamma)$. The label $\gamma$ is related to the partial wave decomposition at cross-section level, organized by collinear spin along the collision axis. To see this, we insert complete states $1=\sum_{J, i} |\Psi_{J}^{(i)}\rangle \langle \Psi_{J}^{(i)} |$ twice
\begin{equation}
    \langle P_1 P_2 \left| \mathcal{E}(n_a) \mathcal{E}(n_b)\right| P_1 P_2\rangle
    = \sum_{J_1, i_1; J_2 , i_2} \langle P_1 P_2 |\Psi_{J_1}^{(i_1)}\rangle \langle \Psi_{J_1}^{(i_1)} \left| \mathcal{E}(n_a) \mathcal{E}(n_b)\right| \Psi_{J_2}^{(i_2)}\rangle \langle \Psi_{J_2}^{(i_2)}| P_1 P_2\rangle\,,
\end{equation}
where $J_{1,2}$ are collinear spins and $i_{1,2}$ are other possible labels. Applying collinear boost $\Lambda_Y = e^{-i Y \hat{z}\cdot \vec{\mathbf{K}}}$, 
 each term in the matrix element is an eigenfunction under such transformation
\begin{align}
    \langle P_1 P_2 |\Lambda_Y^{-1}|\Psi_{J_1}^{(i_1)}\rangle 
    \dots
    \langle \Psi_{J_2}^{(i_2)}|\Lambda_Y| P_1 P_2\rangle 
    =  e^{-(J_1-J_2)Y} \langle P_1 P_2 |\Psi_{J_1}^{(i_1)}\rangle 
    \dots
    \langle \Psi_{J_2}^{(i_2)}| P_1 P_2\rangle\,.
\end{align}
We can regard $|\Psi_{J_1}^{(i_1)}\rangle  \langle \Psi_{J_2}^{(i_2)}|$ as a ``light-ray transition matrix'' with collinear spin $J_1-J_2$ along collision axis, which generalizes the concept ``light-ray density matrix'' $|\Psi\rangle  \langle \Psi|$ in~\cite{Chang:2022ryc}. 

On the other hand, $\Lambda_Y$ keeps $u,v$ invariant and rescales $w \to e^{-2Y} w$. From this, we can conclude that the label $\gamma$ in the celestial block $G_{\delta,j}^{(\gamma)}$ corresponds to interference effect of states with different collinear spins. As an example, in the next section, we will see in the decomposition of pure gluon EEC, the $\gamma$ poles in $c_{\delta,j,\gamma}^{gg}$ are located at integer values, indicating there is no interference between odd and even collinear spins or the odd-spin ``light-ray transition matrices'' $|\Psi_{J_1}^{(i_1)}\rangle  \langle \Psi_{J_2}^{(i_2)}|$ do not contribute.

\section{Application of collider celestial block to gluon scattering}
\label{sec:gluon}
 
This section presents the calculation of the EEC for exclusive gluon scattering processes within the framework of perturbative quantum field theory. While modern collider experiments such as those at the Large Hadron Collider (LHC) achieve proton-proton collisions at center-of-mass energies of multi-TeV, the characteristic energy scale for individual partonic interactions typically resides in the GeV regime. This corresponds to relatively small momentum fraction carried by the interacting parton. In this case, gluonic degrees of freedom predominantly constitute the initial state for perturbative QCD processes due to the enhanced PDFs associated with gluons at small $x$ values \cite{Dulat:2015mca,NNPDF:2017mvq}.

Moreover, in the large $N_c$ limit, the scattering process is dominated by pure gluon scattering due to the enhancement of the color factors~\cite{tHooft:1973alw,Witten:1979kh}. In reality, \(N_c = 3\) and the presence of multiple quark flavors enhances the contribution of quark-involving processes, but pure gluon scattering still captures many essential features of QCD dynamics and provides a clean setup for theoretical understanding of observables.
We therefore focus exclusively on the pure gluon process in what follows, keeping in mind that it serves as a good approximation and highlights important QCD dynamics.

\begin{figure}
    \centering
    \includegraphics[width=\textwidth]{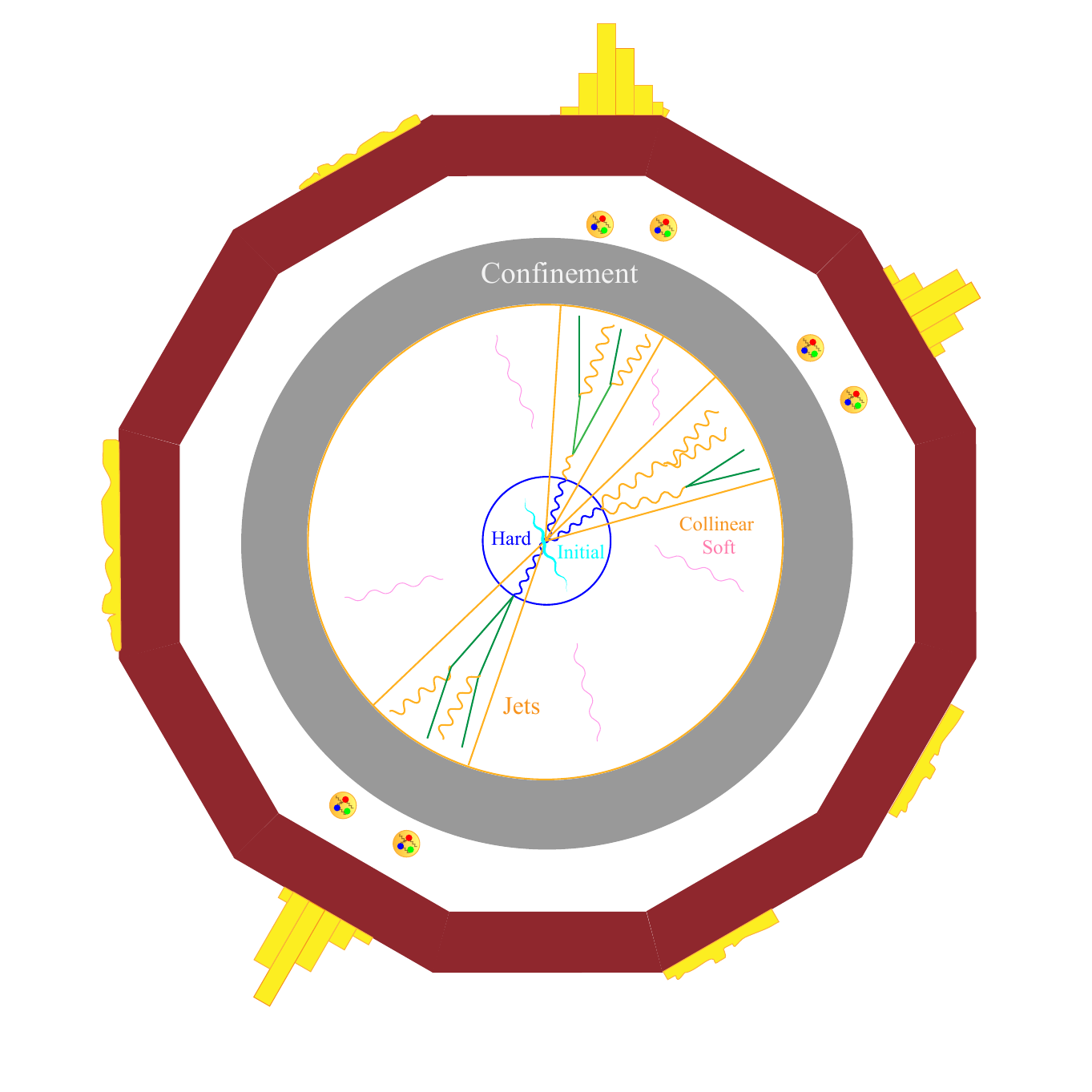}
    \caption{Schematic cross-sectional view of the scattering process inside a hadron collider. The light blue wavy lines labeled "Initial" represent incoming partons (e.g., gluons) originating from the colliding hadrons and moving perpendicular to the plane of the figure. Through a hard scattering event, these initial partons produce final-state partons (such as gluons), which subsequently undergo jet formation via collinear splitting and soft radiation. Their energies gradually decrease until they reach the QCD confinement scale, \(\Lambda_{\text{QCD}}\), at which point hadronization occurs. The resulting hadrons ultimately reach the detector, where their energies are measured, as illustrated by the yellow bars in the figure.}
    \label{fig:cross_plot}
\end{figure}

A typical collision experiment is illustrated in Fig.~\ref{fig:cross_plot}. Initial-state gluons scatter at a relatively high energy scale (e.g., several hundred GeV). Due to the asymptotic freedom of QCD~\cite{Gross:1973id,Gross:1973ju,Gross:1974cs,Politzer:1973fx}, the coupling constant \(\alpha_s(\mu)=\frac{g(\mu)^2}{4\pi}\) is sufficiently small at this scale, ensuring the reliability of perturbative calculations. At leading order in the coupling \(\alpha_s\), the dominant scattering process is \(gg \to gg\), where the final-state gluons are produced exactly back-to-back in the parton center of mass frame. These energetic gluons then initiate collinear splitting and soft radiation, leading to the formation of a dijet structure. Throughout this evolution, although the energy of particles within the jets gradually decreases, it remains within a regime that is accessible to perturbative methods.
Collinear splitting and soft radiations often generate large logarithmic terms in perturbation theory, which require proper resummation. This can be systematically achieved using Soft-Collinear Effective Theory (SCET)~\cite{Bauer:2000ew,Bauer:2000yr,Bauer:2001ct,Bauer:2001yt}.
As the parton shower progresses, the energy scale continues to decrease until it reaches the scale of \(\Lambda_{\text{QCD}}\), at which point perturbative methods are no longer applicable. At this scale, partons undergo a dynamic confinement transition (hadronization), forming hadrons that are subsequently measured by the detectors. 

The EEC in hadron colliders, as defined in Eq.~\eqref{HEEC}, can be viewed as a function of the solid angles of two detectors, thereby capturing the correlation structure between particles moving along these two directions. The angle \(\chi\) between the two detectors is essentially related to the time and energy scales at which the interactions between the particles took place. This idea was proposed for tomography final-state QCD dynamics using EEC~\cite{Komiske:2022enw} and can be used for $\alpha_s$ measurement and studying hadronization transition in jet substructures~\cite{CMS:2024mlf,Lee:2024esz}.
In particular, when \(\chi\) is extremely small, the dominant correlation between the two hadrons occurs after confinement, where only residual nuclear interactions are present, allowing the hadrons to be approximated as independent particles. Under these conditions, the EEC \(\langle\mathcal{E}(\vec{n}_1)\mathcal{E}(\vec{n}_2)\rangle\) is approximately factorized into \(\langle\mathcal{E}(\vec{n}_1)\rangle \langle\mathcal{E}(\vec{n}_2)\rangle\), reflecting a very small correlation (beyond weak residual nuclear forces) between particles in the two directions. As \(\chi\) increases to the order of \(\Lambda_{\text{QCD}}/Q\), the interaction time approaches the hadronization process, thus probing the associated non-perturbative dynamics where the perturbative QCD framework is not applicable.

With a further increase in \(\chi\), such that \(1 \gg \chi \gg \Lambda_{\text{QCD}}/Q\), the relevant timescale of the correlation enters the parton shower regime, and the corresponding energy scale remains much above the confinement scale \(\Lambda_{\text{QCD}}\). This is the perturbative collinear regime, where collinear factorization and resummation techniques are applicable~\cite{Dixon:2019uzg,Konishi:1979cb,Konishi:1978yx,Konishi:1978ax,Kalinowski:1980wea,Richards:1982te,Chen:2023zzh}. We will briefly discuss this limit in Sec.~\ref{sec:collinear} and present the LP result obtained via factorization approach.

For a generic angle $0<\chi <\pi$, away from the end points, this correlation mainly probes the dynamics of hard scattering processes. For the $gg\to gg$ process, the final-state gluons are exactly back-to-back in the parton center of mass frame, which do not generate a non-vanishing correlation in the bulk region $\chi\in (0,\pi)$. Consequently, the leading-order contribution comes from $gg\to ggg$ for pure gluon scattering. Increasing the precision in the bulk region requires higher-loop calculations.

Finally, as \(\chi\) approaches \(\pi\), the EEC enters the back-to-back regime, which will be discussed in Sec.~\ref{sec:b2b}. In this regime, collinear splitting and soft radiation once again dominate the EEC contribution. An alternative approach to understanding this limit is through large spin physics~\cite{Korchemsky:2019nzm, Chen:2023wah}. This behavior parallels that of the EEC in $e^+ e^-$ colliders, where factorization and resummation are well-understood~\cite{Moult:2018jzp} and has achieved $\mathrm{N^4LL}$ resummation in~\cite{Duhr:2022yyp}. 

However, unlike the EEC in spherically symmetric sources, the EEC at hadron colliders preserves only the azimuthal rotation symmetry around the beam axis. As a result, it exhibits an additional dependence on the rapidities of the two detectors. This leads to two other interesting limits: the opposite coplanar limit and the Regge limit, which will be discussed in Sec.~\ref{sec:coplanar} and Sec.~\ref{sec:Regge}, respectively.

In the following subsections, we focus on analyzing the EEC in the vicinity of the hard scattering event. For purposes of experimental comparison, one may first reconstruct the hard scattering process from the data and then compare the resulting observables with theoretical predictions.

\subsection{Analytic calculation of EEC for gluon scattering}
\label{sec:result}

The EEC at hadron colliders defined previously in Eq.~\eqref{HEEC} is formulated for the complete proton-proton scattering process. In this paper, however, we restrict our focus to the pure gluon scattering component and perform perturbative calculations in the parton center-of-mass frame. As mentioned, for generic detector configurations, considering only the \( gg \to gg \) hard scattering process is insufficient, as it results in a nonzero EEC distribution only when the two directions are strictly coincident or back-to-back. Consequently, the first nontrivial contribution to the EEC arises in the \( gg \to ggg \) tree-level process, whose scattering amplitude can be described by the Parke-Taylor formula~\cite{Parke:1986gb,Elvang:2013cua}.

In the language of the spinor-helicity formalism, the color-ordered 5-gluon tree amplitude can be expressed as
\begin{equation}
    \label{mhv}
    A_5[1,2,3,4,5] = \frac{
        \langle i j \rangle^4
    }{
        \langle 1 2 \rangle 
        \langle 2 3 \rangle 
        \langle 3 4 \rangle 
        \langle 4 5 \rangle 
        \langle 5 1 \rangle
    }.
\end{equation}
Here, we consider all gluons to be outgoing, with \( i \) and \( j \) denoting the gluons of helicity \(-1\), while the remaining three gluons have helicity \(+1\).

Using the color decomposition formula, the full 5-gluon tree amplitude can be expressed as:
\begin{equation}
    \label{amp}
    \mathcal{A}_5^{\text{full,tree}} = g^{3}\sum_{\sigma\in S_{4}}
    \text{Tr}(T^{a_1}T^{\sigma(a_2)}T^{\sigma(a_3)}T^{\sigma(a_4)}T^{\sigma(a_5)}) A_5[1,\sigma(2,3,4,5)],
\end{equation}
where \( a_i \) denotes the color index of the \(i\)-th gluon, and \( T^{a_i} \) are the corresponding color matrices normalized such that \( \text{Tr}[T^a T^b] = \delta^{ab} \).

By summing over all color and helicity configurations and averaging over the color indices and helicities of the initial particles 1 and 2, we obtain the averaged squared amplitude for five-gluon scattering. After some calculations, we find that this can be expressed in the following compact form:
\begin{equation}
    \label{sqaured_amp}
    \overline{\sum_{\text{h,c}}
        \left| \mathcal{A}_5^{\text{full,tree}} \right|^2}        
    =  \frac{27g^6}{16} 
        \frac{ \sum\limits_{1\leq i<j \leq 5} s_{ij}^4}{\prod\limits_{1\leq i<j \leq 5}s_{ij}} 
        \sum_{\sigma \in S_4} 
        \left( 
            s_{1 \sigma(2)}
            s_{\sigma(2)\sigma(3)}
            s_{\sigma(3)\sigma(4)}
            s_{\sigma(4)\sigma(5)}
            s_{\sigma(5)1}
        \right),
\end{equation}
where \( s_{ij} = (p_i + p_j)^2 \), and \( p_i \) denotes the momentum of the \(i\)-th gluon. The notation `h' and `c' refers to the helicity and color configurations of the gluons, respectively, and the overline denotes averaging over the color and helicity configurations of the initial particles 1 and 2.

Based on the averaged squared amplitude for five-gluon scattering, we replace the differential cross section for complete proton scattering, \( d\sigma_{pp\to i+j+X} \), in Eq.~\eqref{HEEC} with the tree-level differential cross section for five-gluon scattering, \( d\sigma_{gg\to ggg} \). The LO EEC for gluon scattering can then be expressed as
\begin{equation}
    \label{gluon_eec_formula}
    \frac{d^2\Sigma}{d\Omega_a d\Omega_b} =
    \frac{1}{2Q^2} \int d\Pi_3 \,\frac{1}{3!}
    \overline{\sum_{\text{h,c}} \left| \mathcal{A}_5^{\text{full,tree}} \right|^2}
    \sum\limits_{i,j=3}^{5}
    p_i^0 p_j^0 
    \delta^{(2)}(\Omega_a - \Omega_{p_i}) \delta^{(2)}(\Omega_b - \Omega_{p_j}),
\end{equation}
where \( \Omega_{a,b} \) and \( \Omega_{p_i} \) represent the solid angles of the two detectors and the \(i\)-th gluon, respectively, and \( Q^\mu \) denotes the parton center-of-mass momentum. The factor of $3!$ in front of the squared scattering amplitude arises due to the indistinguishability of the three final-state gluons. The notation \( \int d\Pi_3 \) refers to the phase space integration for the final state gluons, given by
\begin{equation}
    \label{PhaseSpace}
    \int d\Pi_3 = \left( \prod\limits_{i=3}^{5} \int \frac{d^3 p_i}{(2\pi)^3 2p_i^0} \right)
    (2\pi)^4 \delta^{(4)}\left( Q - \sum\limits_{i=3}^{5} p_i \right).
\end{equation}

\begin{figure}
    \centering
    \includegraphics[width=\textwidth]{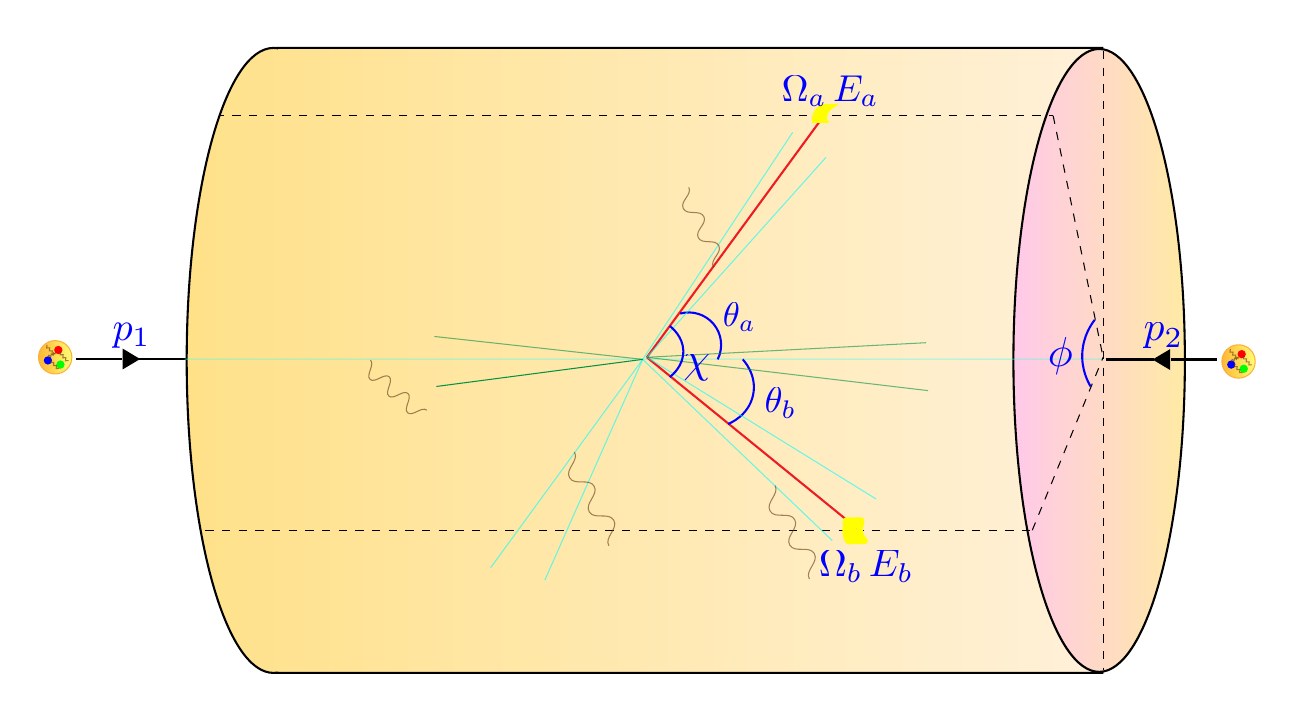}
    \caption{Schematic illustration of the energy correlation function at the hadron collider. Two hadrons with momenta \( p_1 \) and \( p_2 \), moving in opposite directions, collide and produce outgoing particles. Detectors positioned on the cylindrical surface record the energies of particles passing through them. The EEC is defined as the correlation function between a pair of such calorimeters, located at solid angles \( \Omega_a \) and \( \Omega_b \). Using the rotation symmetry around the beam axis, a configuration is specified by two polar angles \( \theta_a \) and \( \theta_b \), and a relative azimuthal angle \( \phi \). The parameter \( \chi \) denotes the opening angle between the two detectors. In this section, we perform perturbative QCD calculations at the parton level in the parton center of mass frame; hence, all geometric parameters shown in the figure correspond to this frame.}
    \label{fig.collider}
\end{figure}

After calculation, we find that Eq.~(\ref{gluon_eec_formula}) can be reduced to a single-variable integral:
\begin{equation}
    \label{eec_int}
    \frac{d^2\Sigma}{d\Omega_a d\Omega_b} =
    \frac{Q^2}{16384 \pi^5} \int_0^1  d x \, \frac{(1 - x)^2 x^2}{(1 - x \zeta)^3}
    \overline{\sum_{\text{h,c}} \left| \mathcal{A}_5^{\text{full,tree}} \right|^2},
\end{equation}
where we define \( \displaystyle x = \frac{2p_3^0}{Q^0}   \), and \(\displaystyle  \zeta =\frac{1 - \cos \chi}{2}  \), with \( \chi \) denoting the angle between the two detectors. In the above calculation, we have assumed that \( p_3 \) is aligned with the direction of \( \Omega_a \) and \( p_4 \) with \( \Omega_b \), which cancels out the symmetry factor \( \frac{1}{3!} \). See Fig.~\ref{fig.collider} for an illustration.

The explicit form of the averaged squared amplitude can be expressed in terms of \( x \), \( Q \), and other kinematic variables illustrated in Fig.~\ref{fig.collider}. Substituting this into Eq.~(\ref{eec_int}) and performing the integration, we obtain the following expression for the EEC in pure gluon scattering at hadron colliders:
\begin{equation}
        \begin{aligned}
        \label{full result}
        \frac{d^2\Sigma}{d\Omega_a d\Omega_b}&=\frac{9 g^6}{(8 \pi) ^5 (c_{\Delta Y}-c_{\phi }){}^5(c_{\Delta Y}+c_Y) (c_Y+c_{\phi }) (c_{\Delta Y} c_{\phi }+c_{\Delta Y} c_Y-c_Y c_{\phi }-1)}\\ 
        &\times\left[
        C_0+C_1\, \ln \left(\frac{c_Y+c_{\phi }}{c_{\Delta Y}+c_Y}\right)+C_2\, \ln \left(\frac{\left(c_{\Delta Y}+s_{\Delta Y}\right) \left(c_Y+s_Y\right)+1}{c_{\Delta Y}+c_Y+s_{\Delta Y}+s_Y}\right)s_{\Delta Y} s_Y\right.\\
        &\qquad\left.+C_3\, \phi  \csc (\phi )+C_4\,\Delta Y s_{\Delta Y}+C_5\,  \arccos\left(\frac{c_Y c_{\phi }+1}{c_Y+c_{\phi}}\right)\csc (\phi )s_Y\right].
    \end{aligned}
\end{equation}
Here, \( 0 \leq \phi \leq \pi \) denotes the azimuthal angle separation between the two detectors in the transverse plane. We also introduce the shorthand notations \( c_{\phi} = \cos(\phi) \), \( c_{Y} = \cosh(Y) \), \( c_{\Delta Y} = \cosh(\Delta Y) \), \( s_{Y} = \sinh(Y) \), and \( s_{\Delta Y} = \sinh(\Delta Y) \), where \( Y \) and \( \Delta Y \) denote the rapidity sum and rapidity difference of the two detectors, respectively, and are defined such that \( Y \geq 0 \) and \( \Delta Y \geq 0 \):
\begin{equation}
    \label{Y&DeltaY}
    Y = \frac{1}{2}\ln\left(\frac{1+\cos\theta_a}{1-\cos\theta_a}
    \frac{1+\cos\theta_b}{1-\cos\theta_b}\right),
    \quad
    \Delta Y = \frac{1}{2}\ln\left(\frac{1+\cos\theta_a}{1-\cos\theta_a}
    \frac{1-\cos\theta_b}{1+\cos\theta_b}\right).
\end{equation}
The coefficients \(C_0\) through \(C_5\) that appear in Eq.~\eqref{full result} are complicated polynomials in \(c_{\phi}\), \(c_{Y}\), and \(c_{\Delta Y}\), which are given in Appendix~\ref{sec:appendixA}.

Another parametrization of this result is useful for discussing the redundancy in the rational terms:
\begin{equation}
    \begin{aligned}
        \label{eq:full result in other form}
        \frac{d^2\Sigma}{d\Omega_a d\Omega_b}&=\frac{9 g^6}{8 \left(8\pi\,(1-y_a^2)\,(1-y_b^2)\,\zeta\right) ^5 (1-\zeta)    \left(\left(y_a-y_b\right)^2+4 \zeta\, y_a\, y_b\right)
       \left(\left(y_a-y_b\right)^2-4 \zeta\, (1-y_a\, y_b-\zeta)\right)}\\
       &\times\left[A_0\ln(1-\zeta)+A_1\ln\left(\frac{1-y_b}{1-y_a}\right)+A_2\arccos\left(\frac{2-y_a-y_b-2\zeta}{2\sqrt{(1-y_a)(1-y_b)(1-\zeta)}}\right)+A_3\right]\\
       &+\left(y_a\leftrightarrow y_b\right)+\left(y_a\rightarrow -y_a,\,y_b\rightarrow -y_b\right)+\left(y_a\rightarrow -y_b,\,y_b\rightarrow -y_a\right).
    \end{aligned}
\end{equation}
Here, \(y_a = \cos\theta_a\) and \(y_b = \cos\theta_b\), where \(\theta_a\) and \(\theta_b\) represent the angles between the detectors \(a\) and \(b\) and the beam axis in the center-of-mass frame of parton, respectively, as illustrated in Fig.~\ref{fig.collider}. The coefficients \(A_0\) through \(A_3\) are again complicated polynomials of \(y_a\), \(y_b\), and \(\zeta\), whose explicit expressions are provided in Appendix~\ref{sec:appendixA}.

An intriguing property of our EEC result becomes more evident when written in the form of Eq.~\eqref{eq:full result in other form}, which we now explain in detail. The coefficients of the three transcendental terms, $A_0$, $A_1$, and $A_2$, although algebraically complicated, are in fact directly related to the residues of simple poles of the integrand. As a result, they can be extracted in a relatively straightforward manner. By contrast, the rational term $A_3$ is considerably more difficult to determine, making it both natural and worthwhile to ask whether it can be bootstrapped once the transcendental pieces are fixed. This question is further motivated by the case of the $e^+e^-$ EEC, where the rational part is uniquely determined by the coefficients of the transcendental terms together with the limiting behaviors at $\zeta \to 0,1,\infty$. Although the physical range of $\zeta$ is $[0,1]$, the asymptotics at $\zeta \to \infty$ can be understood by analytic continuation, with the leading behavior obtained by approximating the integrand before integration. Inspired by this analogy, we explore whether a similar bootstrap mechanism applies to the hadron-collider case. We will see that while $A_3$ cannot be uniquely fixed by the available constraints, its structure is nonetheless highly restricted, and a large subset of its coefficients can indeed be determined.

Some of the constraints on $A_3$ come from the fact that the EEC is free of unphysical singularities; for example, in the collinear limit no poles stronger than $1/\zeta$ appear. Moreover, the singular behavior in unphysical regions obtained via analytic continuation provides even stronger constraints on the polynomial $A_3$. The physical region for $y_a$, $y_b$, and $\zeta$ is given by
\begin{equation}
    \label{eq:physical region}
    -1 \leq y_a \leq 1, \quad -1 \leq y_b \leq 1, \quad \frac{1 - y_a y_b - \sqrt{1 - y_a^2} \sqrt{1 - y_b^2}}{2} \leq \zeta \leq \frac{1 - y_a y_b + \sqrt{1 - y_a^2} \sqrt{1 - y_b^2}}{2}.
\end{equation}
We can analytically continue these variables beyond the physical region to impose further constraints. In the limit $\zeta \to \infty$, the EEC scales as $1/\zeta$, while in the limits $y_a \to \infty$ or $y_b \to \infty$ it scales as $y_a^3$ or $y_b^3$, respectively. These asymptotics can be inferred from approximating the integrand prior to integration, and they motivate the following rational ansatz for $A_3$, consistent with the limiting behaviors:
\begin{equation}
    A_3 = \zeta (1 - y_a^2) (1 - y_b^2) \left( (y_a - y_b)^2 + 4 \zeta y_b y_a \right) \left( (y_a - y_b)^2 - 4 \zeta (1 - \zeta - y_a y_b) \right) \sum\limits_{n_1,n_2,n_3} c_{n_1,n_2,n_3} \zeta^{n_1} y_a^{n_2} y_b^{n_3},
\end{equation}
where the summation runs over integers $n_1 \in [0,4]$ and $n_2, n_3 \in [0,11]$. For simplicity, we have incorporated into our ansatz for $A_3$ the symmetrized terms under the exchange $y_a \leftrightarrow y_b$, etc. Consequently, the coefficients $c_{n_1,n_2,n_3}$ are symmetric under $n_2 \leftrightarrow n_3$, and vanish when $n_2+n_3$ is odd. These relations reduce the naive count of independent coefficients from 720 to 210.

The most stringent constraint comes from the $\zeta \to 0$ limit. In the physical region this corresponds to the collinear configuration $y_a \to y_b$. Remarkably, the EEC remains free of singularities stronger than $1/\zeta$ even if $y_a$ and $y_b$ are treated as independent variables rather than being set equal. Imposing this condition uniquely determines the coefficients $c_{n_1,n_2,n_3}$ for all terms with $n_1 \leq 2$.

This leaves 84 coefficients undetermined, which can be further constrained by examining additional limiting behaviors. In the limits $y_a \to \pm 1$ (or $y_b \to \pm 1$), imposing the physical condition $y_b \to \pm (1 - 2\zeta)$ (or $y_a \to \pm (1 - 2\zeta)$) ensures that no singularities stronger than $1/(1 \mp y_a)$ (or $1/(1 \mp y_b)$) arise. Further constraints are obtained by requiring that our ansatz for $A_3$ reproduces the same residues of the leading divergence in both the physical collinear limit $\zeta \to 0$ and the back-to-back limit $\zeta \to 1$ (where "physical" refers to the region defined in Eq.~\eqref{eq:physical region}). Likewise, we demand that the residues of the leading divergence in the physical limits $y_a \to \pm 1$ (or $y_b \to \pm 1$) match those of the full result. Analogous conditions can also be applied in the Regge limit, characterized by $y_a \to 1$, $y_b \to -1$, and $z \to 1$. Collectively, these physical constraints yield 70 additional conditions. 
Although they are not sufficient to fully bootstrap the functional form of $A_3$, we find that the rational structure of \( A_3 \) is subject to strong restrictions.

\subsection{Expansion of EEC in collider celestial block}
\label{sec:block_expansion}

Using the celestial block defined in Sec.~\ref{sec:ccb_hadron_collider}, we can expand our EEC result Eq.~\eqref{full result} in the OPE limit (collinear limit). This expansion serves as an alternative representation of our result, which can also provide an independent consistency check once we identify the operators appearing in the OPE.

The EEC defined in Eq.~\eqref{gluon_eec_formula} is a concrete example of the general definition used in Sec.~\ref{sec:ccb_hadron_collider}, $\langle \alpha(p_1) \beta(p_2) \left| \mathcal{E}(n_a) \mathcal{E}(n_b)\right| \alpha(p_1) \beta(p_2)\rangle$, where $\alpha$ and $\beta$ now denote the incoming gluons. To expand the EEC in the collider celestial block, we first need to change the variables in Eq.~\eqref{full result} to the Lorentz invariant variables defined in Eq.~\eqref{eq: uvw} via
\begin{equation}
    u = 1 - 2e^{\Delta Y}\cos\phi + e^{2\Delta Y}, \quad v = e^{2\Delta Y}, \quad w_p = e^{-Y - \Delta Y}.
\end{equation}
Here, the subscript in $w_p$ is an abbreviation for "parton", emphasizing that $w$ is defined by the parton momentum.

Before performing the celestial block decomposition, we first strip off overall kinematic factor and define:
\beq
    \label{eq:eec_relate_Fgg}
    \tilde{F}^{gg}(r,t,w_p) = (n_a \cdot n_b)^3 \langle g(p_1) g(p_2) \left| \mathcal{E}(n_a) \mathcal{E}(n_b) \right| g(p_1) g(p_2) \rangle
\eeq
Note that we have parametrized \( z = r t \) and \( \bar{z} = r/t \) to analyze the collinear limit $r\sim \chi\to 0$ more conveniently.
Following Eq.~\eqref{eq:parton block}, \( \tilde{F}^{gg}(r,t,w_p) \) admits the following celestial block decomposition:
\beq
    \tilde{F}^{gg}(r,t,w_p) = \sum_{\delta,j}\tilde{F}_{\delta,j}^{gg}(r,t,w_p) = \sum_{\delta,j} \int \frac{d\gamma}{2\pi i}\; \tilde{c}^{gg}_{\delta,j,\gamma} F_{\delta,j,\gamma}(r,t,w_p).
\eeq
To extract the OPE coefficients \( \tilde{c}^{gg}_{\delta,j,\gamma} \), we first expand \( \tilde{F}^{gg}(r,t,w_p) \) as a Laurent series around \( r=0 \), which corresponds to the OPE (collinear) limit. To handle the potential Dirac delta function in the block coefficients \( \tilde{c}^{gg}_{\delta,j,\gamma} \), we decompose \( \tilde{F}^{gg}(r,t,w_p) \) into two parts: one that vanishes as \( w_p \to \infty \) and the other that remains finite. 

For the first part, we perform a Mellin transformation with respect to \( w_p \) and match its leading term in the \( r \to 0 \) limit to the leading power behavior of \( G_{\delta,j}^{(\gamma)}(r,t) \sim 2r^{\delta} \cos(j\psi) \), where \( \psi \) is the phase of \( t = e^{i\psi} \). This procedure allows us to extract the leading twist OPE coefficient. After subtracting the leading twist block contribution from \( \tilde{F}^{gg}(r,t,w_p) \), we repeat the process for the next-to-leading twist, and so on. In this way, we can iteratively extract the OPE coefficients for arbitrary twist.

As for the second part, which remains finite as \( w_p \to \infty \), we directly match it to \( G_{\delta,j}^{(\gamma)}(r,t) \) order by order in the \( r \to 0 \) expansion, without performing a Mellin transformation. This leads to the following block expansion for Eq.~\eqref{full result}:
\begin{equation}
    \begin{aligned}
        &\frac{d^2\Sigma}{d\Omega_a d\Omega_b}=-\frac{27 g^6}{2(8\pi )^5(n_a \cdot n_b)^3}\bigg[\int \frac{d\gamma}{2\pi i} \;\pi \csc(\pi\gamma)\frac{14\gamma (\gamma^4 + 55 \gamma^2 + 304)}{75}  w_p^\gamma G^{(\gamma)}_{4,0}(r,t)-\frac{112}{5}G^{(0)}_{4,0}(r,t)\\
        &\quad +\int \frac{d\gamma}{2\pi i} \;\pi \csc(\pi\gamma)\frac{ \gamma (122 \gamma^6 + 8897 \gamma^4 + 157493 \gamma^2 + 612168)}{11025} w_p^\gamma G^{(\gamma)}_{6,0}(r,t)-\frac{664}{35}G^{(0)}_{6,0}(r,t)\\
        &\quad +\int \frac{d\gamma}{2\pi i} \;\pi \csc(\pi\gamma)\frac{ \gamma (641 \gamma^6 + 50456 \gamma^4 + 875819 \gamma^2 + 2994204)}{110250} w_p^\gamma G^{(\gamma)}_{6,2}(r,t)-\frac{1544}{175}G^{(0)}_{6,2}(r,t)\\
        &\quad+\cdots\bigg]. 
        \label{eq:gg_celestial_block_expansion}
    \end{aligned}
\end{equation}
It is worth emphasizing that the boundary term in the \( w_p \to \infty \) limit admits an expansion in terms of 2d conformal blocks for identical scalar operators.

For concreteness, we also list coefficients for higher twist contributions here:
\begin{equation}
    \begin{aligned}
        \tilde{c}^{gg}_{8,0,\gamma} &= \frac{\pi\gamma\csc(\pi\gamma)}{714420000} \Big(120109 \gamma^8 + 15371718 \gamma^6 + 355716165 \gamma^4 + 1217888312 \gamma^2 \\
        &\quad+ 3550010256\Big)-\frac{5924}{1125}i\pi\delta(\gamma), \\
        \tilde{c}^{gg}_{8,2,\gamma} &= \frac{ \pi\gamma\csc(\pi\gamma) }{25004700}\Big(4489 \gamma^8 + 667257 \gamma^6 + 20534871 \gamma^4 + 175103003 \gamma^2 + 416708100\Big)\\
        &\quad-\frac{23392}{2205}i\pi\delta(\gamma), \\
        \tilde{c}^{gg}_{8,4,\gamma} &= \frac{\pi\gamma\csc(\pi\gamma)}{28576800}\Big(629 \gamma^8 + 118398 \gamma^6 + 4409853 \gamma^4 + 44932672 \gamma^2 + 90973008\Big)\\
        &\quad-\frac{584}{315}i\pi\delta(\gamma), 
        \label{eq:gg_celestial_block_expansion_data1}
    \end{aligned}
\end{equation}
\begin{equation}
    \begin{aligned}
        \tilde{c}^{gg}_{10,0,\gamma} &= \frac{\pi\gamma\csc(\pi\gamma)}{3388636944000} \Big(3280573 \gamma^{10} + 952734530 \gamma^8 + 40225339869 \gamma^6 \\
        &\quad + 430721770820 \gamma^4 + 2857348218208 \gamma^2 + 4231389744000\Big) -\frac{33184}{33957}i\pi\delta(\gamma), \\
        \tilde{c}^{gg}_{10,2,\gamma}  &= \frac{\pi\gamma\csc(\pi\gamma)}{4356818928000} \Big(6236911 \gamma^{10} + 1847649617 \gamma^8 + 77931465117 \gamma^6 \\
        &\quad + 602372032295 \gamma^4 + 2138931170332 \gamma^2 + 5855881729248\Big)-\frac{1449956}{1091475}i\pi\delta(\gamma), \\
        \tilde{c}^{gg}_{10,4,\gamma}  &= \frac{\pi\gamma\csc(\pi\gamma)}{69155856000} \Big(31893 \gamma^{10} + 11177210 \gamma^8 + 628406229 \gamma^6 + 10457371980 \gamma^4 \\
        &\quad + 70437030928 \gamma^2 + 117590938560\Big)-\frac{3488}{3465}i\pi\delta(\gamma), \\
        \tilde{c}^{gg}_{10,6,\gamma}  &= \frac{\pi\gamma\csc(\pi\gamma)}{899026128000} \Big(34647 \gamma^{10} + 16226705 \gamma^8 + 1133481261 \gamma^6 + 24907422375 \gamma^4 \\
        &\quad + 182768988292 \gamma^2 + 265225763520\Big)-\frac{652}{4095}i\pi\delta(\gamma).
        \label{eq:gg_celestial_block_expansion_data2}
    \end{aligned}
\end{equation}
We have suppressed an overall constant $\displaystyle -\frac{27 g^6 }{2(8\pi )^5}$ for the above coefficients.

\begin{figure}[ht]
    \centering
    \includegraphics[width=0.8\textwidth]{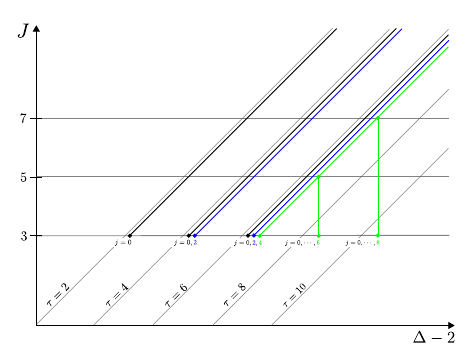}
    \caption{Chew–Frautschi plot relevant to the collinear limit of the EEC at hadron colliders. Compared to the generic operator structure in a general quantum field theory, at leading order in the perturbative expansion, only \( j = 0 \) operators contribute at twist-2, and there are no contributions from \( j = 4 \) operators at twist-4. A more detailed discussion is provided in the main text.}
    \label{fig:chewf}
\end{figure}

Based on the celestial block decomposition above, we now discuss the possible operators that contribute in the OPE limit. To visualize the (free) operator spectrum relevant to the tree-level \( gg \to ggg \) process, we use a Chew–Frautschi plot, as shown in Fig.~\ref{fig:chewf}.

Since our calculation is performed at leading order in a perturbative gauge theory, the resulting structure is significantly simpler than that of a generic quantum field theory. We recall that the general form of the OPE is given by~\cite{Chang:2020qpj}
\begin{equation}
    \begin{aligned}
        \label{eq:EEOPE}
        \mathcal{E}(\vec n_1) \mathcal{E}(\vec n_2) 
        &\sim\sum\limits_i \big( \mathbb{O}_{i,J=3,j=0}(\vec n) + \mathbb{O}_{i,J=3,j=2}(\vec n)  
        + \mathbb{O}_{i,J=3,j=4}(\vec n)   \big) \\
        &+\sum\limits_{n,i} \mathcal{D}_{2n} \mathbb{O}_{i,J=3+2n,j=4}(\vec n)   \,,
    \end{aligned}
\end{equation}
and we will frame our discussion within this formalism. Here, $\mathbb{O}_{i,J,j}$ denotes different light-ray operators labeled by spin $J$ and transverse spin $j$, with explicit examples for the leading twist operators to be given later. The $\mathcal{D}_{2n}$ are differential operators which convert $2n$ units of spin into transverse spin. These differential operators are responsible for the appearance of higher transverse spin primaries in our result. Note that, compared to Eq.~(2.1), the OPE coefficients and the explicit angular dependence (such as the $\theta^{\delta-6}$ factors) have been suppressed in Eq.~(3.18) for brevity.

The leading power behavior of the EEC in the collinear limit scales as \( r^{-2} \), corresponding to a twist-2, transverse spin-0 operator. This is consistent with the coefficient data extracted previously, where the leading term arises from the block with \( \delta = 4 \) and \( j = 0 \). Since the celestial dimension \( \delta \) is related to the bulk scaling dimension \( \Delta \) by \( \Delta = \delta + 1 \), the corresponding operator has twist \( \tau = \Delta - J = 2 \). 

In weakly coupled gauge theories, twist-2 operators may carry transverse spin \( j = 0 \) or \( j = 2 \). Notably, the transverse spin-2 contribution admits a physical interpretation as arising from the interference between intermediate gluon states with different helicities~\cite{Chen:2020adz,Chen:2021gdk}. However, in the tree-level \( gg \to ggg \) process, the transverse spin-2 contribution vanishes, as it probes the interference between amplitudes \( \mathcal{M}_{gg \to gg}^{h_1 h_2; h_3 +} \) and \( \mathcal{M}_{gg \to gg}^{h_1 h_2; h_3 -} \), which is vanishing due to the tree-level MHV selection rule~\cite{Parke:1986gb}.

The leading twist operators in QCD that contribute to the EEC in the collinear limit have been described in detail in~\cite{Chen:2021gdk}. Since we are considering pure gluon scattering now, we need only the gluon twist-2 operator:
\begin{equation}
    \mathcal{O}_g^{[J],ij}=-\frac{1}{2^J}F_c^{\text{(}i+}(i D^{+})^{J-2}F_c^{j\text{)}+}=-\frac{1}{2^J}(\partial^{+}A_c^i)(i \partial^{+})^{J-2}(\partial^+ A_c^j)+\cdots,
\end{equation}
where $i$, $j$ are the transverse indices and $\cdots$ represents terms that are higher order in gauge coupling or vanish after some projections given below. 

With the following convention for the mode expansion of massless free gluon field,
\begin{equation}
    A^{\mu}_{c}(x)=\sum_{\lambda} \int\frac{dp^+ d^2p_{\perp}}{(2\pi)^3 2p^+}
    \left(\epsilon^{\mu}_{\lambda}(p) a_{p,\lambda,c} e^{-ip\cdot x} 
        +{\epsilon^{*}_{\lambda}}^{\mu}(p) a^{\dagger}_{p,\lambda,c} e^{ip\cdot x} \right)\,,
\end{equation}
We can obtain the corresponding light-ray operator:
\begin{equation}
    \begin{aligned}
        \mathbb{O}_g^{[J],ij}(\vec n)=-\frac{1}{2}\sum_{\lambda,\lambda',c}\int\frac{E^2 dE}{(2\pi)^3 2E} E^{J-1}\left(\epsilon_{\lambda}^{*i}\epsilon_{\lambda'}^j a_{p,\lambda,c}^{\dagger}a_{p,\lambda',c}+(-1)^J\epsilon_{\lambda}^i\epsilon_{\lambda'}^{*j}a_{p,\lambda',c}^{\dagger}a_{p,\lambda,c}\right),
    \end{aligned}
\end{equation}
where $\vec n$ denotes the direction of $\vec p$ and $\epsilon_\lambda$ is a polarization vector with helicity $\lambda$. The $(-1)^J$ indicates we should analytically continue the $J$ separately for odd and even collinear spin. For gluon operator, choosing odd spin branch, i.e. setting $(-1)^J\to -1$, results in an identically zero outcome. Therefore, in this paper, we choose the analytic continuation from the even spin branch.

The gluon twist-2 operator $\mathbb{O}_g^{[J],ij}(\vec n)$ can be further decomposed into a scalar and a traceless symmetric operator in the transverse plane:
\begin{align}
    \mathbb{O}_g^{[J]}(\vec n)&=g_{ij}^{\perp} \mathbb{O}_g^{[J],ij}(\vec n)=\sum_{\lambda,c}\int \frac{E^2 dE}{(2\pi)^3 2E} E^{J-1} a_{p,\lambda,c}^{\dagger} a_{p,\lambda,c},\\
    \mathbb{O}_{\tilde g,\lambda}^{[J]}(\vec n)&=\epsilon_{\lambda,i}(p)\epsilon_{\lambda,j}(p)\mathbb{O}_g^{[J],ij}(\vec n)=-\sum_{c}\int \frac{E^2 dE}{(2\pi)^3 2E} E^{J-1} a_{p,\lambda,c}^{\dagger} a_{p,-\lambda,c} .
\end{align}
In general cases, both $\mathbb{O}_g^{[J=3]}$ and $\mathbb{O}_{\tilde g,\lambda}^{[J]}$ can appear in the OPE of $\mathcal E(\vec{n}_1) \mathcal E(\vec{n}_2)$. However, for the tree-level $gg\to ggg$ process, only the transverse spin-$0$ operator $\mathbb{O}_g^{[J=3]}$ contributes a non-vanishing matrix element due to the tree-level MHV selection rule, as mentioned previously.

We now turn to the twist-4 operators. It can be seen from the Chew-Frautschi plot that the twist-4 operators can only carry transverse spin-0 or transverse spin-2 at leading order in the perturbative expansion, consistent with the discussion in \cite{Chen:2021gdk}.

At leading order in the perturbative expansion, only operators involving up to four fields can be probed. There are two possible schematic constructions of such operators, as illustrated below:
\begin{gather}
    F\partial^{J-2}  \Box F,\label{twist4 1}\\ 
    F\partial^{J_1} F \partial^{J_2} F\partial^{J_3} F,\label{twist4 2}
\end{gather}
with $J_1+J_2+J_3+4=J$. Since $\Box F$ vanishes in the free theory due to the equation of motion, it seems that the only viable form of gluon twist-4 operators is that of Eq.~\eqref{twist4 2}. Here we expect that its corresponding light-ray operator exists for $J=3$ as an analytic continuation from $J\geq 4$. Recently, the spectra of multi-field operators have been studied in other theories and shown to exhibit interesting analytic structures~\cite{Homrich:2024nwc,Homrich:2022mmd,Henriksson:2023cnh,Kravchuk:2024wmv,Fardelli:2024heb}.
Moreover, for the tree-level OPE, there should be no twist-4 operators with transverse spin-4, which implies that at least one pair of the four transverse indices in Eq.~\eqref{twist4 2} should be contracted. 

We end this discussion of the Chew-Frautschi plot by noting that the primary operators contributing to the EEC in the collinear limit at leading order in the perturbative expansion share the same structure as those contributing to the EEEC in $\mathcal{N}=4$ SYM theory. This coincidence may indicate that our EEC result, Eq.~\eqref{full result}, obtained from pure gluon scattering at hadron colliders, is somehow related to the EEEC in $\mathcal{N}=4$ SYM theory.

\begin{figure}[ht]
    \begin{center}
        \makebox[\textwidth][c]{
            \subfloat[]{
                \includegraphics[width=0.6\textwidth]{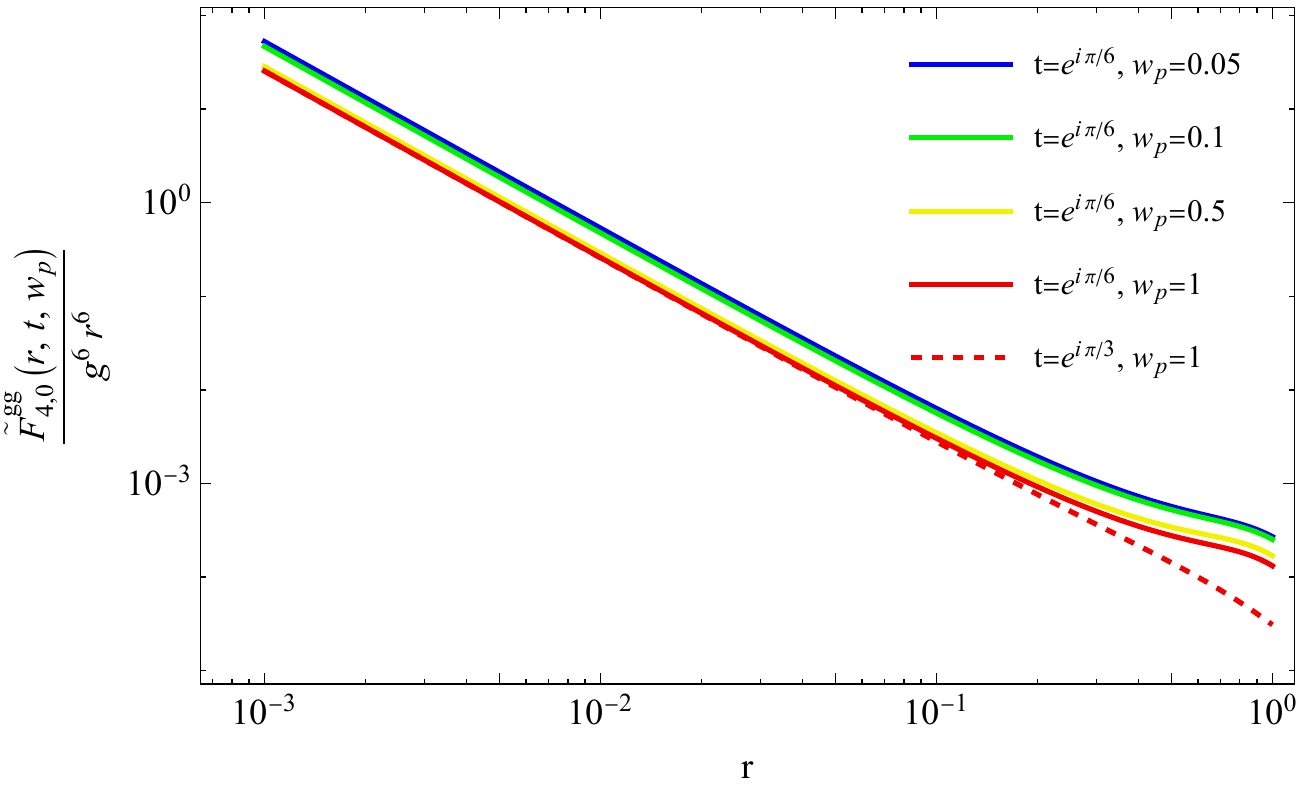}\label{Figure block.a}
                }\quad
            \subfloat[]{
                \includegraphics[width=0.62\textwidth]{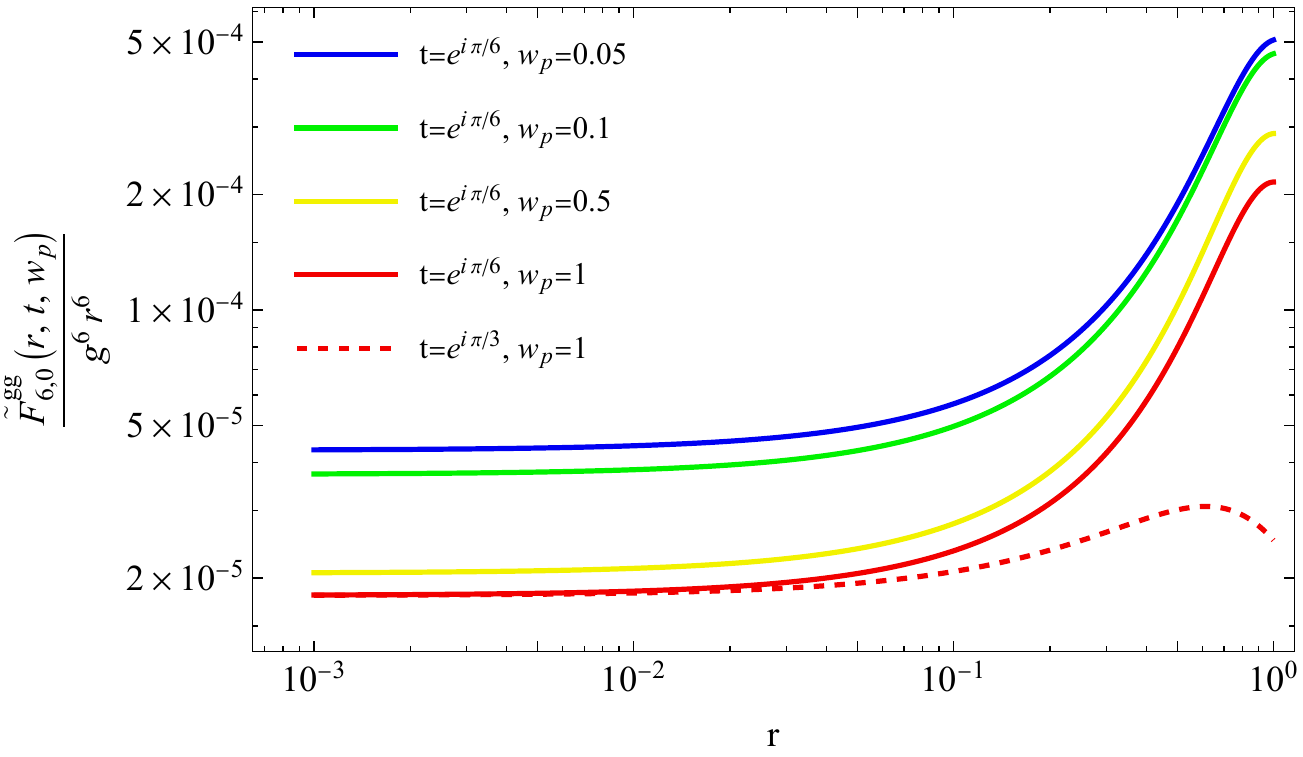}\label{Figure block.b}
                }
        }\\
        \makebox[\textwidth][c]{
            \subfloat[]{
                \includegraphics[width=0.64\textwidth]{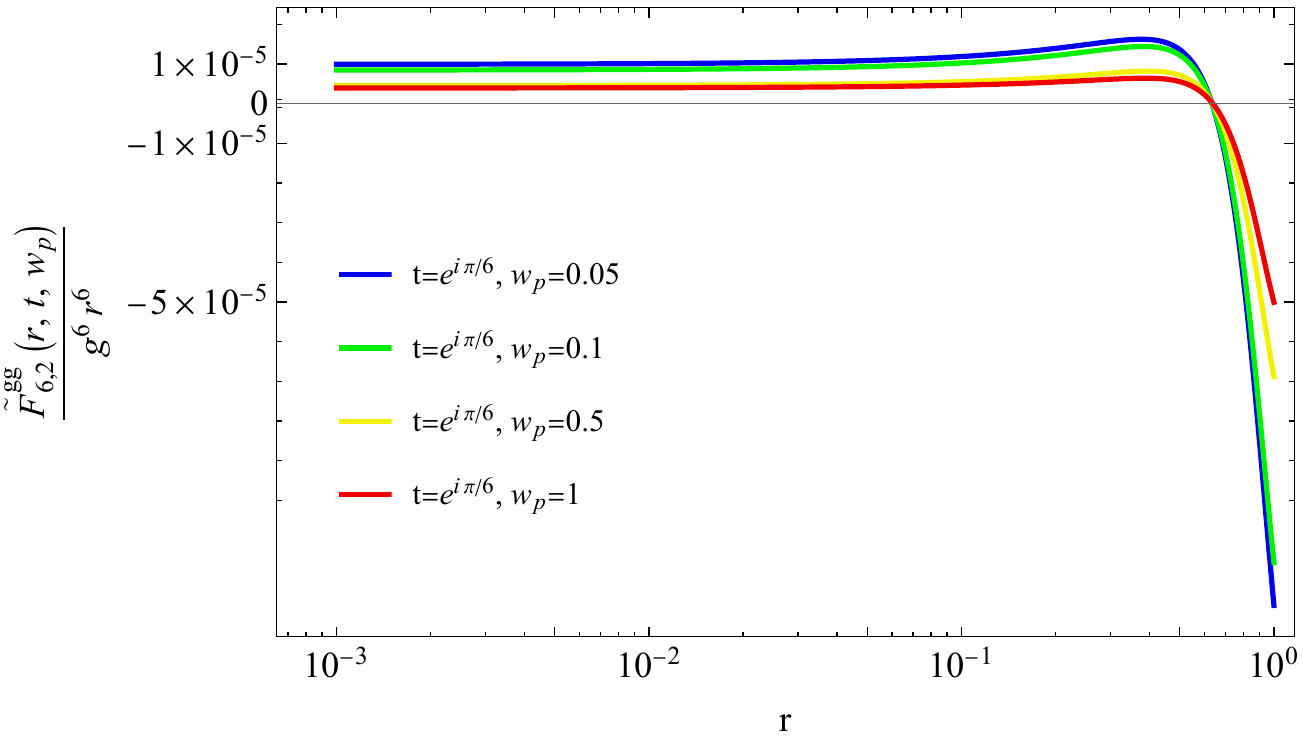}\label{Figure block.c}
                }\;
            \subfloat[]{
                \includegraphics[width=0.63\textwidth]{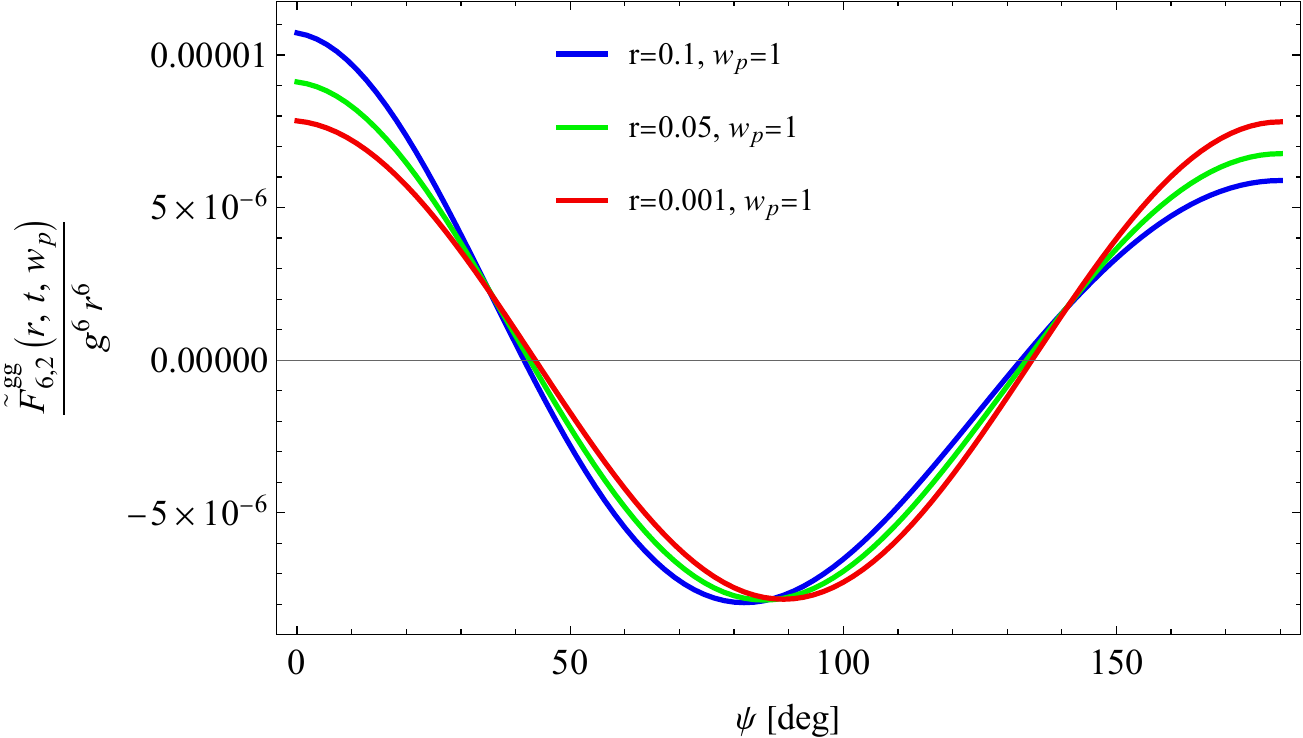}\label{Figure block.d}
                }\qquad
        }
    \end{center}
    \caption{
    Panels (a), (b), and (c) show the dependence of the first two leading-twist blocks on \( r \), for different values of \( w_p \) and \( t \), plotted on a double-logarithmic scale. Each of these plots exhibits clear linear behavior in the collinear limit and illustrates the global dilation effect induced by \( w_p \). Panel (d) displays the dependence of the twist-4, transverse spin-2 block on \( \psi \), with \( w_p = 1 \) and varying \( r \), revealing the characteristic \( \cos(2\psi) \) structure.
    }
    \label{Figure block}
\end{figure}
To gain an intuitive understanding of the celestial block decomposition of the EEC, we illustrate the first two leading twist contributions of \(\displaystyle \tilde{F}_{\delta,j}^{gg}(r,t,w_p) \) to highlight their dependence on the three kinematic variables and to emphasize their utility compared to a naive power expansion. In the collinear limit, where \( r \to 0 \), \(\displaystyle \tilde{F}_{\delta,j}^{gg}(r,t,w_p) \) differs from the EEC by an overall factor proportional to \( r^6 \), as indicated in Eq.~\eqref{eq:eec_relate_Fgg}, where  \(\displaystyle (n_a \cdot n_b)^3 \sim r^6 \). To better visualize their contributions to the EEC, we divide \(\displaystyle \tilde{F}_{\delta,j}^{gg}(r,t,w_p) \) by \( r^6 \). Additionally, we also divide by \( g^6 \) in order to present the numerical result.

For the twist-2 block contribution, which involves only the transverse spin-0 primary operator, its dependence on \( t \) arises from the descendants of the primary operator. Since the descendants are associated with higher twists, their contributions are suppressed by a factor of \( r^{\tau - 2} \) and can thus be neglected in the collinear limit (\( r \to 0 \)). The same argument applies to the twist-4 transverse spin-0 block contribution; however, it does not hold for the transverse spin-2 block contribution. As shown in Figs.~\ref{Figure block.a},~\ref{Figure block.b} and~\ref{Figure block.c}, each of the first three blocks exhibits clear scaling behavior in the collinear limit, consistent with the expectation that a block with twist \( \tau \) scale as \( r^{\tau - 4} \) in this limit. These blocks also reveal a global dilation effect as \( w_p \) becomes small. For the twist-4 transverse spin-2 block contribution, Fig.~\ref{Figure block.d} shows a characteristic \( \cos(2\psi) \) dependence when \( r \) is sufficiently small, as expected for a transverse spin-2 block. 

\begin{figure}[ht]
    \begin{center}
        \makebox[\textwidth][c]{
            \subfloat[]{
                \includegraphics[width=0.6\textwidth]{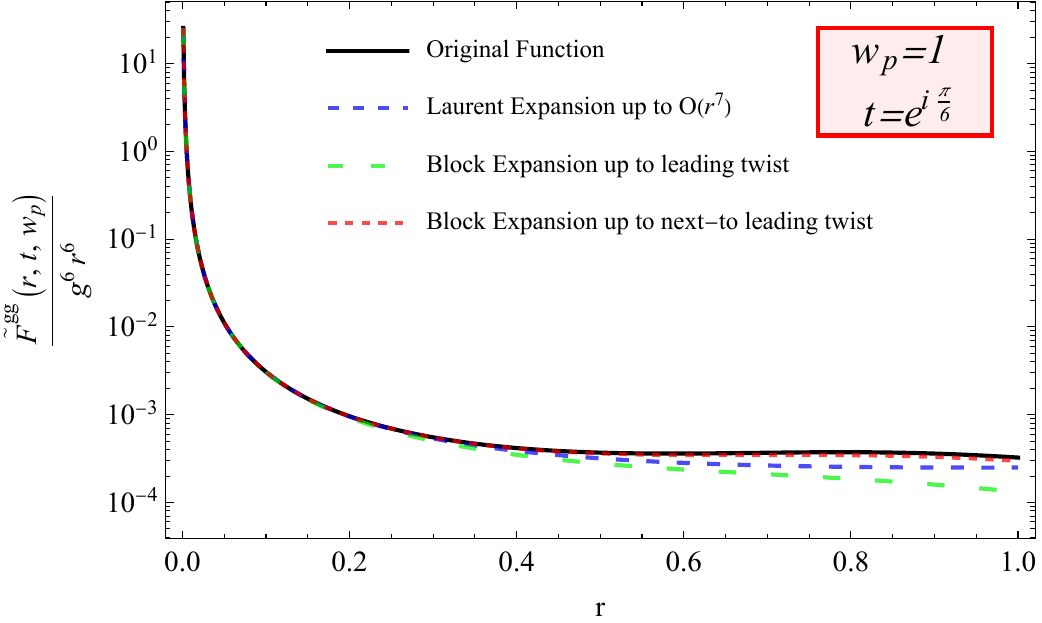}\label{Figure comparison.a}
                }\quad
            \subfloat[]{
                \includegraphics[width=0.6\textwidth]{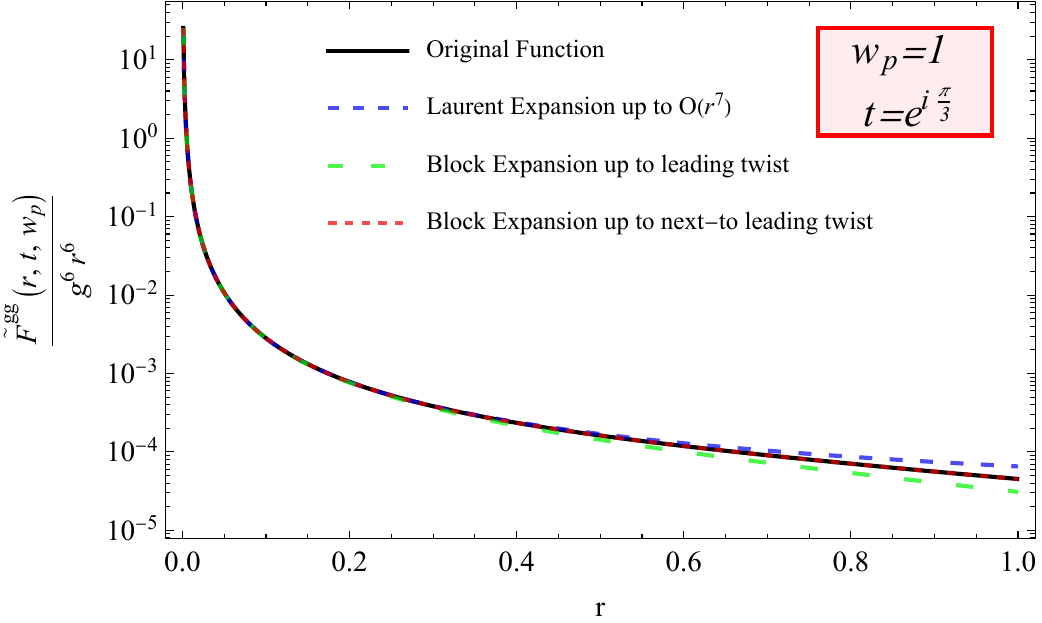}\label{Figure comparison.b}
                }
        }\\
        \makebox[\textwidth][c]{
            \subfloat[]{
                \includegraphics[width=0.6\textwidth]{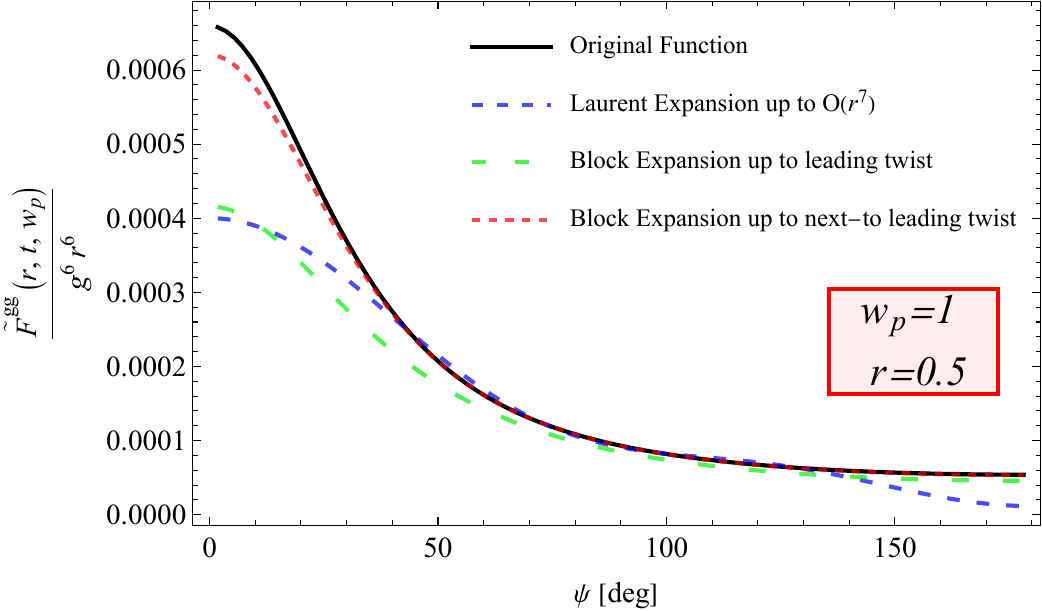}\label{Figure comparison.c}
                }\quad
            \subfloat[]{
                \includegraphics[width=0.6\textwidth]{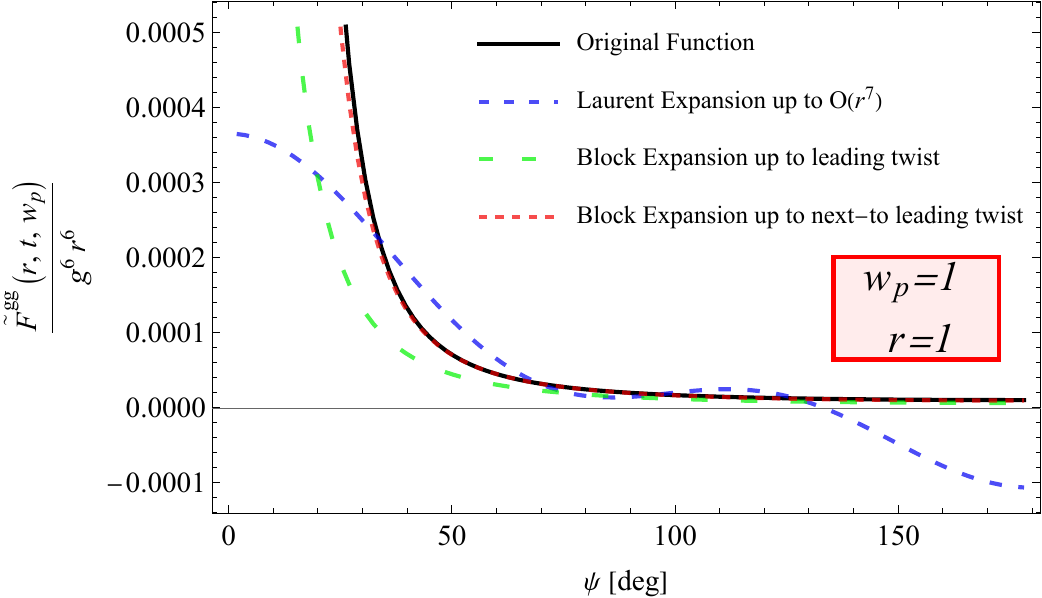}\label{Figure comparison.d}
                }
        }
    \end{center}
    \caption{    
    Comparison of the original function \(\tilde{F}^{gg}(r,t,w_p)/(g^6 r^6)\) (black solid line) with three approximations: the Laurent expansion up to \(\mathcal{O}(r^7)\) (blue dashed line), the block expansion with leading twist only (green dashed line), and the block expansion including next-to-leading twist (red dashed line). Panels (a) and (b) show results for \(w_p = 1\) with \(t = e^{i\pi/6}\) and \(t = e^{i\pi/3}\), respectively, as \(r\) varies from 0 to 1. Panels (c) and (d) show results for \(w_p = 1\) with fixed \(r = 0.5\) and \(r = 1\), respectively, as \(\psi\) varies from \(0^\circ\) to \(180^\circ\). Inclusion of the next-to-leading twist improves the agreement with the original function across the parameter space.}
    \label{Figure comparison}
\end{figure}
In Figs.~\ref{Figure comparison.a} and \ref{Figure comparison.b}, the Laurent series expansion (up to \(\mathcal{O}(r^7)\)) is compared with the block expansions retaining only the leading twist and including up to the next-to-leading twist, for \(t = e^{i\pi/6}\) and \(t = e^{i\pi/3}\), respectively. Over the range \(0 \leq r \leq 1\), the block expansion including the next-to-leading twist (red line) exhibits better agreement with the original function (black line) than both the leading-twist-only approximation (green line) and the truncated Laurent series (blue line), particularly at larger values of \(r\).

In Figs.~\ref{Figure comparison.c} and \ref{Figure comparison.d}, \(r\) is fixed at 0.5 and 1, respectively, with \(w_p = 1\), and the horizontal axis represents \(\psi\), ranging from \(0^\circ\) to \(180^\circ\). The block expansion that includes the next-to-leading twist provides a significantly improved match to the original function across the parameter space when compared to the other approximations. 

As discussed at the beginning of Sec.~\ref{sec:gluon}, hadronization effects dominate in the extremely small \( r \) regime, rather than the perturbative parton-level correlations. To see the perturbative collinear dynamics, we are interested in moderately small values of \( r \), where the advantage of a power series expansion — despite its rapid convergence in the strict \( r \to 0 \) limit — is diminished. By contrast, the block expansion — which organizes contributions by the twist of primary operators and resums all their descendants — provides both a symmetry-respecting representation and a better numerical approximation in this regime.

\subsection{Analyticity in transverse spin}

From the celestial block decomposition \eqref{eq:gg_celestial_block_expansion}-\eqref{eq:gg_celestial_block_expansion_data2},
one interesting feature is that we find more non-vanishing high transverse spin blocks as celestial dimension $\delta$ increases, which is similar to the celestial block decomposition for three-point energy correlator~\cite{Chang:2022ryc,Chen:2022jhb}. For three-point energy correlator, it was shown that, given a fixed celestial twist $\delta-j$, the OPE coefficients are analytic with respect to transverse spin $j$. This conclusion follows from the discussion of analyticity in spin and Lorentzian inversion formula in conformal field theories~\cite{Caron-Huot:2017vep,Simmons-Duffin:2017nub}.

As we have shown in Sec.~\ref{sec:ccb_hadron_collider}, the building blocks of celestial blocks $F_{\delta,j,\gamma}$ are 2d conformal blocks, which implies that Lorentzian inversion formula technique is applicable in the celestial block decomposition. Let us start with \eqref{eq:celestial_block_decomposition} and deform the contour of $\gamma$. For tree-level result \eqref{eq:gg_celestial_block_expansion}, we observe that all the $\gamma$-poles are located at integer values, which leads to
\begin{equation}
    F_{gg\to ggg}(z,\bar{z},w) = \sum_{m=0}^\infty w^m \sum_{\delta,j}  \left[-\underset{\gamma=m}{\mathrm{Res}} \tilde{c}^{gg}_{\delta, j,\gamma}\right] G^{(m)}_{\delta,j}(z,\bar{z})\,.
\end{equation}
This means we can first expand the tree-level result in the $w\to 0$ limit as a series of the form $\sum_m w^m F^{(m)}(z,\bar{z})$ and then apply Lorentzian inversion formula to each term $F^{(m)}(z,\bar{z})$ to extract the coefficients $-\underset{\gamma=m}{\mathrm{Res}} \tilde{c}^{gg}_{\delta, j,\gamma}$ analytically. The OPE coefficients $-\underset{\gamma=m}{\mathrm{Res}} \tilde{c}^{gg}_{\delta, j,\gamma}$ may differ from the outcome of Lorentzian inversion formula for low lying transverse spins, depending on the behavior of $F^{(m)}(z,\bar{z})$. Thorough discussion of Lorentzian inversion formula can be found in~\cite{Caron-Huot:2017vep,Simmons-Duffin:2017nub,Alday:2017vkk}.

\begin{figure}[ht]
    \begin{center}
        \makebox[\textwidth][c]{
            \subfloat[]{
                \includegraphics[width=0.6\textwidth]{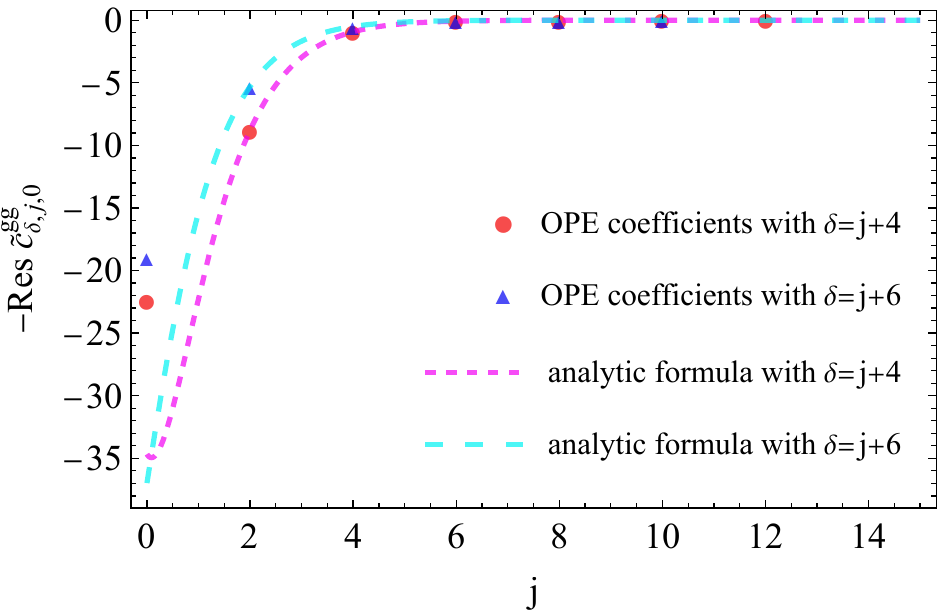}\label{Figure tran_spin_4.a}
                }\qquad
            \subfloat[]{
                \includegraphics[width=0.6\textwidth]{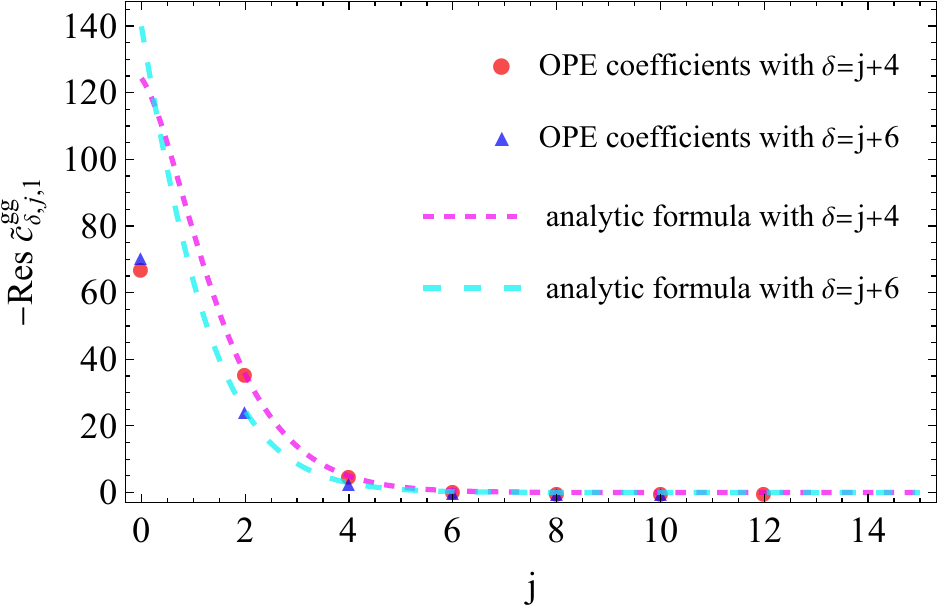}\label{Figure tran_spin_4.b}
                }\qquad
        }\\
        \makebox[\textwidth][c]{
            \subfloat[]{
                \includegraphics[width=0.6\textwidth]{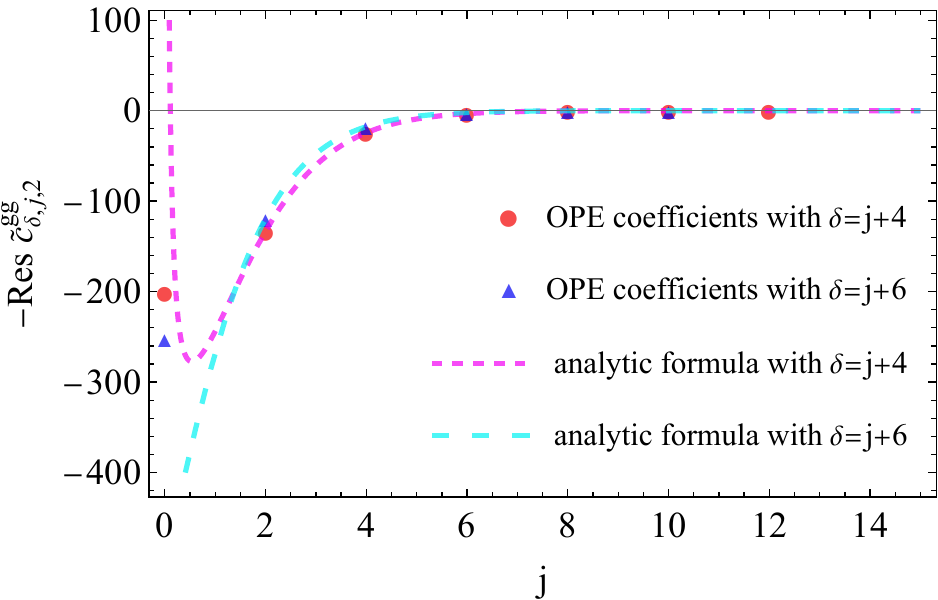}\label{Figure tran_spin_4.c}
                }\qquad
            \subfloat[]{
                \includegraphics[width=0.6\textwidth]{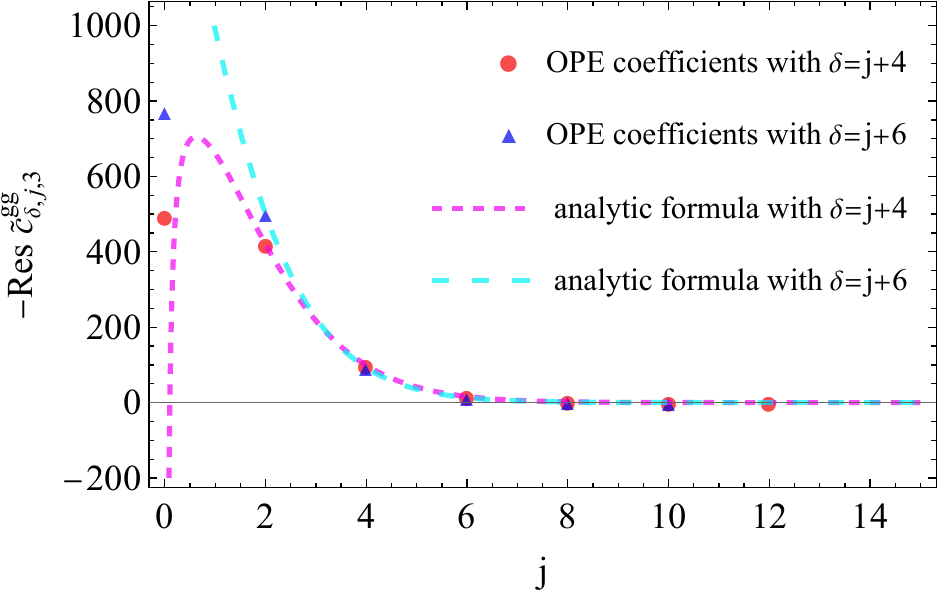}\label{Figure tran_spin_4.d}
                }\qquad
            }
    \end{center}
    \caption{
    Panels (a)–(d) compare the negative residues of the OPE coefficients (shown as discrete points) with the corresponding analytic formulas (depicted as dashed lines) for various transverse spin values \( j \). Red and blue points denote the OPE coefficients with celestial twist 4 and 6, respectively, while the magenta and cyan dashed lines represent the corresponding analytic formulas extracted using Lorentzian inversion formula. Each panel corresponds to a different value of \( \gamma \), illustrating the agreement (or discrepancy) between the extracted OPE coefficients and their analytic formulas.
    }
    \label{Figure tran_spin_4}
\end{figure}

As an example, we consider the OPE coefficients $\tilde{c}^{gg}_{\delta, j,\gamma}$ at the pole $\gamma=0$, which corresponds to the contribution of $w^0$ in the $w\to 0$ limit. For tree-level $gg\to ggg$ process, the leading celestial twist is $\delta-j=4$.
At leading and subleading celestial twist $\delta-j=4,6$, we can obtain the following analytic results from Lorentzian inversion formula
\begin{align}
&-\underset{\gamma=0}{\mathrm{Res}}\, \tilde{c}^{gg}_{\delta = j+4, j,\gamma}\Bigg|_{j\geq 4} = -\frac{16}{3} \frac{\Gamma(j+2)^2}{\Gamma(2j+3)} (24 H_{j+1}-11) \,,\nonumber\\
&-\underset{\gamma=0}{\mathrm{Res}}\, \tilde{c}^{gg}_{\delta = j+6, j,\gamma}\Bigg|_{j\geq 4} = -\frac{128}{15} \frac{\Gamma(j+3)^2}{\Gamma(2j+5)} (30 H_{j+2}-19) \,,
\end{align}
where $H_n$ is the $n$-th harmonic number. By comparing with OPE coefficients from direct block expansion, we find these analytic expressions are valid for $j\geq 4$, demonstrating the analyticity in transverse spin. We also obtain similar results at the poles $\gamma=1,2,3,4$
{\footnotesize
\begin{align}
    &-\underset{\gamma=1}{\mathrm{Res}}\, \tilde{c}^{gg}_{\delta = j+4, j,\gamma}\Bigg|_{j\geq 4} =
    \frac{8}{15}\frac{\Gamma(j+2)\Gamma(j+3)}{\Gamma(2j+3)} (120 H_{j+1}+113)\,,\nonumber\\
    &-\underset{\gamma=1}{\mathrm{Res}}\, \tilde{c}^{gg}_{\delta = j+6, j,\gamma}\Bigg|_{j\geq 4} = 
    \frac{4}{5} \frac{\Gamma(j+3)\Gamma(j+4)}{\Gamma(2j+5)}(120 H_{j+2}-\frac{80}{j+2}+\frac{80}{j+3}+183)\,,\nonumber\\
    & -\underset{\gamma=2}{\mathrm{Res}}\, \tilde{c}^{gg}_{\delta = j+4, j,\gamma}\Bigg|_{j\geq 4} =
    \frac{\Gamma(j+2)\Gamma(j+4)}{\Gamma(2j+3)}\left(-32 H_{j+1} +\frac{48}{j(j+3)} -\frac{80}{(j+1)(j+2)} -\frac{1088}{15}\right)\,,\nonumber\\
    & -\underset{\gamma=2}{\mathrm{Res}}\, \tilde{c}^{gg}_{\delta = j+6, j,\gamma}\Bigg|_{j\geq 4} =
    \frac{\Gamma(j+3)\Gamma(j+5)}{\Gamma(2j+5)}\left(-32 H_{j+2} + \frac{96}{(j+1)(j+4)}-\frac{64}{(j+2)(j+3)}-\frac{22412}{105}\right)\,,\nonumber\\
    &-\underset{\gamma=3}{\mathrm{Res}}\, \tilde{c}^{gg}_{\delta = j+4, j,\gamma}\Bigg|_{j\geq 4} =
    \frac{\Gamma(j+2)\Gamma(j+5)}{\Gamma(2j+3)}\left(\frac{32}{3}H_{j+1} -\frac{24}{j(j+3)} + \frac{40}{(j+1)(j+2)} + \frac{15566}{315} \right) \,,\nonumber\\
    &-\underset{\gamma=3}{\mathrm{Res}}\, \tilde{c}^{gg}_{\delta = j+6, j,\gamma}\Bigg|_{j\geq 4} =
    \frac{\Gamma(j+3)\Gamma(j+6)}{\Gamma(2j+5)} \left(\frac{16}{3}H_{j+2} -\frac{28}{(j+1)(j+4)}+\frac{4}{(j+2)(j+3)}+\frac{7069}{45}\right)\,,\nonumber\\
    &
    -\underset{\gamma=4}{\mathrm{Res}}\, \tilde{c}^{gg}_{\delta = j+4, j,\gamma}\Bigg|_{j\geq 4} =
    \frac{\Gamma(j+2)\Gamma(j+6)}{\Gamma(2j+3)}\left(
    -\frac{8}{3}H_{j+1} +\frac{8/15}{(j-2)(j+5)}-\frac{16/3}{(j-1)(j+4)}+\frac{124/5}{j(j+3)}-\frac{76/3}{(j+1)(j+2)} - \frac{14183}{630}
    \right)\,,\nonumber\\
    &-\underset{\gamma=4}{\mathrm{Res}}\, \tilde{c}^{gg}_{\delta = j+6, j,\gamma}\Bigg|_{j\geq 4} =
    \frac{\Gamma(j+3)\Gamma(j+7)}{\Gamma(2j+5)}\left(
    \frac{16/15}{(j-1)(j+6)} - \frac{16/3}{j(j+5)} + \frac{88/5}{(j+1)(j+4)} - \frac{8/3}{(j+2)(j+3)}
    -\frac{146033}{1890}
    \right)\,.
\end{align}
}

\begin{figure}[ht]
    \begin{center}
        \makebox[\textwidth][c]{
            \subfloat[]{
                \includegraphics[width=0.6\textwidth]{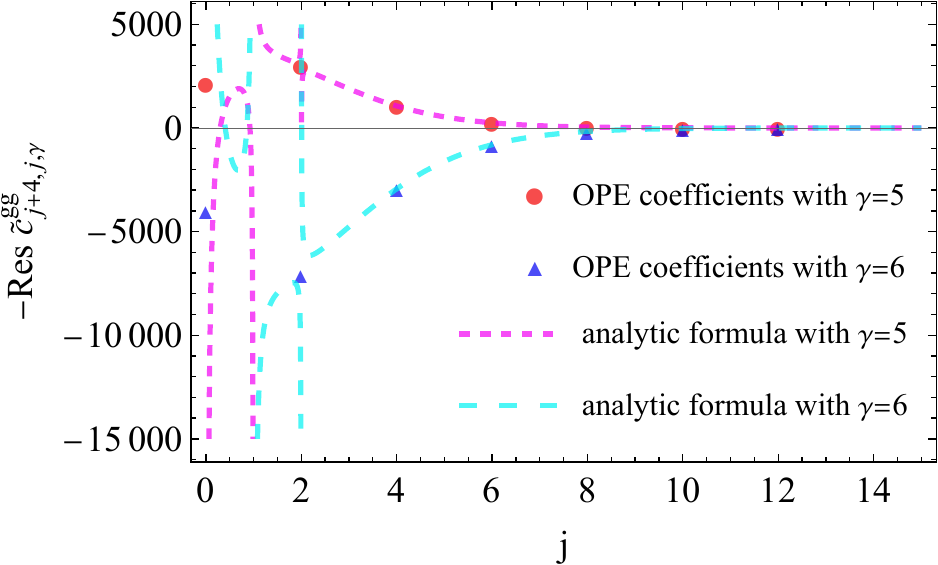}\label{Figure tran_spin_5.a}
                }\qquad
            \subfloat[]{
                \includegraphics[width=0.6\textwidth]{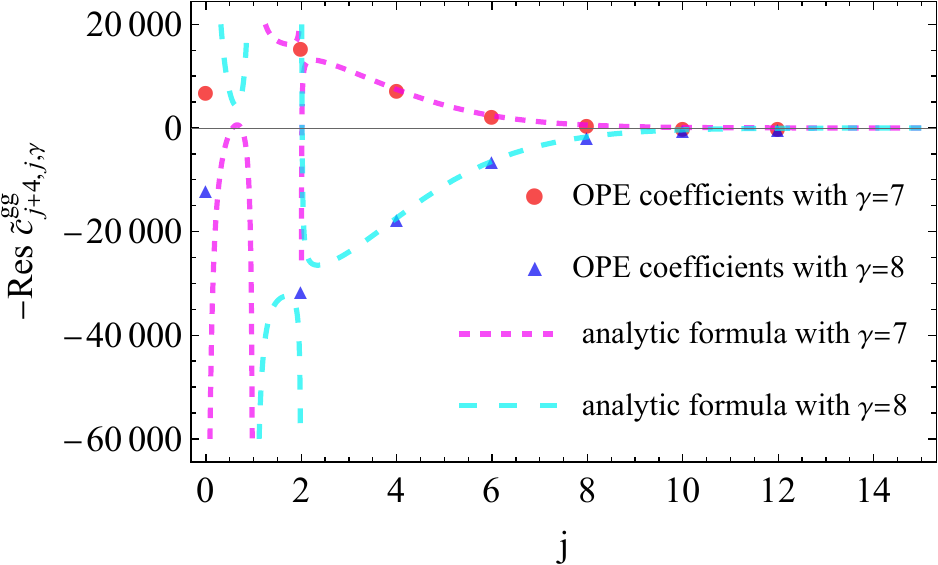}\label{Figure tran_spin_5.b}
                }\qquad
        }
    \end{center}
    \caption{
    Panels (a) and (b) compare the negative residues of the OPE coefficients (represented by discrete points) with the corresponding analytic formulas (depicted by dashed lines) for various spin values \(j\). The red and blue points represent the OPE coefficients with \(\gamma=5\) and \(\gamma=6\) in (a), and \(\gamma=7\) and \(\gamma=8\) in (b), respectively. The magenta and cyan dashed lines correspond to the analytic formulas under the same conditions. These plots demonstrate the level of agreement between the OPE coefficients and the analytic formulas, with notable deviations observed at smaller values of \(j\).
    }
    \label{Figure tran_spin_5}
\end{figure}

The leading celestial twist coefficients $\underset{\gamma=m\in \mathbb{Z}}{\mathrm{Res}}\, \tilde{c}^{gg}_{\delta = j+4, j,\gamma}$ are relatively simple.
Using Lorentzian inversion formula, we extract these coefficients up to $m=15$ and find
an analytic formula that can describe all examples for $m \geq 5$
\begin{align}
-\underset{\gamma=m}{\mathrm{Res}}\, \tilde{c}^{gg}_{\delta = j+4, j,\gamma}\Bigg|_{m\geq 5,\; j\geq m-1}
= (-1)^m \frac{\Gamma(j+2)\Gamma(j+2+m)}{\Gamma(m+1)\Gamma(2j+3)}\Bigg(
-64 H_{j+1} + \frac{4m(m-1)(m-2)(m-3)}{15(j-2)(j+5)}\nonumber\\
- \frac{4m(m-1)^2(m-2)}{3(j-1)(j+4)} + \frac{4m(m-1)(3m^2-5m +28)}{5j(j+3)}
-\frac{4m(m^3-2m^2+17m+8)}{3(j+1)(j+2)}\nonumber\\
-\frac{3 m^8+40 m^7+364 m^6+2278 m^5+8185 m^4+16306 m^3+18520 m^2+13056 m+6336}{9 (m+1) (m+2) (m+3)
   (m+4)}
\Bigg)\,.
\end{align}
Given that $\gamma$ is closely related to the collinear spin along collision axis, this expression provides an interesting example that is both analytic in collinear spin and transverse spin. 

We have checked that the above analytic formulas are consistent with the block coefficients given in Eqs.~\eqref{eq:gg_celestial_block_expansion_data1}, \eqref{eq:gg_celestial_block_expansion_data2} and higher $j$ expressions that are not explicitly given in this paper. Fig.~\ref{Figure tran_spin_4} and Fig.~\ref{Figure tran_spin_5} illustrate a comparison between the analytic formulas and the negative residues of the OPE coefficients. 
The comparison shows that analyticity in transverse spin holds when $j$ is greater than a certain threshold $j_*(\delta-j, \gamma)$, similar to the observation in \cite{Chen:2022jhb,Chang:2022ryc}. 
The validity threshold of Lorentzian inversion formula is determined by the Regge behavior of a correlator\footnote{To reach the Regge limit, one should first independently analytic continue $\bar{z}$ around $1$, and then take the limit $z,\bar{z}\to 0$.}. For example, in unitary CFTs, the formula holds for spins $J>1$~\cite{Caron-Huot:2017vep}. Here, however, the physical interpretation of the threshold in transverse spin is not yet clear. General discussions of the transverse spin analyticity beyond tree level and its validity threshold may necessitate further investigation in the future.

\section{Singular behaviors of EEC and LO factorization}
\label{sec:factorization}

To gain a deeper understanding of the EEC results presented in Sec.~\ref{sec:result}, we now derive a leading-power factorization approximation for several key limits of the EEC. These approximations also provide an independent verification of our results. By analyzing the singular behavior of the EEC in various limits, including the collinear limit, the opposite coplanar limit, the back-to-back limit, and the Regge limit, we can uncover universal features inherent to high-energy scattering processes.
\subsection{Collinear limit}
\label{sec:collinear}

First, we discuss the singular behavior of the EEC in the collinear limit. 
Recall from Eq.~\eqref{eec_int} that the LO EEC can be computed via a single-variable integration over the averaged squared amplitude of the 5-gluon scattering process, denoted as $\displaystyle \overline{\sum\limits_{\text{h,c}} \left| \mathcal{A}_5^{\text{full,tree}} \right|^2}$, combined with the integration kernel $\displaystyle\frac{(1-x)^2 x^2}{(1-x \zeta)^3}$, which encodes the detector configurations. In the collinear limit, where $\zeta \to 0$, the squared amplitude factorizes into the product of a four-point squared amplitude and a collinear factor as follows:
\begin{equation}
    \label{eq: amp_col}
    \overline{\sum\limits_{\text{h,c}}\left| \mathcal{A}_5^{\text{full,tree}} \right|^2}
    = \overline{\sum\limits_{\text{h,c}}\left| \mathcal{A}_4^{\text{full,tree}} \right|^2} 
    C_{gg}(x).
\end{equation}
From the discussion below Eq.~\eqref{eec_int}, $x$ is defined as $x = \frac{2p_3^0}{Q^0}$, representing the energy fraction of the third gluon splitting from a slightly off-shell gluon with momentum $p_c$. Note that the collinear factor $C_{gg}(x)$ is related to the traditional DGLAP splitting function $P_{gg}(x)$ through the following relation:
\begin{equation}
    \label{eq: Pgg}
    C_{gg}(x) = \frac{2g^2 p_\perp^2}{x(1-x)p_c^4} P_{gg}(x),
\end{equation}
where $p_\perp$ is the transverse momentum of the third gluon relative to $p_c$. These quantities are related by $p_\perp^2 = x(1-x)p_c^2$, with $p_c^2 = Q^2 (1-x)x\zeta$.

By substituting Eq.~\eqref{eq: amp_col} and~\eqref{eq: Pgg} into Eq.~\eqref{eec_int}, we obtain the factorized LO EEC in the collinear limit at leading power:
\begin{equation}
    \label{eq: col_fac}
    \frac{d^2\Sigma}{d\Omega_a d\Omega_b } \xrightarrow{\zeta \to 0}
    \frac{Q^2}{16384 \pi^5}\overline{\sum_{\text{h,c}}
    \left| \mathcal{A}_4^{\text{full,tree}} \right|^2}J_{gg},
\end{equation}
where the jet function $J_{gg}$ is:
\begin{equation}
        J_{gg}=\int_0^1  d x (1-x)^2 x^2 C_{gg}(x)
        =\frac{42 g^2 (\mathcal{Y} +1)^2}{5Q^2\mathcal{Y} {\Delta R}^2}.
\end{equation}
Here, ${\Delta R}^2$=$\phi^2+{\Delta Y}^2$ and $\mathcal{Y}=e^Y$. Note that $\Delta R^2\to 0$ in the collinear limit $\zeta \to 0$. Consequently, the LO EEC in the collinear limit at leading power is expressed as:
\begin{equation}
    \frac{d^2\Sigma}{d\Omega_a d\Omega_b } \xrightarrow{\zeta \to 0}
    \frac{189 g^6 \left(\mathcal{Y} ^2+\mathcal{Y} +1\right)^3}{81920 \pi ^5 \text{$\Delta $R}^2 \mathcal{Y} ^3},
\end{equation}
where we have used the four-point squared amplitude:
\begin{equation}
    \label{eq: A4}
    \overline{\sum_{\text{h,c}}
    \left| \mathcal{A}_4^{\text{full,tree}} \right|^2}
    =\frac{9 g^4 \left(\mathcal{Y} ^2+\mathcal{Y} +1\right)^3}{2 \mathcal{Y} ^2 (\mathcal{Y} +1)^2}
\end{equation}
This analysis demonstrates how the LO EEC can be expressed in terms of a four-point squared amplitude and an associated collinear jet function. While beyond the scope of this work, we believe that an all-order (in $\alpha_s$) collinear factorization formula for the EEC exists at hadron colliders, generalizing the $e^+e^-$ collider case derived in~\cite{Dixon:2019uzg,Chen:2023zzh}.

\subsection{Opposite coplanar limit}
\label{sec:coplanar}

The LO EEC exhibits the $\delta \phi^{-1}$ singular behavior in the opposite coplanar limit $\delta \phi = \pi - \phi\to 0$. This is a consequence of the splitting of an incoming gluon into two collinear gluons, one of which interacts with the other incoming gluon.

Eqs.~\eqref{eq: amp_col} and~\eqref{eq: Pgg} remain valid in the opposite coplanar limit, provided that the transverse momenta \( p_\perp \) and \( p_c \) are replaced by their corresponding expressions appropriate to this limit, and the energy fraction \( x = \frac{2p_3^0}{Q^0} \) is substituted by the energy fraction \( y \) of the slightly off-shell gluon resulting from the splitting of the incoming gluon. Since the definition of \( Y \) requires \( Y \geq 0 \), we only need to consider the case where \( p_2^\mu \) splits into \( p_c^\mu \) and \( p_5^\mu \), with the following momentum components:
\begin{align}
    p_2^\mu &= \left( \frac{Q^0}{2}, \, \vec{0}, \, -\frac{Q^0}{2} \right), \\
    p_c^\mu &= \left( \frac{y Q^0}{2}, \, \vec{p}_\perp, \, -\frac{y Q^0}{2} - \frac{p_\perp^2}{(1-y) Q^0} \right), \\
    p_5^\mu &= \left( \frac{(1-y) Q^0}{2}, \, -\vec{p}_\perp, \, -\frac{(1-y) Q^0}{2} + \frac{p_\perp^2}{(1-y) Q^0} \right),
\end{align}
where $p_\perp$ is treated as a small parameter, and the expressions for $p_c^\mu$ and $p_5^\mu$ are given up to $\mathcal{O}(p_\perp^2)$ accuracy. An analogous approximation will also be employed in the following subsection when discussing the back-to-back limit. Here $p_\perp^2$ and $y$ are related to $x$ via momentum conservation through a complicated relation. However, for LP contribution in the LO EEC, we only need the leading power expansions in $\delta \phi$ of $p_\perp^2$ and $y$. These are given by:
\begin{equation}
    p_\perp^2 = \frac{\Delta_{\mathcal Y} \delta \phi^2 Q^2 \left( a^2 (\Delta_{\mathcal Y} +1)^4 \mathcal{Y}^2 + (\mathcal{Y} \Delta_{\mathcal Y} +1)^2 \right)}{(\Delta_{\mathcal Y} + 1)^2 \mathcal{Y} (\mathcal{Y} \Delta_{\mathcal Y} + 1)^2} + \mathcal{O}(\delta \phi^3), \quad y = \frac{1}{\mathcal{Y}} + \mathcal{O}(\delta \phi),
\end{equation}
where $\Delta_{\mathcal Y} = e^{\Delta Y}$, and $a$ is the coefficient of $x$ at the order of $\delta \phi^1$. To be explicit, a similar momentum conservation calculation gives the following relation for $x$:
\begin{equation}
    x = \frac{\mathcal{Y} \Delta_{\mathcal Y} + 1}{\mathcal{Y} \Delta_{\mathcal Y} + \mathcal{Y}} + a \delta \phi + \mathcal{O}(\delta \phi^2).
\end{equation}
Substituting these expressions into Eq.~\eqref{eq: amp_col} and~\eqref{eq: Pgg}, we obtain the factorized form of the LO EEC in the opposite coplanar limit, analogous to the collinear factorization:
\begin{equation}
    \label{eq: op_co_fac}
    \frac{d^2\Sigma}{d\Omega_a d\Omega_b} \xrightarrow{\phi \to \pi} \frac{Q^2}{16384 \pi^5} \overline{\sum_{\text{h,c}} \left| \mathcal{A}_4^{\text{full,tree}} \right|^2} B_{gg},
\end{equation}
where the beam function $B_{gg}$ is given by:
\begin{equation}
  \begin{aligned}
    \label{eq: Bgg}
    B_{gg} &= \int_0^1 dx \frac{(1-x)^2 x^2}{(1-x\zeta)^3} C_{gg}(y) \\
    &= \int_{-\infty}^{+\infty} da \frac{12 g^2 \left( \mathcal{Y}^2 - \mathcal{Y} + 1 \right)^2 \left( \Delta_{\mathcal Y} + \mathcal{Y} \right)^3 \left( \mathcal{Y} \Delta_{\mathcal Y} + 1 \right)^4}{Q^2 \delta \phi \Delta_{\mathcal Y} \left( \Delta_{\mathcal Y} + 1 \right)^2 \left( \mathcal{Y} - 1 \right) \mathcal{Y}^5 \left( a^2 (\Delta_{\mathcal Y} + 1)^4 \mathcal{Y}^2 + (\mathcal{Y} \Delta_{\mathcal Y} + 1)^2 \right)} \\
    &= \frac{12 \pi g^2 \left( \mathcal{Y}^2 - \mathcal{Y} + 1 \right)^2 \left( \Delta_{\mathcal Y} + \mathcal{Y} \right)^3 \left( \mathcal{Y} \Delta_{\mathcal Y} + 1 \right)^3}{Q^2 \delta \phi \Delta_{\mathcal Y} \left( \Delta_{\mathcal Y} + 1 \right)^4 \left( \mathcal{Y} - 1 \right) \mathcal{Y}^6}.
  \end{aligned}
\end{equation}
At LP accuracy in the opposite coplanar limit, the LO EEC factorizes as:
\begin{equation}
    \frac{d^2\Sigma}{d\Omega_a d\Omega_b} \xrightarrow{\phi \to \pi} \frac{27  g^6\left( \Delta_{\mathcal Y}^2 + \Delta_{\mathcal Y} + 1 \right)^3 \left( \mathcal{Y}^2 - \mathcal{Y} + 1 \right)^2 \left( \Delta_{\mathcal Y} + \mathcal{Y} \right)^3 \left( \mathcal{Y} \Delta_{\mathcal Y} + 1 \right)^3}{8192 \pi^4 \delta \phi \Delta_{\mathcal Y}^3 \left( \Delta_{\mathcal Y} + 1 \right)^6 \left( \mathcal{Y} - 1 \right) \mathcal{Y}^6},
\end{equation}
where we have used the result:
\begin{equation}
    \label{eq: A4_delta}    
    \overline{\sum_{\text{h,c}} \left| \mathcal{A}_4^{\text{full,tree}} \right|^2} = \frac{9 g^4 \left( \Delta_{\mathcal Y}^2 + \Delta_{\mathcal Y} + 1 \right)^3}{2 \Delta_{\mathcal Y}^2 \left( \Delta_{\mathcal Y} + 1 \right)^2}.
\end{equation}
Note that the four-point squared amplitude in this limit differs from Eq.~\eqref{eq: A4} only by the replacement $\mathcal{Y}\to \Delta_{\mathcal Y}$.

\subsection{Back-to-back limit}
\label{sec:b2b}

In the back-to-back limit, the LO EEC factorization becomes more intricate compared to the previous two cases. This complexity arises from the fact that there are two different contributions: the collinear splitting of one of the two outgoing gluons, and the emission of a soft gluon from one of the four hard gluons. 

At LP LL accuracy, the fixed-order factorization simplifies significantly, primarily resulting from the splitting of a soft-collinear gluon from one of the two outgoing gluons. Thus, the LO EEC in the back-to-back limit at LP LL accuracy follows a structure analogous to the previous cases:
\begin{equation}
    \label{eq: b2b_fac}
    \frac{d^2\Sigma}{d\Omega_a  d\Omega_b } 
    \xrightarrow{\delta r \to 0}
    \frac{Q^2}{16384 \pi^5} \overline{\sum_{\text{h,c}} 
    \left| \mathcal{A}_4^{\text{full, tree}} \right|^2} 
    \mathcal{CS}_{gg},
\end{equation}
where $\delta r = \sqrt{\delta \phi^2 + (\mathcal{Y} - 1)^2}$ parametrizes the deviation from the back-to-back configuration, and $\mathcal{CS}_{gg}$ denotes the collinear-soft factor. In order to compute this factor, it is crucial to first analyze the kinematics in the back-to-back limit.

Consider a slightly off-shell gluon with momentum $p_c^\mu$ that splits into two collinear gluons. One of these gluons, with momentum $p_3^\mu$ (or $p_4^\mu$), is detected by a detector aligned along the direction of $\vec{n}_a$ (or $\vec n_b$), while the other gluon, with momentum $p_5^\mu$, is soft. The momenta of these three gluons can be more conveniently expressed in a frame where the z-axis is aligned with the direction of the slightly off-shell gluon. In this frame, the momenta of the three gluons are given by:
\begin{align}
    p_c^\mu &= \left(p_c, \vec{0}, p_c - \frac{p_\perp^2}{2 p_c y (1 - y)}\right), \\
    p_3^\mu &= \left(y p_c, \vec{p}_\perp, y p_c - \frac{p_\perp^2}{2 y p_c}\right), \\
    p_5^\mu &= \left((1 - y) p_c, -\vec{p}_\perp, (1 - y) p_c - \frac{p_\perp^2}{2 (1 - y) p_c}\right),
\end{align}
where $\vec{p}_\perp$ is of order $\delta r^1$, and $y$ deviates slightly from $1$ by an amount of order $\delta r^1$. Similar to the opposite coplanar limit, we can use momentum conservation to relate $p_c$, $p_\perp^2$, and $x$ to $y$:
\begin{align}
    p_c &= \frac{Q^0}{2} \left(1 + \frac{y \Delta_{\mathcal Y} {\delta r}^2}{(1 - y)(1 + \Delta_{\mathcal Y})^2}\right), \\
    p_\perp^2 &= \frac{Q^2 y^2 \Delta_{\mathcal Y} {\delta r}^2}{(1 + \Delta_{\mathcal Y})^2}, \\
    x &= y \left(1 + \frac{y \Delta_{\mathcal Y} {\delta r}^2}{(1 - y)(1 + \Delta_{\mathcal Y})^2}\right).
\end{align}

Substituting these expressions into the following definition of the collinear-soft factor $\mathcal{CS}_{gg}$, where the factor of 2 accounts for the two jets in the back-to-back limit,
\begin{equation}
    \mathcal{CS}_{gg} = 2 \int_0^1 dx \frac{(1 - x)^2 x^2}{(1 - x \zeta)^3} C_{gg}(y),
\end{equation}
and changing the integration variable from $x$ to $y$, we obtain the collinear-soft factor:
\begin{equation}
        \label{eq: csgg_LL}
        \mathcal{CS}_{gg} \approx \int_0^{1 - \frac{\sqrt{\Delta_{\mathcal Y}} \delta r}{\Delta_{\mathcal Y} + 1}} 
        dy \frac{24 g^2 (\Delta_{\mathcal Y} + 1)^2 \left(y^2 - y + 1\right)^2}{\Delta_{\mathcal Y} Q^2 (1 - y){\delta r}^2} 
        = -\frac{24 (\Delta_{\mathcal Y} + 1)^2 g^2 \ln \delta r}{\Delta_{\mathcal Y} Q^2 {\delta r}^2} + \cdots.
\end{equation}
Here the ellipsis represents terms beyond LP LL accuracy. Thus, the LO EEC  at LP LL accuracy in the back-to-back limit is:
\begin{equation}
    \label{eq: b2b_LL}
    \frac{d^2\Sigma}{d\Omega_a  d\Omega_b } 
    \xrightarrow{\delta r \to 0} -\frac{27 g^6 \left(1 + \Delta_{\mathcal Y} + \Delta_{\mathcal Y}^2\right)^3  
    \ln \delta r}{4096 \pi^5 \Delta_{\mathcal Y}^3 {\delta r}^2},
\end{equation}
where we have used Eq.~\eqref{eq: A4_delta}.

The calculation of the non-logarithmically enhanced term requires separate treatment of soft gluon emission and collinear splitting contributions.
Referring back to Eq.~\eqref{eec_int}, the LO EEC can be computed through a single-variable integration over \( x \), the energy fraction of the third gluon. In the back-to-back limit, \( x \) ranges from 0 to 1, covering the full kinematic region, which includes both soft gluon emission and collinear splitting. By applying momentum conservation, we can directly relate \( p_5^\mu \) to \( x \). In the back-to-back limit where \( \delta r \to 0 \), this relation simplifies accordingly.

However, it is important to note that this simplification is sensitive to the value of $x$. Specifically, when $x$ deviates from 1 by an amount of order $\delta r^0$, the relation at leading order in $\delta r$, is given by:
\begin{equation}
    p_5^\mu = \left(\frac{Q^0}{2}(1 - x), \frac{Q^0 (1 - x) \sqrt{\Delta_{\mathcal Y}}}{1 + \Delta_{\mathcal Y}}, 
    0, \frac{Q^0 (1 - x)(\Delta_{\mathcal Y} - 1)}{2 (1 + \Delta_{\mathcal Y})}\right),
\end{equation}
which is collinear to the momentum of the third gluon. Similarly, when $x$ deviates from 1 by an amount of order $\delta r^1$, the fifth gluon becomes soft. Conversely, when $x$ deviates by an amount of order ${\delta r}^2$ or less, the fifth gluon becomes collinear with the fourth gluon.

Therefore, the LO EEC at LP accuracy in the back-to-back limit can be separated into two terms based on the value of $x$:
\begin{equation}
    \label{eq: b2b_fac_cs}
    \frac{d^2\Sigma}{d\Omega_a  d\Omega_b } 
    \xrightarrow{\delta r \to 0}
    \frac{Q^2}{16384 \pi^5} 
    \left(\overline{\sum_{h,c} 
    \left| \mathcal{A}_4^{\text{full, tree}} \right|^2} \mathcal{C}_{gg}
    + \overline{\sum 
    \mathcal{H}_{4g}\mathcal{S}_{gg}} \right).
\end{equation}
In the first term, due to the exchange symmetry between the third and fourth gluons in the back-to-back limit, the contributions from the collinear splitting of the two gluons are identical. We have absorbed the factor of 2 arising from the exchange symmetry into the collinear factor $\mathcal{C}_{gg}$ to maintain consistency with the collinear-soft factor $\mathcal{CS}_{gg}$. In the second term, the summation runs over all possible configurations. Here, $\mathcal{H}_{4g}$ denotes the hard function, which depends on the color configurations of the four hard gluons, while $\mathcal{S}_{gg}$ represents the soft factor. Since a soft gluon can be emitted from any of the four hard gluons, it is crucial to consider the color configurations in order to properly factorize the scattering amplitude.

The collinear factor $\mathcal{C}_{gg}$ is straightforward to compute, as it corresponds to the NLL term in Eq.~\eqref{eq: csgg_LL}:
\begin{equation}
    \mathcal{C}_{gg} = -\frac{2g^2(\Delta_{\mathcal Y} +1)^2  
    (6\ln\Delta_{\mathcal Y} -12 \ln (\Delta_{\mathcal Y} +1)+11)}{Q^2 \Delta_{\mathcal Y}  {\delta r}^2 },
\end{equation}
which encapsulates the contribution from both situations where the gluon with momentum \( p_3 \) or \( p_4 \) is collinearly split from a slightly off-shell gluon.

In contrast, the soft factor $\mathcal{S}_{gg}$ requires more careful treatment, as it accounts for the soft gluon emission, which is sensitive to the color configurations of the gluons and the specific leg from which the emission occurs. We begin by considering the singular factorization approximation of the 5-gluon scattering amplitude in the presence of soft gluon emission:
\begin{equation}
    \label{eq: A5_soft}
    \mathcal{A}_5^{\text{full,tree}}(\{c\},\{h\})\xrightarrow{\text{soft emission}}
    \sum_{i,c^\prime} \mathcal{A}_4^{\text{full,tree}}(i,\{c\},\{h\},c^\prime)
    S_{gg}(i,h_5,\{c\},c^\prime),
\end{equation}
where $i$ denotes the $i$-th leg from which the soft gluon is emitted, and $c^\prime$ is the color index of the slightly off-shell gluon. The sets $\{h\}$ and $\{c\}$ represent the helicities and colors of the five gluons, respectively. The dependence on $i$ in $\mathcal{A}_4^{\text{full,tree}}$ is realized by changing the color index of the $i$-th gluon from $c_i$ to $c^\prime$. Note that $S_{gg}$ is different from the soft factor $\mathcal{S}_{gg}$ introduced in Eq.~\eqref{eq: b2b_fac_cs}, and it depends on $i$, $h_5$, $\{c\}$ and $c^\prime$. An explicit calculation shows that the soft factor $S_{gg}$ is given by:
\begin{equation}
    S_{gg}(i,h_5,\{c\},c^\prime)=-2g f^{c^\prime c_i c_5}\frac{p_i\cdot \epsilon_5^*}{p_i\cdot p_5},
\end{equation}
where $c_i$ denotes the color index of the $i$-th gluon, and $f^{c^\prime c_i c_5}$ is the structure constant of the color group $SU(3)$. The fifth gluon is soft, and its polarization vector is denoted by $\epsilon_5^*$. 

Starting from Eq.~\eqref{eq: A5_soft}, we can derive the singular factorization approximation of the squared amplitude: 
\begin{equation}
    \begin{aligned}
        \sum_{\{h\},\{c\}}\left| \mathcal{A}_5^{\text{full, tree}} \right|^2=
        \sum_{\{c\},c^\prime,c^{\prime\prime},i,j}
        &\left(\sum_{\{h\}} \mathcal{A}_4^{\text{full,tree}}(i,\{c\},\{h\},c^\prime)
        \mathcal{A}_4^{*\text{full,tree}}(j,\{c\},\{h\},c^{\prime\prime})\right)\\
        &\quad \times\left(\sum_{c_5,h_5} S_{gg}(i,h_5,\{c\},c^\prime)
        S_{gg}^*(j,h_5,\{c\},c^{\prime\prime})\right).
    \end{aligned}
\end{equation}

By inserting the expressions above into Eq.~\eqref{eec_int}, we can compute the soft factor $\mathcal{S}_{gg}$ as follows:
\begin{equation}
    \begin{aligned}
        \mathcal{S}_{gg} &= \int_0^1 dx \frac{(1 - x)^2 x^2}{(1 - x \zeta)^3} 
        \left(\sum_{c_5,h_5} S_{gg}(i,h_5,\{c\},c^\prime)S_{gg}^*(j,h_5,\{c\},c^{\prime\prime})\right)\\
        &=\int_{-\frac{1}{\delta r}}^{-\delta r}  \frac{da}{a} 
        \left(4g^2\sum_{c_5} f^{c^\prime c_i c_5} f^{c^{\prime\prime} c_j c_5}
        \frac{p_i\cdot p_j}{\left(p_i\cdot p_5\right)\left( p_j\cdot p_5\right)}\right),
    \end{aligned}
\end{equation}
where we have used 
\begin{equation}
        \sum_{h_5} \epsilon_5^{*\mu}\epsilon_5^{*\nu}=-g^{\mu\nu}, \quad x=1+a \delta r.
\end{equation}

Similarly, the hard function $\mathcal{H}_{4g}$ is obtained via the following formula: 
\begin{equation}
    \mathcal{H}_{4g} = \sum_{\{h\}} \mathcal{A}_4^{\text{full,tree}}(i,\{c\},\{h\},c^\prime)
    \mathcal{A}_4^{*\text{full,tree}}(j,\{c\},\{h\},c^{\prime\prime}).
\end{equation}

Inserting all the relevant expressions into Eq.~\eqref{eq: b2b_fac_cs} and combining it with Eq.~\eqref{eq: b2b_LL}, we obtain the LO EEC in the back-to-back limit:
\begin{equation}
    \begin{aligned}
        \frac{d^2\Sigma}{d\Omega_a  d\Omega_b } 
        \xrightarrow{\delta r \to 0}
        &-\frac{9g^6 \left(\Delta_{\mathcal Y} ^2+\Delta_{\mathcal Y} +1\right)^3  
        \left(12 \ln \left(\frac{ {\delta r}\sqrt{\Delta_{\mathcal Y} }}{\Delta_{\mathcal Y} +1}\right)
        -12 \varphi  \tan \varphi+11\right)}
        {16384 \pi ^5 \Delta_{\mathcal Y} ^3 {\delta r}^2}\\
        &+\frac{27g^6 \left(\Delta_{\mathcal Y} ^2+\Delta_{\mathcal Y} +1\right)^2  
        \left(2 \left(\Delta_{\mathcal Y} ^2+1\right) \varphi  \sin 2 \varphi 
        +\left(\Delta_{\mathcal Y} ^2-1\right) \ln \Delta_{\mathcal Y} \cos 2 \varphi \right)}
        {16384 \pi ^5 \Delta_{\mathcal Y} ^2 {\delta r}^2 \left(\Delta_{\mathcal Y} ^2-2 \Delta_{\mathcal Y}  \cos 2 \varphi +1\right)},                
    \end{aligned}
\end{equation}
where $\varphi$ is defined as $\displaystyle \varphi=\arctan{\left(\frac{\mathcal{Y}-1}{\delta\phi}\right)}$.

\subsection{Regge limit}
\label{sec:Regge}

The three limits of the EEC discussed above have been extensively studied in the EEC at $e^+e^-$ colliders or the TEEC at hadron colliders~\cite{Dixon:2019uzg,Moult:2018jzp,Gao:2019ojf,Gao:2024wcg}. However, the EEC at hadron colliders also exhibits an interesting divergent behavior in the limit $\Delta Y \to \infty$ (or $\Delta_{\mathcal Y}=e^{\Delta Y}\to\infty$), i.e. when the two detectors are widely separated in rapidity, which has not been explored in previous studies of energy correlators. This specific limit corresponds to the Regge limit where the momentum transfer $t \to 0$ in the $t$-channel gluon exchange during the scattering process. At higher perturbative orders, this regime is associated with the so-called multi-Regge kinematics, in which multiple particles are emitted with strongly ordered rapidities between the two detectors.

While the physics in the Regge limit has been well-studied in the phenomenology such as Mueller-Navelet jets, it remains to be investigated in the context of energy correlators. This is particularly interesting because the light-ray operators and energy correlators have nice connections with the Regge theory~\cite{Kravchuk:2018htv,Kologlu:2019bco,Chen:2024iuv}. Therefore, studying high-energy scattering in the Regge limit from the perspective of energy correlators may yield many exciting insights.

In this subsection, we study the EEC in the Regge limit and reproduce the LO EEC result at LP LL accuracy. Taking the \(\Delta Y \to \infty\) (or $\Delta_{\mathcal Y}\to\infty$) limit of LO expression in Eq.~\eqref{full result}, we obtain the following result at LP LL accuracy:

\begin{equation} 
    \label{eq: regge_result} 
    \frac{d^2\Sigma}{d\Omega_a d\Omega_b }\xrightarrow{\Delta_{\mathcal Y} \to \infty} \frac{27 g^6\,\mathcal{Y} \,\Delta_{\mathcal Y}^3\,\ln(\Delta_{\mathcal Y})}{8192 \pi ^5 \left(1+2 \mathcal{Y} \cos \phi +\mathcal{Y}^2\right)}. 
\end{equation}

The difficulty in factorizing the EEC in the Regge limit arises from the fact that, unlike the previous limits where factorization typically results from collinear splitting or soft radiation of the external legs in a 2-to-2 hard scatter, here we are dealing with a more complicated 2-to-3 factorization. This is evident from the divergent behavior in Eq.~\eqref{eq: regge_result}, which persists for all values of the azimuthal angle \(\phi\). This divergence is associated with the singularity of the \(t\)-channel propagator, which requires a detailed study to properly factorize. In the language of SCET, this kinematic region is related to the Glauber contribution~\cite{Rothstein:2016bsq}.

\begin{figure}
    \centering
    \includegraphics[width=0.6\linewidth]{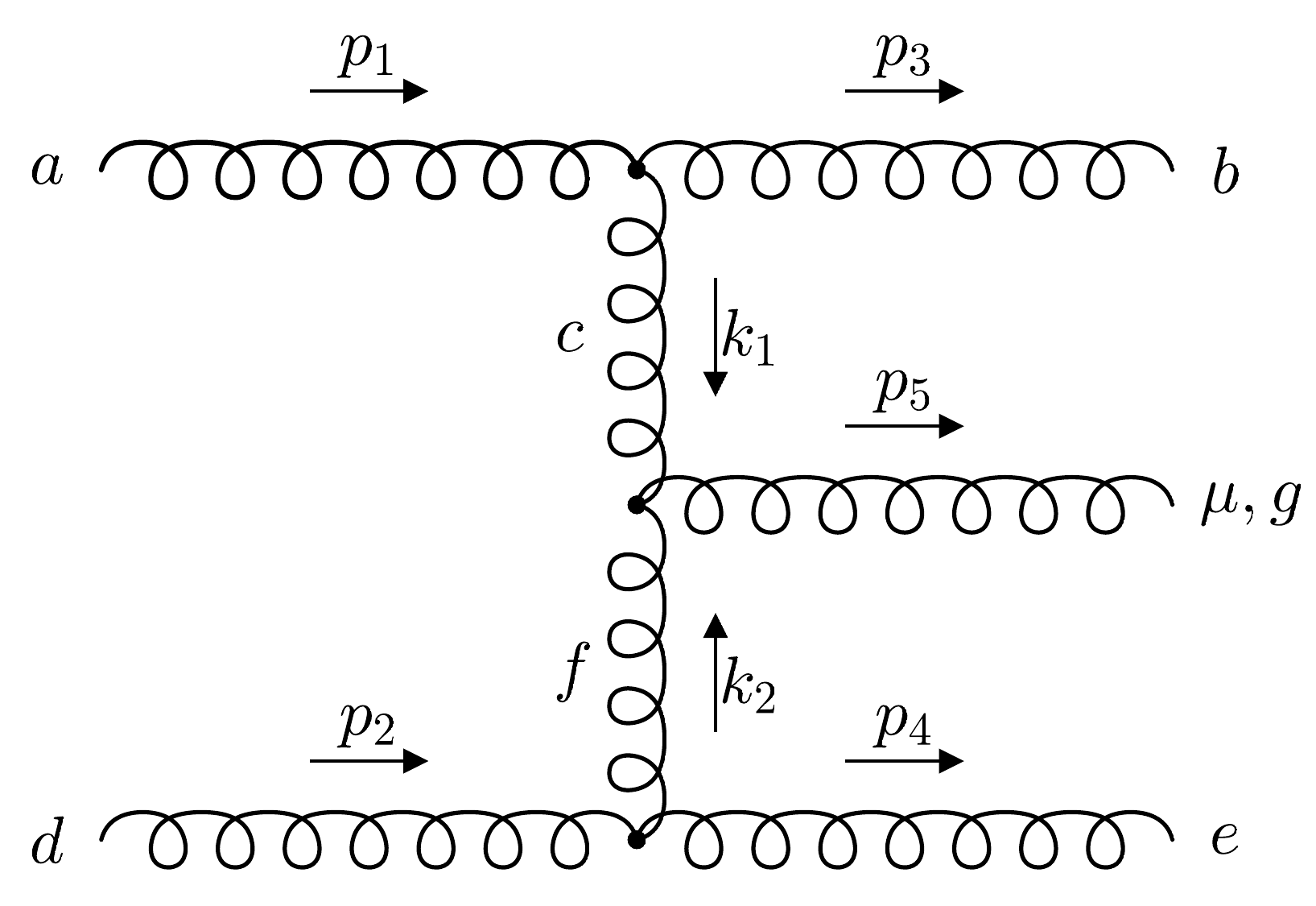}
    \caption{The ladder diagram for gluon scattering. The Latin letters \(a\) through \(g\) denote color indices of the corresponding gluons, while the Greek letter \(\mu\) denotes the Lorentz index of the corresponding gluon. The polarization index \(h_i\) for momentum \(p_i\) has been suppressed for clarity.}
    \label{fig:regge limit a}
\end{figure}

Inspired by Lipatov's calculation of the two-loop corrections for reggeized gluon exchange in the 2-to-2 quark scattering process~\cite{Kuraev:1976ge, Lipatov:1976zz}, we perform a similar calculation for gluon scattering in the Regge limit. We begin by first calculating the most singular contribution to the scattering amplitude shown in Fig.~\ref{fig:regge limit a}.

\begin{figure}
    \centering
    \includegraphics[width=0.45\linewidth]{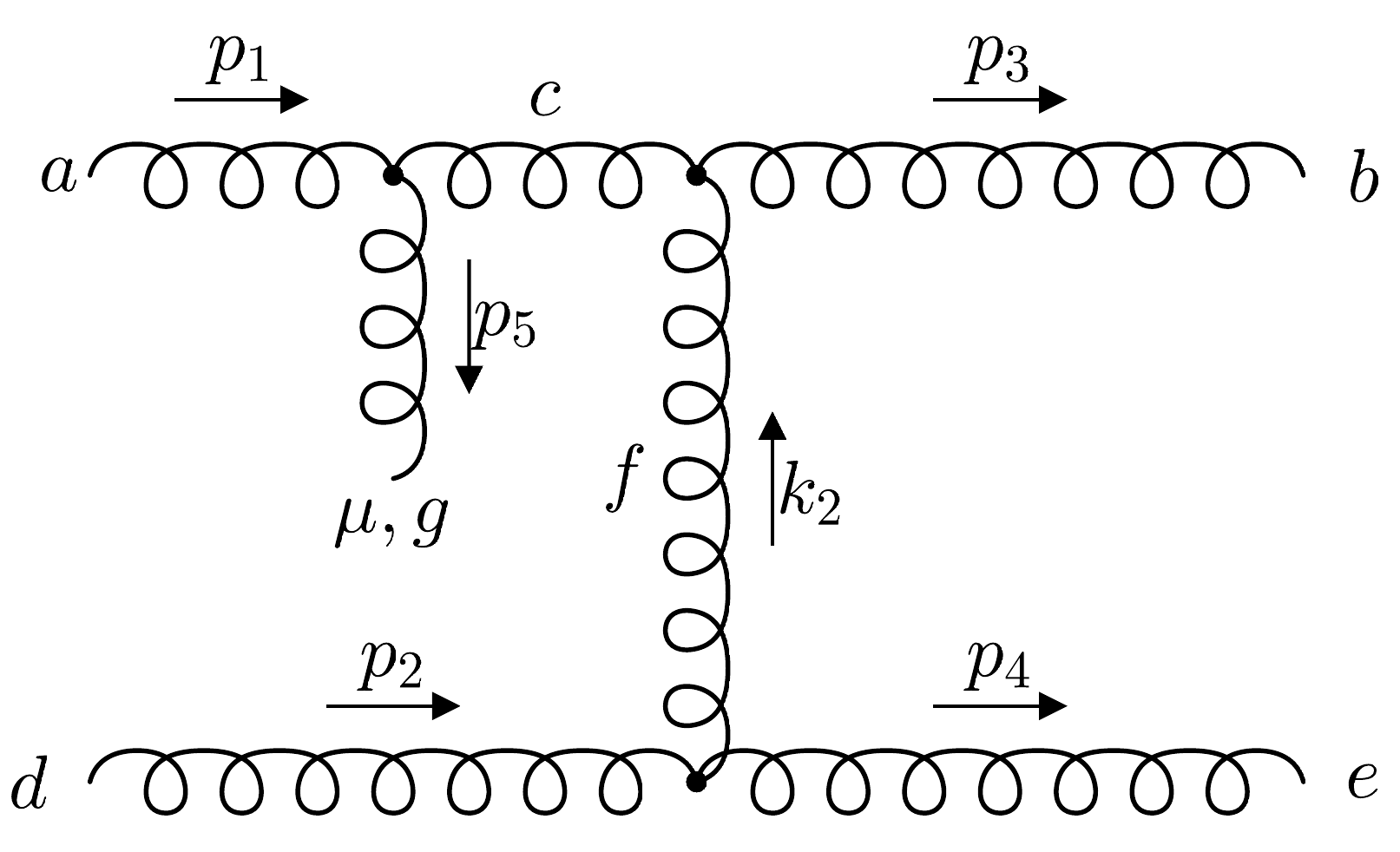}\quad
    \includegraphics[width=0.45\linewidth]{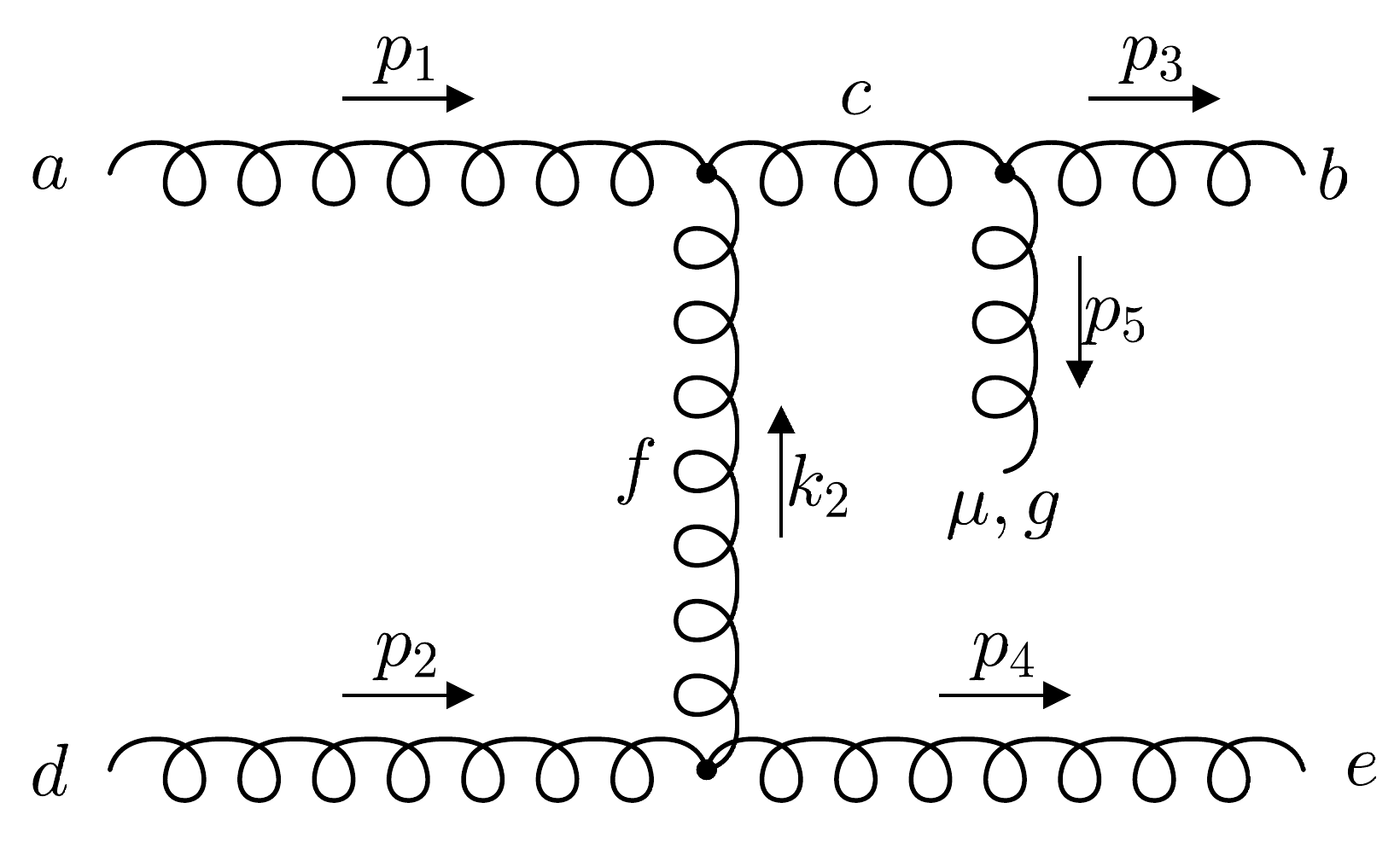}
    \caption{The diagrams for gluon scattering with $p_5$ coming out from the upper gluon line.}
    \label{fig:regge limit b&c}
\end{figure}

\begin{figure}
    \centering
    \includegraphics[width=0.45\linewidth]{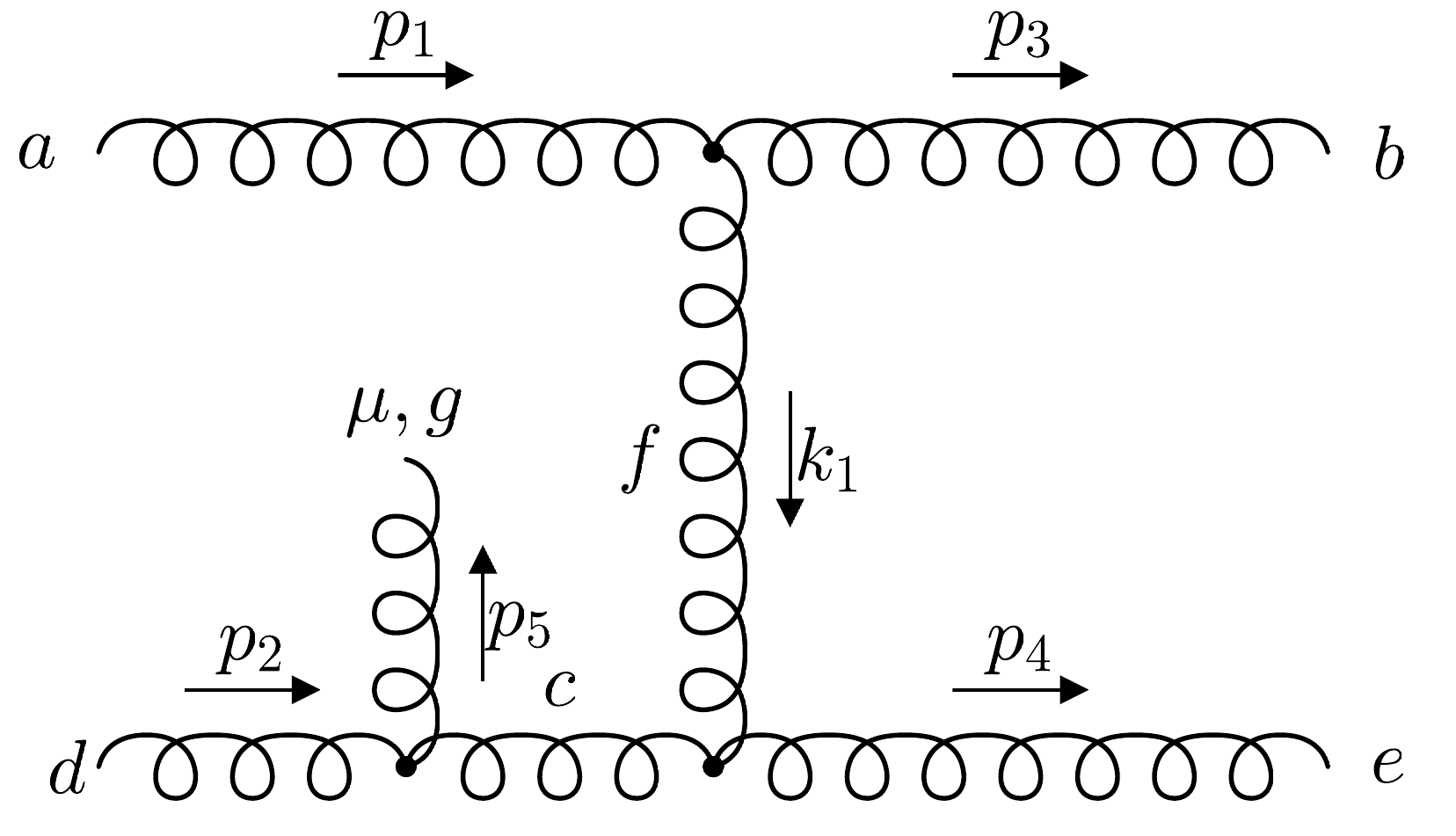}\quad
    \includegraphics[width=0.45\linewidth]{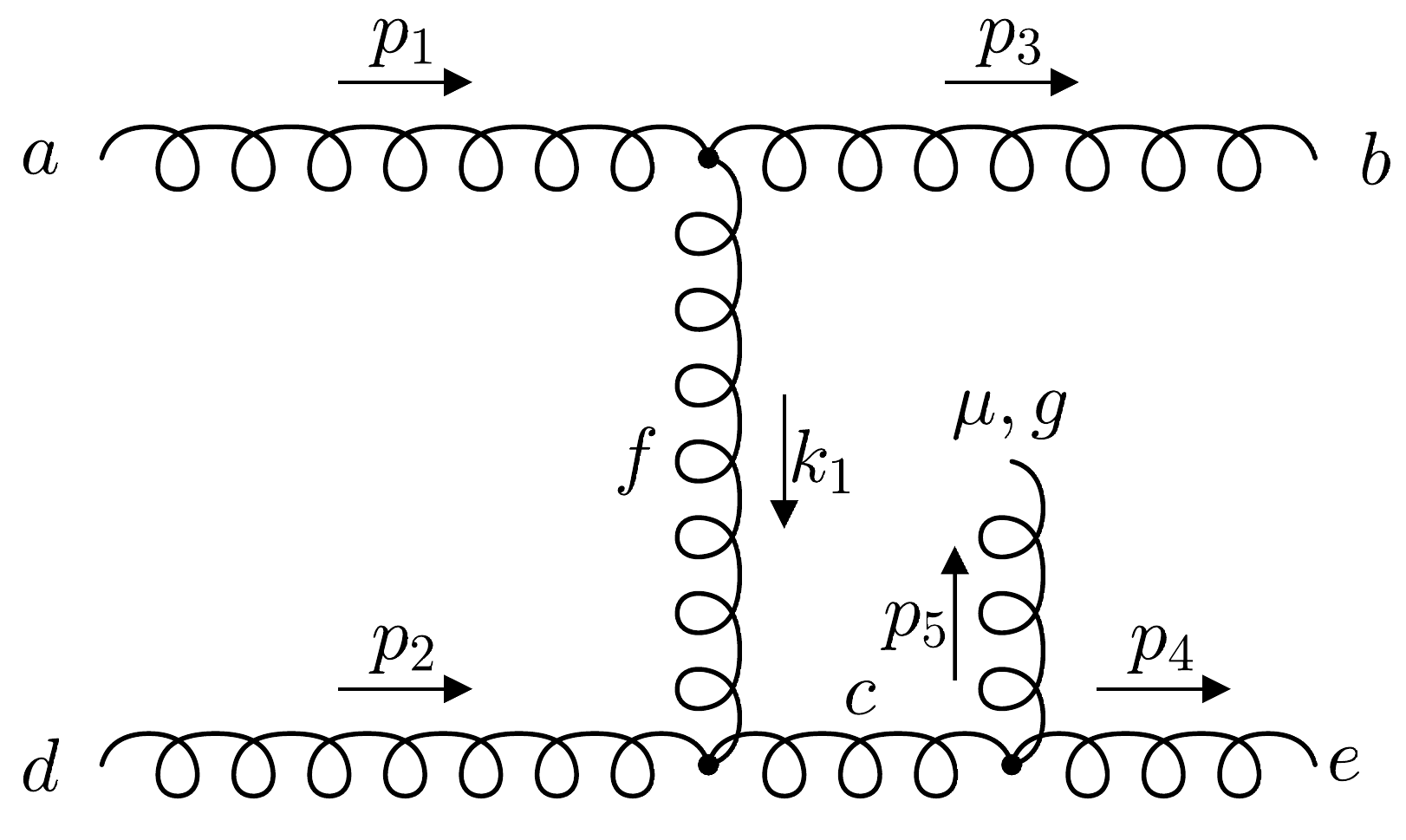}
    \caption{The diagrams for gluon scattering with $p_5$ coming out from the lower gluon line. }
    \label{fig:regge limit d&e}
\end{figure}

Now we parametrize the momenta \(k_1\) and \(k_2\) in terms of Sudakov parameters \(\rho_1\), \(\rho_2\), and \(\lambda_1\), \(\lambda_2\):
\begin{equation}
    \label{eq:Sudakov decomposition}
    \begin{aligned}
        k_1^\mu &= \rho_1 p_1^\mu + \lambda_1 p_2^\mu + k_{1\perp}^\mu, \\
        k_2^\mu &= \rho_2 p_1^\mu + \lambda_2 p_2^\mu + k_{2\perp}^\mu.
    \end{aligned}
\end{equation}
In the Regge limit, the dominant divergence of this amplitude arises from the region where
\begin{equation}
    \label{eq:regge region}
    \begin{aligned}
        1 \gg |\rho_1| \gg |\rho_2|, \\
        1 \gg |\lambda_2| \gg |\lambda_1|.
    \end{aligned}
\end{equation}
Within this kinematic regime, the amplitude can be computed using the eikonal approximation. By suppressing the polarization vector \(\epsilon^{*\mu}(p_5, h_5)\), we derive the leading divergent contribution to the amplitude for Fig.~\ref{fig:regge limit a}:
\begin{equation}
    \label{eq:regge a}
    i M_a^\mu = \frac{2g^3}{\lambda_1 \rho_2 s} \eta^{h_1 h_3}\eta^{h_2 h_4} f_{abc} f_{def} f_{cfg} (\rho_1 p_1^\mu - \lambda_2 p_2^\mu - k_{1\perp}^\mu + k_{2\perp}^{\mu}).
\end{equation}
Here, \( s = (p_1 + p_2)^2 \) is the square of the center-of-mass energy and $\eta^{h_i h_j}$ is the metric in the helicity space. 
There are other diagrams that contribute at the same order of divergence, in which the gluon with momentum \(p_5\) comes from the four external gluon legs, as shown in Fig.~\ref{fig:regge limit b&c} and Fig.~\ref{fig:regge limit d&e}.

Since these Feynman diagrams must be analyzed within the same kinematic configuration, Eq.~\eqref{eq:Sudakov decomposition} remains valid with \(p_5^\mu = k_1^\mu + k_2^\mu\). The two Feynman diagrams in Fig.~\ref{fig:regge limit b&c} can be combined, yielding a compact expression for the leading divergent part of the amplitudes, analogous to the form in Eq.~\eqref{eq:regge a}:

\begin{equation}
    iM_{b,c}^\mu=-\frac{4g^3}{\rho_2\lambda_2 s}\eta^{h_1 h_c}\eta^{h_c h_3} \eta^{h_2 h_4}f_{abc} f_{def} f_{cfg}\, p_1^\mu.
\end{equation}
Here, \(h_c\) denotes the polarization index of the gluon associated with color index \(c\).
Similarly, the leading divergent components of the amplitudes depicted in Fig.~\ref{fig:regge limit d&e} combine into:

\begin{equation}
    iM_{d,e}^\mu=\frac{4g^3}{\rho_1\lambda_1 s}\eta^{h_1 h_3}\eta^{h_2 h_c} \eta^{h_c h_4}f_{abc} f_{def} f_{cfg}\, p_2^\mu.
\end{equation}
By summing all the contributions outlined above, we obtain the scattering amplitude that captures the leading divergence of the EEC in the Regge limit:
\begin{equation}
    \label{eq:regge amp}
    iM^\mu=\frac{2g^3}{s}f_{abc}f_{def}f_{cfg}\eta^{h_1 h_3}\eta^{h_2 h_4}\Gamma^\mu(k_1,k_2),
\end{equation}
where $\Gamma^\mu(k_1, k_2)$ is intrinsically Lipatov's effective vertex for reggeized gluon exchange, which is determined to be:
\begin{equation}
    \Gamma^\mu(k_1,k_2)=\frac{1}{\lambda_1\rho_2}\left(\left(\rho_1+\frac{2\lambda_1}{\lambda_2}\right)p_1^\mu-\left(\lambda_2+\frac{2\rho_2}{\rho_1}\right)p_2^\mu-k_{1\perp}^\mu+k_{2\perp}^\mu\right).
\end{equation}
In other words, by measuring the EEC in the Regge limit, we can directly probe this effective vertex.

To reproduce the LP LL divergence in the Regge limit as shown in Eq.~\eqref{eq: regge_result}, we substitute Eq.~\eqref{eq:regge amp} into Eq.~\eqref{eec_int} and express the Sudakov parameters $\rho_i$ and $\lambda_i$ in terms of the phase space variables $\mathcal{Y}$, $\Delta_{\mathcal Y}$, $\phi$, and $x$. Summing over all final-state color and polarization indices and averaging over the initial-state color and polarization configurations, we obtain:
\begin{equation}
    \frac{d^2\Sigma}{d\Omega_a d\Omega_b } \xrightarrow{\Delta_{\mathcal Y} \to \infty}\frac{27g^6  \Delta_{\mathcal Y}^3\,\mathcal{Y}}{8192\pi^5}\int_{1-\delta}^{1-\frac{1}{\Delta_{\mathcal Y}}} dx\, \frac{x}{(1-x)(\mathcal{Y}^2+2\,x\,\mathcal{Y}\cos(\phi)+x)}.
\end{equation}
The integration region of the integral is determined by Eq.~\eqref{eq:regge region}, with $\delta$ being an arbitrary number of order $\Delta_{\mathcal Y}^0$ between 0 and 1. Consequently, the integration over $x$ yields the LL divergence as:
\begin{equation}
    \frac{d^2\Sigma}{d\Omega_a d\Omega_b }\xrightarrow{\Delta_{\mathcal Y} \to \infty} \frac{27 g^6\,\mathcal{Y} \,\Delta_{\mathcal Y}^3\,\ln(\Delta_{\mathcal Y})}{8192 \pi ^5 \left(1+2 \mathcal{Y} \cos \phi +\mathcal{Y}^2\right)},
\end{equation}
which is exactly what we have extracted from the LO result.

It is worth emphasizing that although we are working in pure Yang–Mills theory, the Regge limit result at LO can be directly extended to QCD, differing only by the color factors associated with different initial parton states. This is because, the Regge limit in QCD is still dominated by gluon exchange in the \( t \)-channel. 
The resummation of the LP LL EEC in the Regge limit is in fact related to the solution of the BFKL equation, which we leave for future work.

\section{Landscape plots of EEC}
\label{sec:plot}
Due to the highly intricate structure of the LO EEC result for pure gluon scattering, as shown in Eq.~\eqref{full result}, we present in this section the corresponding plots both with and without convolution over PDFs. These are displayed as landscape plots in the boost-invariant $\Delta Y$-$\phi$ plane for various values of $Y$, enhancing their practical utility for experimental measurements. These plots offer a more intuitive visualization of the limiting behaviors discussed in Sec.~\ref{sec:factorization}, manifesting various interesting physics out of complicated theoretical results.

\subsection{Plots in the parton frame}
\label{sec: plot for gluons}
\begin{figure}[ht]
    \begin{center}
        \makebox[\textwidth][c]{
            \subfloat[]{
                \includegraphics[width=0.6\textwidth]{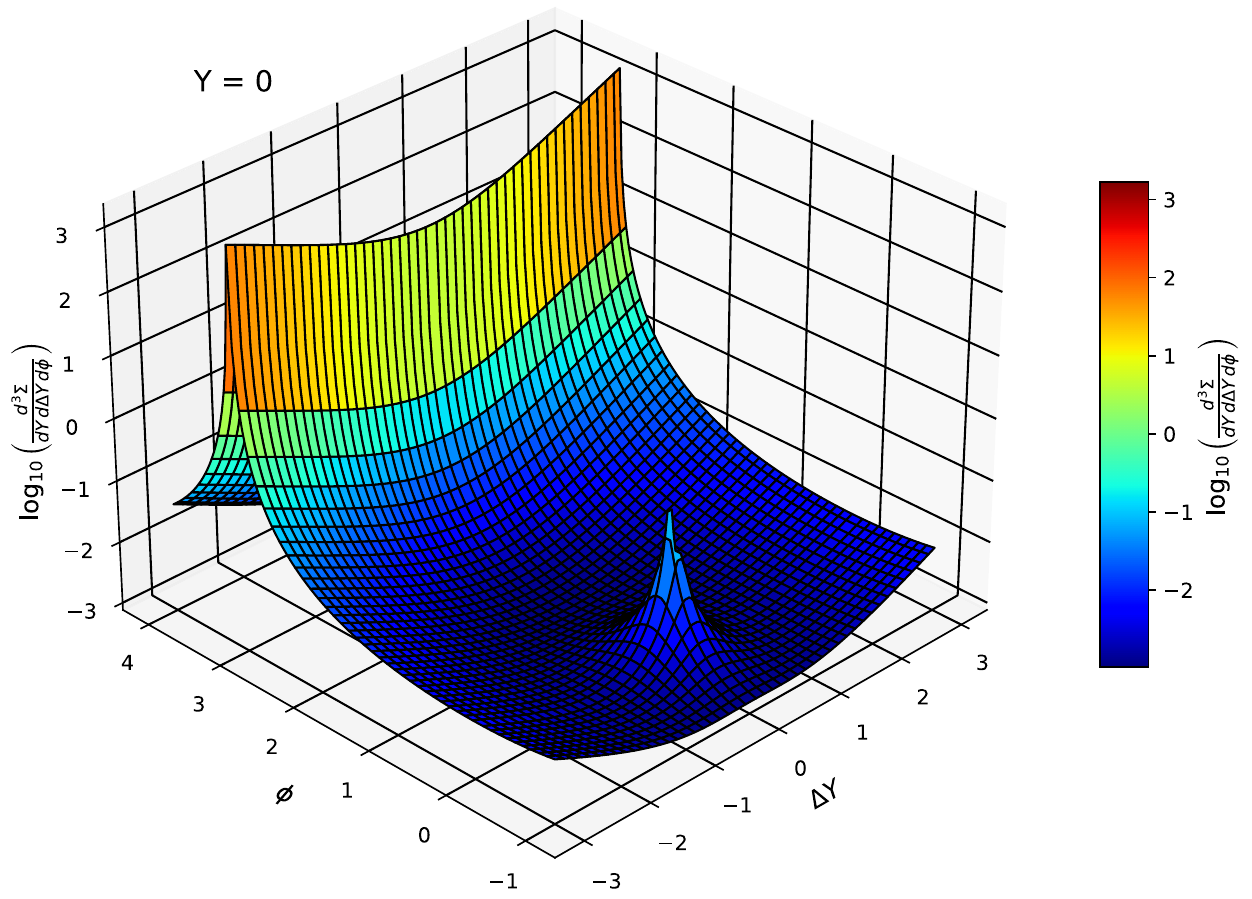}\label{Figure parton_plot.a}
                }
            \subfloat[]{
                \includegraphics[width=0.62\textwidth]{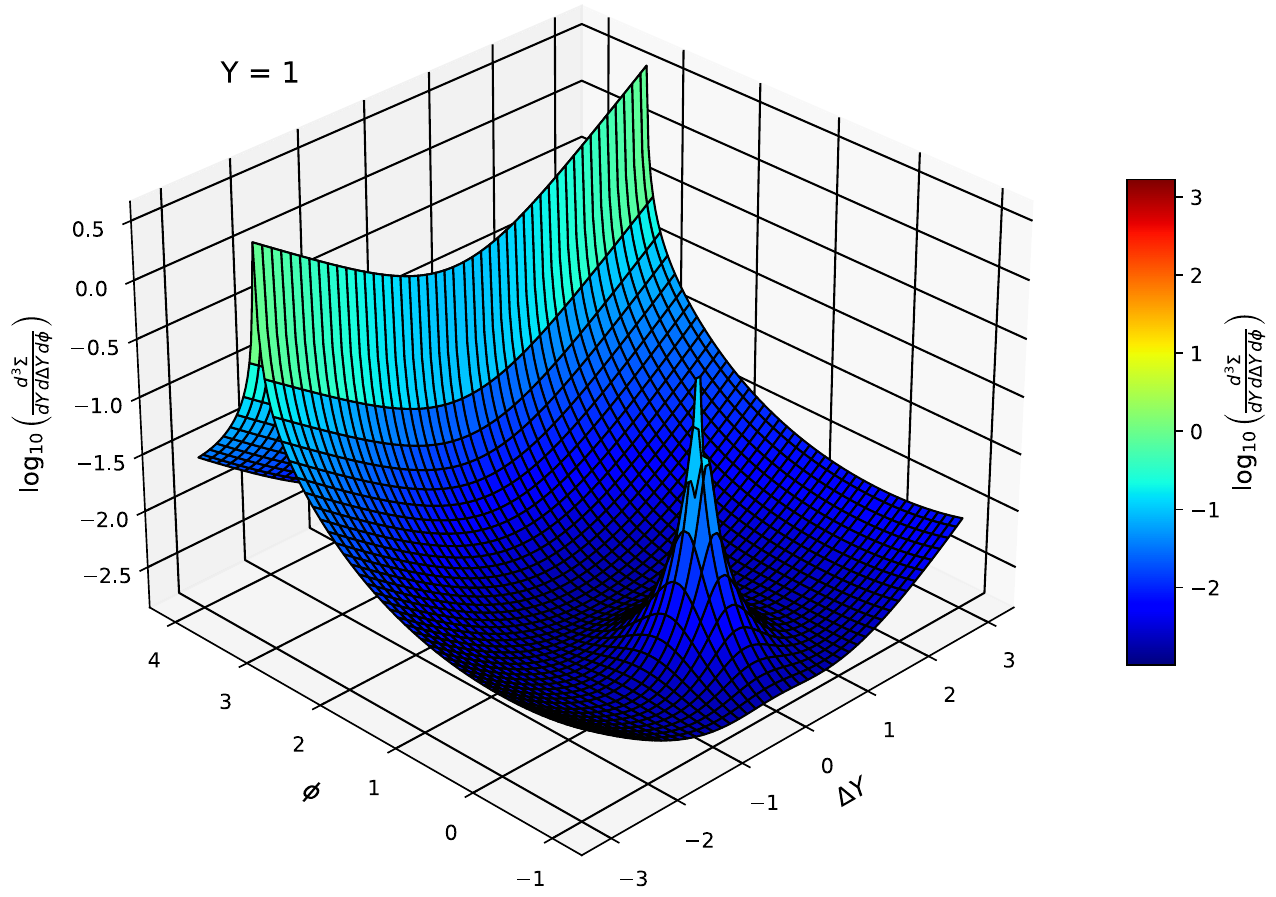}\label{Figure parton_plot.b}
                }
        }\\
        \makebox[\textwidth][c]{
            \subfloat[]{
                \includegraphics[width=0.64\textwidth]{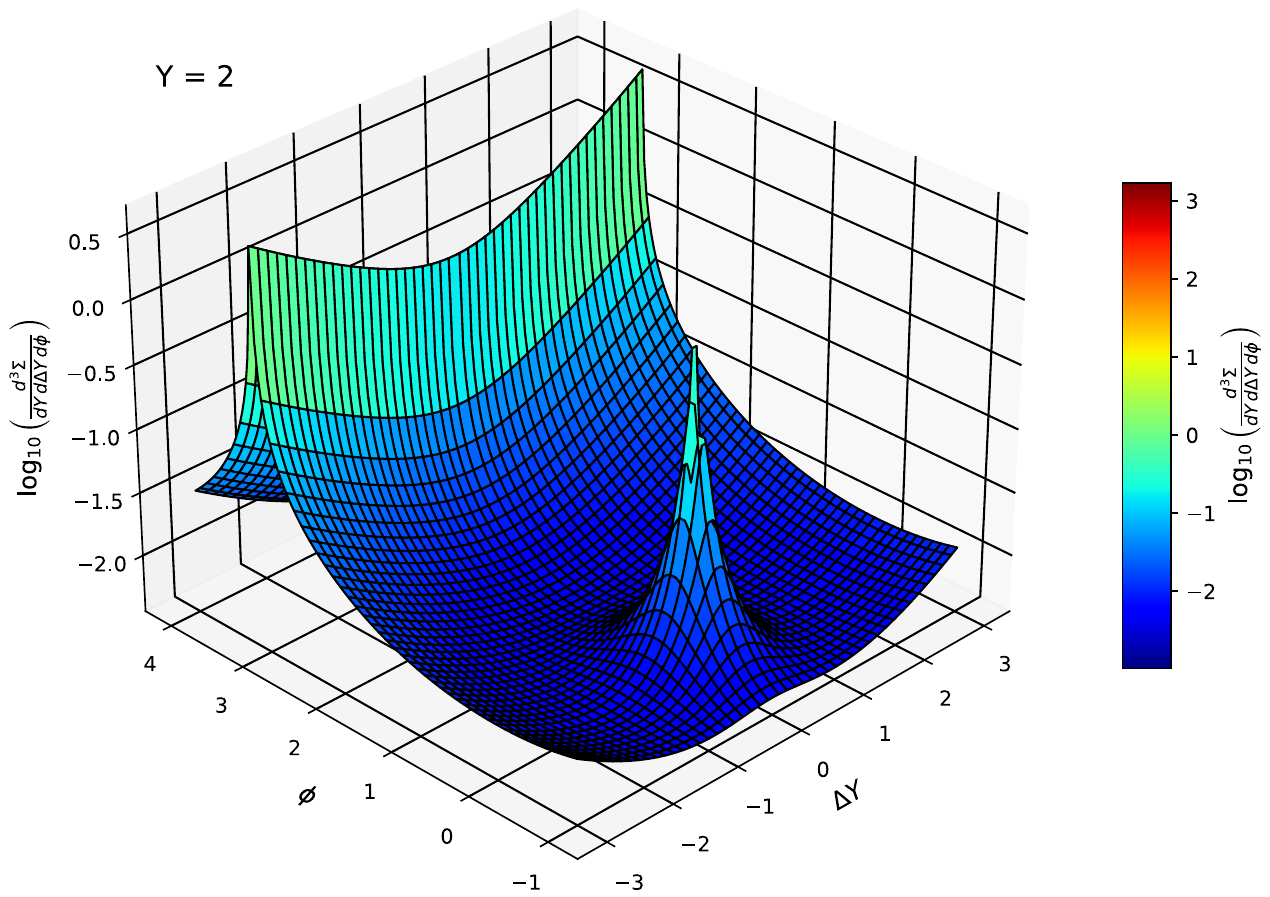}\label{Figure parton_plot.c}
                }
            \subfloat[]{
                \includegraphics[width=0.63\textwidth]{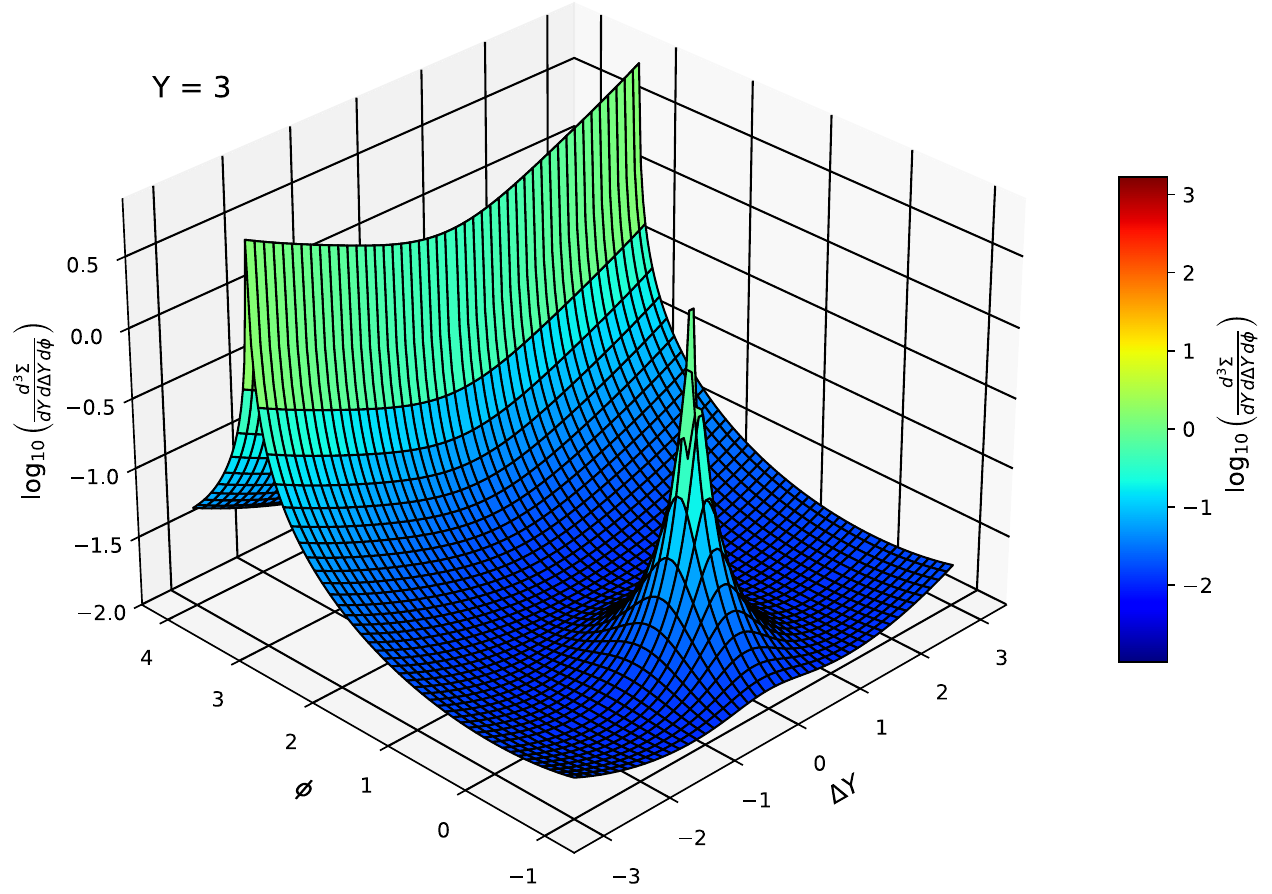}\label{Figure parton_plot.d}
                }
        }
    \end{center}
    \caption{
    Landscape plots for the EEC at the parton level, shown in the \( \Delta Y \)-\( \phi \) plane for values of \( Y \) ranging from 0 to 3, corresponding to panels (a) through (d). The EEC values are displayed on a logarithmic scale (base 10), with \( g = 1 \) used for numerical evaluation. These plots provide a clear visualization of the various limiting behaviors discussed in Sec.~\ref{sec:factorization}.
    }
    \label{Figure plot_parton}
\end{figure}

To facilitate future comparison with experimental data, we express our EEC result as the following distribution:
\begin{equation}
    \label{eq: eec plot formula}
    \frac{d^3\Sigma}{dY d\Delta Yd\phi}=\pi (1-y_a^2)(1-y_b^2)\frac{d^2\Sigma}{d\Omega_a d\Omega_b}=\frac{16\pi e^{2(Y+\Delta Y)}}{(e^Y+e^{\Delta Y})^2(1+e^{Y+\Delta Y})^2}\frac{d^2\Sigma}{d\Omega_a d\Omega_b}.
\end{equation}

As shown in Fig.~\ref{Figure plot_parton}, the EEC exhibits divergent behavior in several limits: the collinear limit, where both \( \Delta Y \) and \( \phi \) are small; the back-to-back limit, where \( Y \to 0 \) and \( \phi \to \pi \); the opposite coplanar limit, where \( \phi \to \pi \) with arbitrary \( Y \); and the Regge limit, where \( \Delta Y \to \infty \) without restrictions on \( \phi \). These plots visualize the LO EEC result, clearly exhibiting its limiting behaviors, and can serve as a valuable observable for precision comparison with experimental data.

As can be clearly seen from the color, the \( \phi \to \pi \) limit in Fig.~\ref{Figure parton_plot.a} is significantly stronger than that in the other subfigures of Fig.~\ref{Figure plot_parton}. This is consistent with the expectation that the back-to-back limit, characterized by \( \phi \to \pi \) and \( Y \to 0 \), exhibits a more pronounced divergence than the opposite coplanar limit, where \( \phi \to \pi \) with general values of \( Y \).

However, the LO result still requires resummation in these divergent kinematic limits in order to facilitate meaningful comparison with experimental data. We leave these resummations for future work.

\subsection{Plots in the hadron frame with PDF convolution}
\label{sec: plot for hardons}

\begin{figure}
    \begin{center}
        \makebox[\textwidth][c]{
            \subfloat[]{
                \includegraphics[width=0.65\textwidth]{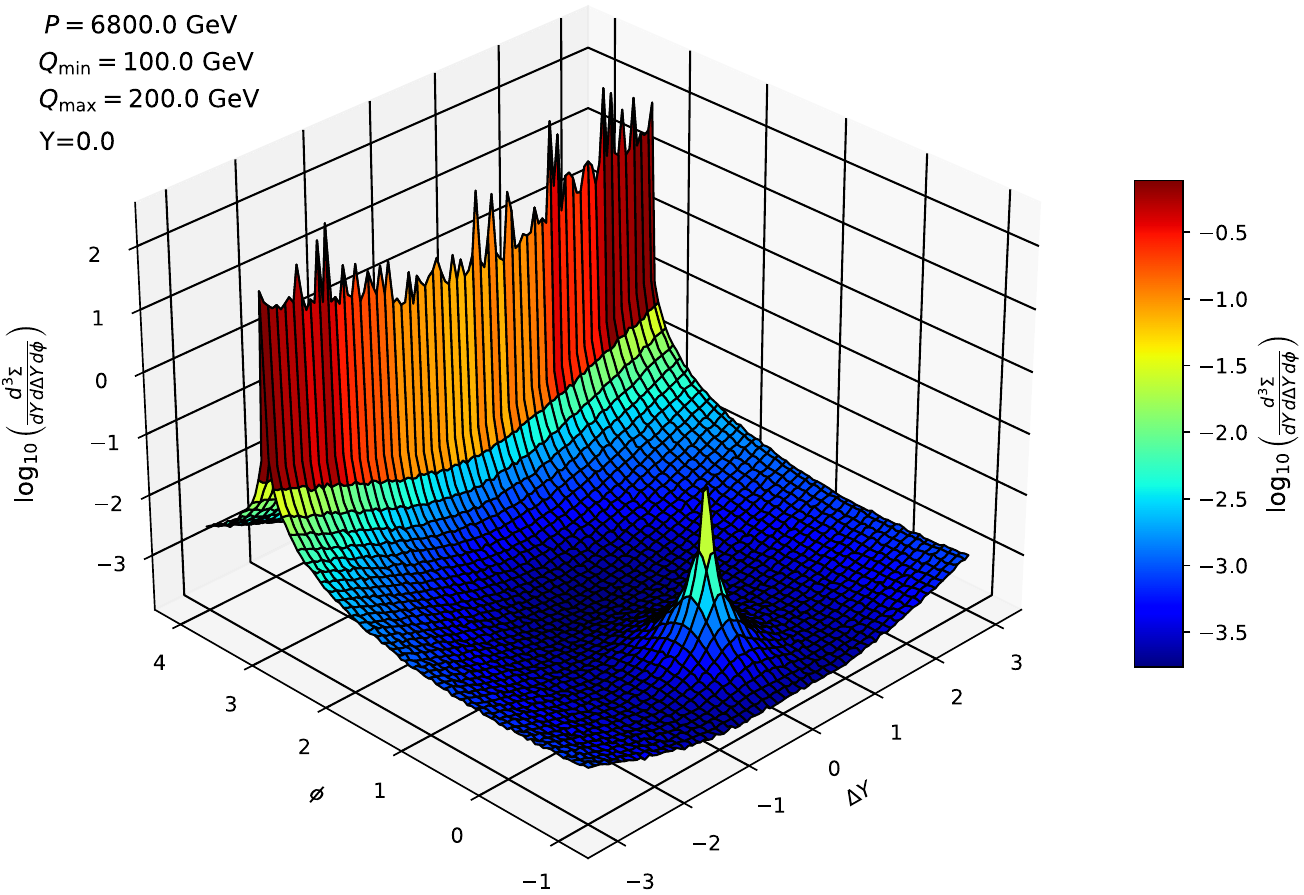}\label{Figure plot_hadron.a}
                }\;
            \subfloat[]{
                \includegraphics[width=0.65\textwidth]{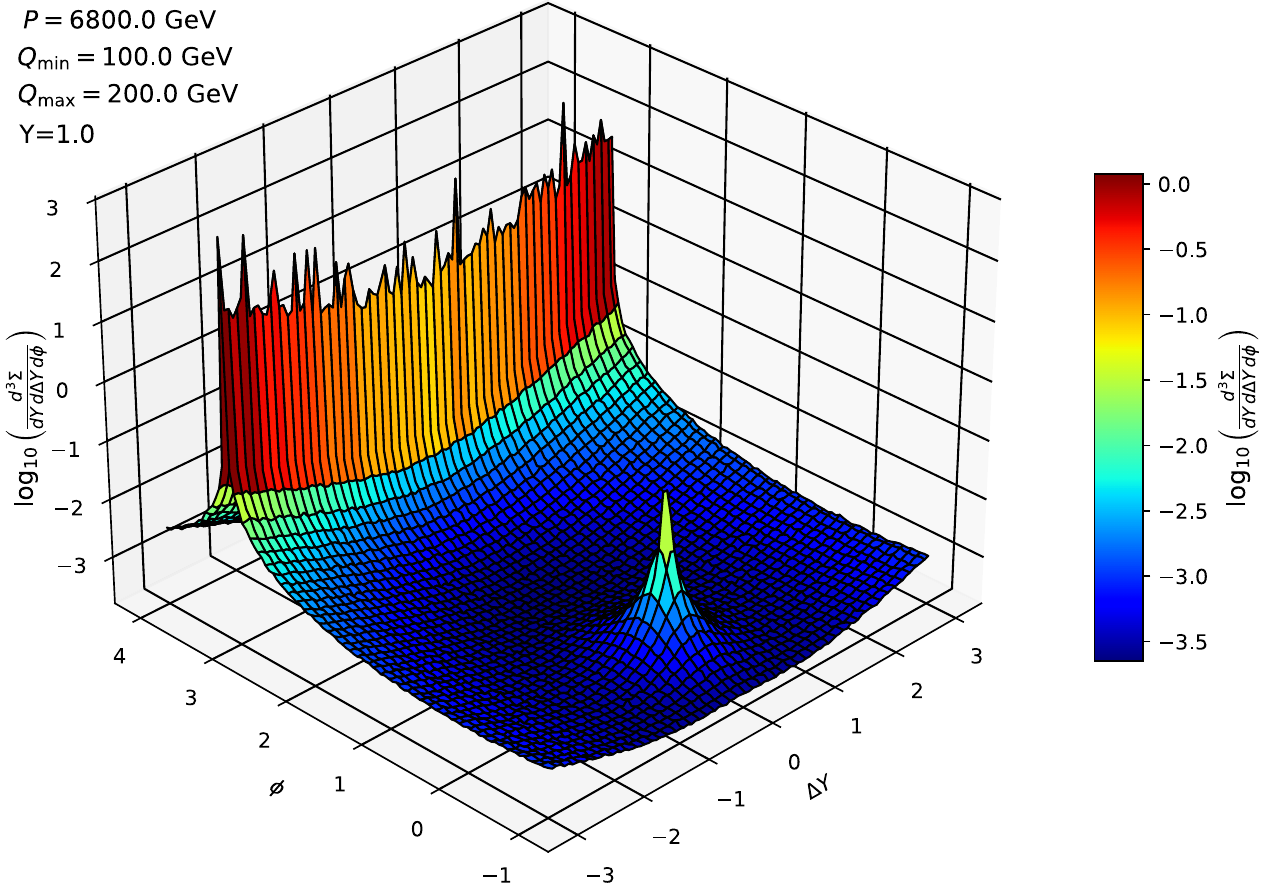}\label{Figure plot_hadron.b}
                }
        }
    \end{center}
    \caption{Landscape plots of the EEC after convolution over PDFs in the Lorentz-invariant $\Delta Y-\phi$ plane, with $P = 6800.0~\text{GeV}$, $Q_{\text{min}} = 100.0~\text{GeV}$, and $Q_{\text{max}} = 200.0~\text{GeV}$. The plots correspond to $Y = 0.0$ and $Y = 1.0$ for Fig.~\ref{Figure plot_hadron.a} and Fig.~\ref{Figure plot_hadron.b}, respectively. The EEC values are shown on a logarithmic scale (base 10) for better visual contrast, with warmer colors indicating higher values. These plots capture the expected limiting behaviors and offer direct comparison between theoretical predictions and experimental data.
    }
    \label{Figure plot_hadron}
\end{figure}

At hadron colliders, the partonic center-of-mass energy cannot be determined directly. Instead, only upper and lower bounds on this energy can be determined. Consequently, the parton-level EEC cannot be directly compared with experimental measurements - the convolution with PDFs in Eq.~\eqref{eq:hadron EEC} must first be included to enable physically meaningful comparisons with experimental data.

In particular, we focus on dijet production events, for which experiments typically report bounds on the parton center-of-mass energy, denoted by \( Q_{\text{min}} \) and \( Q_{\text{max}} \). These bounds also truncate the convolution domain, resolving numerical challenges from integrating over small gluon momentum fractions where the PDF diverges, which causes inefficiency and inaccuracy in numerical integration.

Let us consider the following experimental setup: for any event with incoming hadron energy \( P \), if the parton center-of-mass energy lies within the bounds \( Q_{\text{min}} \) and \( Q_{\text{max}} \), the EEC can be measured and expressed as a function of the three experimental energy scales \( P \), \( Q_{\text{min}} \), and \( Q_{\text{max}} \), along with the three detector configuration variables \( Y \), \( \Delta Y \), and \( \phi \).

We evaluate the PDF convolution using Monte Carlo integration with importance sampling to enhance accuracy and computational efficiency. The resulting landscape plots are shown in Fig.~\ref{Figure plot_hadron}. For enhanced visual clarity, the EEC values are presented on a base-10 logarithmic scale. As illustrated, the hadron-level plots closely resemble their parton-level counterparts, confirming the consistency of the results across different levels of theoretical description.

However, there is a key distinction between the hadron- and parton-level EEC due to the PDF convolution. After convolution with PDF, the divergence in the \( \phi \to \pi \) limit appears similar in Fig.~\ref{Figure plot_hadron.a} and Fig.~\ref{Figure plot_hadron.b}. By contrast, in the parton frame, these two divergences correspond to distinct physical limits: the back-to-back limit and the opposite coplanar limit, respectively. As shown in Fig.~\ref{Figure plot_parton}, the back-to-back divergence is significantly stronger than the opposite coplanar one.

This distinction arises because, for any given value of \( Y \) — the rapidity sum of two detectors in the hadron center-of-mass frame — there always exist partonic collision events where the rapidity sum \( Y_p \) in the parton center-of-mass frame approaches zero. This effectively enhances the opposite coplanar divergence into a back-to-back–like divergence at hadron colliders. Since the hadron-level EEC is obtained by convolving the parton-level result with PDFs, it captures the most singular contributions originating from this back-to-back divergence at the parton level. This explains the similar behavior in the \( \phi \to \pi \) region between Fig.~\ref{Figure plot_hadron.a} and Fig.~\ref{Figure plot_hadron.b}, despite the difference in \( Y \) values (0 and 1, respectively).

\section{Conclusion}\label{sec:conclusion}

In this work, we initiate the study of the full-range EEC at hadron colliders. We generalized the concept of celestial blocks~\cite{Kologlu:2019mfz,Chang:2020qpj} to hadron colliders, which is a natural decomposition basis for studying energy-energy correlation in the collinear limit from the view of light-ray OPE. The advantage of introducing celestial blocks is that it follows from Lorentz symmetry and can clearly separate the dynamic information from kinematics.
To illustrate the celestial block expansion, we first calculate leading order EEC in terms of hadron collider coordinates in the pure Yang-Mills theory. Based on this example, we find that organizing collinear expansion using celestial block yields a more uniform approximation numerically, even a bit away from the collinear region. This could be beneficial for phenomenological applications, as non-perturbative effects, including hadronization, are more significant in the very small angle. Considering larger angles can reduce hadronization effects and improve perturbative control; however, it is necessary to include subleading power corrections for precision prediction, where celestial block expansion demonstrates a better approximation than the ordinary series expansion.

The celestial blocks~\eqref{eq:celestial_blocks_def}, \eqref{eq:celestial_blocks_def1} for hadron colliders have more interesting theoretical features compared with celestial blocks \eqref{eq:ee_celestial_block} for unpolarized $e^+ e^-$ colliders. 
The presence of incoming collision axis probes the transverse spin physics in the light-ray OPE and hence allows contribution of celestial blocks with arbitrary integer transverse spin. We derive the celestial block for EEC at hadron colliders in 4d spacetime. Its explicit functional form contains the 2d conformal block structure, which reflects the equivalence of Lorentz symmetry and conformal symmetry on the celestial sphere. Parallel to the story of collinear three-point energy correlator~\cite{Chang:2022ryc,Chen:2022jhb}, such a celestial block structure allows the application of recently developed conformal collider techniques. As an example, we apply Lorentzian inversion formula to the tree-level EEC result and find the OPE coefficients are analytic with respect to transverse spin. We believe that analyticity in transverse spin can extend beyond tree-level correlators and may hold at higher orders in perturbation theory. Beyond serving as the reference direction for transverse spin, the collision axis also act as a defect which breaks the boost symmetry along the beam direction, and the collinear spin of the ``light-ray transition matrix'' along beam direction is a label in the celestial block. Interestingly, we notice that the OPE coefficients are also analytic with respect to collinear spin and only even collinear spin ``light-ray transition matrices'' contribute to pure gluon EEC at hadron colliders.

While the celestial block expansion has exhibited many theoretical and phenomenological features, many open questions still remain for future work. 
First, it is very important to systematically include running coupling and RG effect in the celestial block expansion. On the one hand, it can enter final-state light-ray OPE in non-conformal theories~\cite{Chen:2023zzh}, which is responsible for resumming large logarithms in the perturbation theory. On the other hand, initial-state radiation or the evolution of parton distribution function may change the simple relation between celestial blocks in hadron frame and parton frame. Second, the physical interpretation of analyticity in transverse spin is still not clear. For local operators in conformal theories, it is understood that the light-ray operators are the key object for explaining analyticity in spin~\cite{Kravchuk:2018htv}. If we follow the similar logic, understanding analyticity in transverse spin may need to consider integral transformation of light-ray operators, which is much less clear than local operator counterpart. Last but not least, it is interesting to see whether celestial block decomposition can be applied to collider data analysis to estimate QCD operator spectrum or other dynamic information. This requires understanding how to perform celestial blocks expansion in the presence of non-perturbative contribution or find a proper way to numerically perform the expansion in subregions where hadronization effects are suppressed.

In addition to the collinear limit, we also discuss various limits of EEC at hadron colliders, including opposite coplanar limit, back-to-back limit and Regge limit. 
This shows studying fully differential EEC at hadron colliders, compared with TEEC, can access larger phase space and probes different interesting QCD dynamics.
Opposite coplanar limit and back-to-back limit are similar to the stories of back-to-back limit of EEC in $e^+e^-$ collider and TEEC in hadron collider~\cite{Moult:2018jzp,Gao:2019ojf}. We only analyze the fixed-order behavior in these limits in this work and have not discussed all-order factorization and resummation. Based on established results in analogous scenarios, it is natural to expect that Sudakov peak also shows up after resummation.
Among all these limits of EEC at hadron colliders, we think Regge limit is the least understood but most interesting kinematic regime, which requires the $2\to 2$ forward scattering as well as $2\to N$ multi-Regge limit at the cross-section level. The resummation of the LP EEC in the Regge limit requires further exploration. But in the LO perturbative expansion, we already notice that the LP result shows a non-trivial dependence on the azimuthal angle $\phi$. It would be very interesting to compare this dependence, after resummation, with experimental data to study the BFKL dynamics.

Last but not least, we proposed an experimental setup for comparisons between experimental data and our theoretical predictions in dijet production. From the theory side, detailed comparisons require future work, including the convolution with PDFs for other partonic channels, resummation in several kinematic limits and adding hadronization corrections. While a complete theoretical understanding remains to be developed, the full-range EEC at hadron colliders is a particularly promising observable that bridges multiple fundamental aspects of QCD dynamics and advances our knowledge of collider physics.

\section*{Acknowledgements}
We would like to thank Qu Cao, Dong-Shan Jian, Yibei Li, Ian Moult, Dao-Ming Mu, Jichen Pan, and Cheng-Tai Tan for helpful discussions. The Feynman diagrams are drawn with the aid of \texttt{FeynGame}~\cite{Bundgen:2025utt}.
H.C. is supported by the U.S. Department of Energy, Office of Science, Office of Nuclear Physics under grant Contract Number DE-SC0011090. H.R. and H.X.Z. are supported by the National Science Foundation of China under contract No. 12425505. H.X.Z. is also supported by Asian Young Scientist Fellowship.

\appendix
\section{Polynomial Coefficients in the EEC}
\label{sec:appendixA}

In this section, we present the complete polynomial coefficients that appear in Eq.~\eqref{full result} and in Eq.~\eqref{eq:full result in other form}. 
The complete LO EEC result is available in the ancillary file \texttt{EEC\_result.m}.

\begin{equation}
    \begin{aligned}
        C_0&=(c_{\Delta Y}-c_{\phi }) (c_Y c_{\Delta Y}+c_{\phi } c_{\Delta Y}-c_Y c_{\phi }-1) \bigg[\Big( c_{\Delta Y}^4(232 c_{\Delta Y}^4-350 c_{\Delta Y}^2+40)+4 c_Y  c_{\Delta Y}^3(56 c_{\Delta Y}^4\\ 
        &-211 c_{\Delta Y}^2+53)+2 c_Y^2c_{\Delta Y}^2 (-96 c_{\Delta Y}^6-814 c_{\Delta Y}^4+197 c_{\Delta Y}^2+161) +6 c_Y^3 c_{\Delta Y}(-112 c_{\Delta Y}^6-888 c_{\Delta Y}^4\\ 
        &+387 c_{\Delta Y}^2+36) +c_Y^4 (-768 c_{\Delta Y}^6-9216 c_{\Delta Y}^4+1740 c_{\Delta Y}^2+54)-6 c_Y^5c_{\Delta Y} (32 c_{\Delta Y}^4+1708 c_{\Delta Y}^2-15)\\ 
        &+24 c_Y^6 (8 c_{\Delta Y}^4-265 c_{\Delta Y}^2-8)+24 c_Y^7c_{\Delta Y} (4 c_{\Delta Y}^2-67)\Big)
        +\textcolor{cyan}{c_{\phi }}\Big(c_{\Delta Y}^3 (12 c_{\Delta Y}^4+77 c_{\Delta Y}^2+24)\\ 
        &+2c_Y c_{\Delta Y}^2 (400 c_{\Delta Y}^6-822 c_{\Delta Y}^4+427 c_{\Delta Y}^2-22)+c_Y^2c_{\Delta Y} (1976 c_{\Delta Y}^6-7216 c_{\Delta Y}^4+1341 c_{\Delta Y}^2-38)\\ 
        &-2c_Y^3 (192 c_{\Delta Y}^8-840 c_{\Delta Y}^6+6512 c_{\Delta Y}^4-213 c_{\Delta Y}^2+6)-2 c_Y^4c_{\Delta Y} (768 c_{\Delta Y}^6-1480 c_{\Delta Y}^4+5817 c_{\Delta Y}^2+234)\\ 
        &+c_Y^5(-2304 c_{\Delta Y}^6+6560 c_{\Delta Y}^4-4668 c_{\Delta Y}^2-246)-24c_Y^6 c_{\Delta Y} (64 c_{\Delta Y}^4-245 c_{\Delta Y}^2+8)\\ 
        &+24 c_Y^7(-16 c_{\Delta Y}^4+74 c_{\Delta Y}^2+11)\Big) 
        +\textcolor{cyan}{c_{\phi }^2}\Big(c_{\Delta Y}^2(-672 c_{\Delta Y}^6+72 c_{\Delta Y}^4+349 c_{\Delta Y}^2-42  )+4 c_Yc_{\Delta Y}\\ 
        & (-530 c_{\Delta Y}^6-25 c_{\Delta Y}^4+138 c_{\Delta Y}^2+4)+c_Y^2 (1344 c_{\Delta Y}^8-1088 c_{\Delta Y}^6-1608 c_{\Delta Y}^4+191 c_{\Delta Y}^2+4)\\ 
        & +2 c_Y^3 c_{\Delta Y}(1728 c_{\Delta Y}^6+1992 c_{\Delta Y}^4-1390 c_{\Delta Y}^2-93)+c_Y^4 (384 c_{\Delta Y}^6+7936 c_{\Delta Y}^4-1386 c_{\Delta Y}^2-66)\\ 
        & +c_Y^5 c_{\Delta Y}(-6144 c_{\Delta Y}^4+6712 c_{\Delta Y}^2+276 )+c_Y^6 (-6336 c_{\Delta Y}^4+2976 c_{\Delta Y}^2+264)+192 c_Y^7c_{\Delta Y} (-10 c_{\Delta Y}^2+3)\Big) \\ 
        &+\textcolor{cyan}{c_{\phi }^3}\Big(c_{\Delta Y} (720 c_{\Delta Y}^6-472 c_{\Delta Y}^4-177 c_{\Delta Y}^2+26)-4 c_Y(288 c_{\Delta Y}^8-928 c_{\Delta Y}^6+399 c_{\Delta Y}^4+74 c_{\Delta Y}^2+1) \\ 
        &-c_Y^2c_{\Delta Y} (6048 c_{\Delta Y}^6-8200 c_{\Delta Y}^4+1306 c_{\Delta Y}^2+173)+2c_Y^3 (768 c_{\Delta Y}^8-6528 c_{\Delta Y}^6+4848 c_{\Delta Y}^4+344 c_{\Delta Y}^2+3)\\ 
        & +4 c_Y^4c_{\Delta Y} (1536 c_{\Delta Y}^6-3696 c_{\Delta Y}^4+1456 c_{\Delta Y}^2+339) +8c_Y^5 (1152 c_{\Delta Y}^6-1152 c_{\Delta Y}^4+92 c_{\Delta Y}^2+57)\\  
        &+24 c_Y^6c_{\Delta Y} (256 c_{\Delta Y}^4-124 c_{\Delta Y}^2-41)+384c_Y^7 (4 c_{\Delta Y}^4-c_{\Delta Y}^2-1)    \Big) 
        +\textcolor{cyan}{c_{\phi }^4}\Big((-144 c_{\Delta Y}^6+120 c_{\Delta Y}^4+35 c_{\Delta Y}^2-6)\\ 
        &+2 c_Y c_{\Delta Y} (288 c_{\Delta Y}^6-560 c_{\Delta Y}^4+256 c_{\Delta Y}^2+29) +c_Y^2(2592 c_{\Delta Y}^6-3184 c_{\Delta Y}^4+778 c_{\Delta Y}^2+35) \\ 
        &-64 c_Y^3c_{\Delta Y} (12 c_{\Delta Y}^6-81 c_{\Delta Y}^4+69 c_{\Delta Y}^2-8)-8c_Y^4 (384 c_{\Delta Y}^6-792 c_{\Delta Y}^4+398 c_{\Delta Y}^2-15) \\ 
        &-32c_Y^5 c_{\Delta Y} (144 c_{\Delta Y}^4-162 c_{\Delta Y}^2+35)-48c_Y^6 (64 c_{\Delta Y}^4-54 c_{\Delta Y}^2+3)+192c_Y^7 c_{\Delta Y} (-4 c_{\Delta Y}^2+3)    \Big) \bigg].
    \end{aligned}
\end{equation}
\begin{equation}
    \begin{aligned}
        C_1&=6 (c_{\Delta Y}+c_Y)^4 (c_{\Delta Y} c_{\phi }+c_{\Delta Y} c_Y-c_Y c_{\phi }-1) \bigg[\Big(c_{\Delta Y}(-40 c_{\Delta Y}^4+106 c_{\Delta Y}^2-57 )\\ 
        &+c_Y(76 c_{\Delta Y}^4-149 c_{\Delta Y}^2+38) +c_Y^2c_{\Delta Y}(32 c_{\Delta Y}^4-312 c_{\Delta Y}^2+250 )-4c_Y^3 (24 c_{\Delta Y}^4+c_{\Delta Y}^2+17)\\ 
        & +16 c_Y^4c_{\Delta Y} (3 c_{\Delta Y}^2-19)  \Big)+\textcolor{cyan}{c_{\phi }} \Big(6 (-20 c_{\Delta Y}^4+6 c_{\Delta Y}^2+5)-4 c_Yc_{\Delta Y}(52 c_{\Delta Y}^4-70 c_{\Delta Y}^2+1)\\ 
        &+c_Y^2(460 c_{\Delta Y}^4-743 c_{\Delta Y}^2-70)+16  c_Y^3c_{\Delta Y} (12 c_{\Delta Y}^4-35 c_{\Delta Y}^2-6) + c_Y^4(-240 c_{\Delta Y}^4+540 c_{\Delta Y}^2+8) \Big)\\ 
        &+\textcolor{cyan}{c_{\phi }^2} \Big(18 c_{\Delta Y}(8 c_{\Delta Y}^4-6 c_{\Delta Y}^2+1) +c_Y(-284 c_{\Delta Y}^4-85 c_{\Delta Y}^2+22)+c_Y^2c_{\Delta Y}(-480 c_{\Delta Y}^4+704 c_{\Delta Y}^2-217 )\\ 
        &+ c_Y^3(528 c_{\Delta Y}^4+40 c_{\Delta Y}^2-12)+4 c_Y^4c_{\Delta Y} (64 c_{\Delta Y}^4-212 c_{\Delta Y}^2+67)  \Big)
        +\textcolor{cyan}{c_{\phi }^3}\Big ((-56 c_{\Delta Y}^4+30 c_{\Delta Y}^2-19)\\ 
        &+4c_Y c_{\Delta Y} (96 c_{\Delta Y}^4-5 c_{\Delta Y}^2-41)+ c_Y^2(-416 c_{\Delta Y}^4+266 c_{\Delta Y}^2+138) -4 c_Y^3 c_{\Delta Y} (128 c_{\Delta Y}^4+24 c_{\Delta Y}^2-65)\\ 
        &+8c_Y^4 (88 c_{\Delta Y}^4-40 c_{\Delta Y}^2-13) \Big)
        +\textcolor{cyan}{c_{\phi }^4} \Big(-8 c_{\Delta Y} (8 c_{\Delta Y}^4-11 c_{\Delta Y}^2+2)+c_Y(-336 c_{\Delta Y}^4+266 c_{\Delta Y}^2+20)\\ 
        &+2c_Y^2 c_{\Delta Y} (256 c_{\Delta Y}^4-160 c_{\Delta Y}^2-61) +16  c_Y^3(28 c_{\Delta Y}^4-19 c_{\Delta Y}^2-1)+16  c_Y^4c_{\Delta Y} (-32 c_{\Delta Y}^4+12 c_{\Delta Y}^2+11) \Big)\\ 
        &+\textcolor{cyan}{c_{\phi }^5} \Big((32 c_{\Delta Y}^4-44 c_{\Delta Y}^2+14)+8c_Y c_{\Delta Y} (6 c_{\Delta Y}^2-5)-4c_Y^2 (64 c_{\Delta Y}^4-70 c_{\Delta Y}^2+11)+c_Y^3c_{\Delta Y}(-64 c_{\Delta Y}^2+48)\\ 
        &+32 c_Y^4(8 c_{\Delta Y}^4-8 c_{\Delta Y}^2+1) \Big)\bigg].
    \end{aligned}
\end{equation}
\begin{equation}
    \begin{aligned}
        C_2&=6 (c_{\Delta Y}+c_Y)^4 (c_{\Delta Y} c_{\phi }+c_{\Delta Y} c_Y-c_Y c_{\phi }-1)\bigg[\Big( c_{\Delta Y}(-76 c_{\Delta Y}^2+55)+c_Y(-32 c_{\Delta Y}^4+128 c_{\Delta Y}^2-42)\\ 
        &+12  c_Y^2c_{\Delta Y} (8 c_{\Delta Y}^2-13)+c_Y^3(72-48 c_{\Delta Y}^2) \Big)+\textcolor{cyan}{c_{\phi }} \Big((112 c_{\Delta Y}^4-56 c_{\Delta Y}^2-34)+c_Yc_{\Delta Y}( -340 c_{\Delta Y}^2+121)\\ 
        & +c_Y^2(-192 c_{\Delta Y}^4+464 c_{\Delta Y}^2+96) +60c_Y^3c_{\Delta Y}(4 c_{\Delta Y}^2-7) \Big)
        +\textcolor{cyan}{c_{\phi }^2} \Big(c_{\Delta Y} (20 c_{\Delta Y}^2+37)\\
        &+c_Y(352 c_{\Delta Y}^4-104 c_{\Delta Y}^2-3) -16c_Y^2 c_{\Delta Y} (33 c_{\Delta Y}^2+19) + c_Y^3(-256 c_{\Delta Y}^4+720 c_{\Delta Y}^2+60)\Big)\\ 
        &+\textcolor{cyan}{c_{\phi }^3 }\Big(4 (-32 c_{\Delta Y}^4+c_{\Delta Y}^2+5)+ c_Yc_{\Delta Y}(64 c_{\Delta Y}^2-74 )+4c_Y^2 (128 c_{\Delta Y}^4+88 c_{\Delta Y}^2-5) -32c_Y^3 (22 c_{\Delta Y}^3+c_{\Delta Y}) \Big)\\ 
        &+\textcolor{cyan}{c_{\phi }^4} \Big(2 c_{\Delta Y} (56 c_{\Delta Y}^2-29)+c_Y(-256 c_{\Delta Y}^4+96 c_{\Delta Y}^2+50) +c_Y^2c_{\Delta Y}(-448 c_{\Delta Y}^2+80) \\
        &+16c_Y^3 (32 c_{\Delta Y}^4+4 c_{\Delta Y}^2-5) \Big)
        +\textcolor{cyan}{c_{\phi }^5} \Big(8(-2 c_{\Delta Y}^2+1)+8c_Y c_{\Delta Y} (16 c_{\Delta Y}^2-11) + c_Y^2(64 c_{\Delta Y}^2-16)\\ 
        &+128c_Y^3 c_{\Delta Y} (1-2 c_{\Delta Y}^2) \Big)\bigg].
    \end{aligned}
\end{equation}
\begin{equation}
    \begin{aligned}
        C_3&=6 (c_Y+c_{\Delta Y})^4 (c_Y c_{\Delta Y}+c_{\phi } c_{\Delta Y}-c_Y c_{\phi }-1)\bigg[\Big((-68 c_{\Delta Y}^4+81 c_{\Delta Y}^2-13)+c_Yc_{\Delta Y} (-40 c_{\Delta Y}^4\\ 
        &+174 c_{\Delta Y}^2-127 )+c_Y^2 (100 c_{\Delta Y}^4-193 c_{\Delta Y}^2+78)+4 c_Y^3 c_{\Delta Y} (8 c_{\Delta Y}^4-44 c_{\Delta Y}^2+39)\\ 
        &+c_Y^4 (-16 c_{\Delta Y}^4+84 c_{\Delta Y}^2-72)\Big)+\textcolor{cyan}{ c_{\phi }}\Big(2 c_{\Delta Y} (52 c_{\Delta Y}^4-55 c_{\Delta Y}^2-4)+ c_Y(-368 c_{\Delta Y}^4+274 c_{\Delta Y}^2+78)\\ 
        &+c_Y^2c_{\Delta Y}(-208 c_{\Delta Y}^4+620 c_{\Delta Y}^2-329 )+c_Y^3(416 c_{\Delta Y}^4-412 c_{\Delta Y}^2-96)+c_Y^4c_{\Delta Y}(64 c_{\Delta Y}^4-448 c_{\Delta Y}^2+420 )   \Big)\\ 
        &+\textcolor{cyan}{c_{\phi }^2}\Big((56 c_{\Delta Y}^4-30 c_{\Delta Y}^2+17)+ c_Yc_{\Delta Y} (400 c_{\Delta Y}^4-284 c_{\Delta Y}^2-147)+c_Y^2(-700 c_{\Delta Y}^4+549 c_{\Delta Y}^2-6)\\ 
        &-56 c_Y^3c_{\Delta Y} (8 c_{\Delta Y}^4-9 c_{\Delta Y}^2-4)+12 c_Y^4(52 c_{\Delta Y}^4-61 c_{\Delta Y}^2-2)   \Big)
        +\textcolor{cyan}{c_{\phi }^3}\Big(-4 c_{\Delta Y} (44 c_{\Delta Y}^4-40 c_{\Delta Y}^2+7)\\ 
        &+c_Y(116 c_{\Delta Y}^4+67 c_{\Delta Y}^2-61) +c_Y^2c_{\Delta Y} (736 c_{\Delta Y}^4-864 c_{\Delta Y}^2+223) -4 c_Y^3 (76 c_{\Delta Y}^4+48 c_{\Delta Y}^2-17)\\ 
        &-4 c_Y^4 c_{\Delta Y} (128 c_{\Delta Y}^4-236 c_{\Delta Y}^2+45)\Big)
        +\textcolor{cyan}{c_{\phi }^4}\Big( (72 c_{\Delta Y}^4-70 c_{\Delta Y}^2+11)+ c_Yc_{\Delta Y}(-384 c_{\Delta Y}^4+44 c_{\Delta Y}^2+210 )\\ 
        & +2 c_Y^2(144 c_{\Delta Y}^4-63 c_{\Delta Y}^2-59) +4 c_Y^3c_{\Delta Y} (128 c_{\Delta Y}^4+16 c_{\Delta Y}^2-59)+24c_Y^4 (-24 c_{\Delta Y}^4+8 c_{\Delta Y}^2+5)\Big)\\ 
        & +\textcolor{cyan}{c_{\phi }^5}\Big(4 c_{\Delta Y} (16 c_{\Delta Y}^4-22 c_{\Delta Y}^2+7)+14 c_Y(24 c_{\Delta Y}^4-19 c_{\Delta Y}^2-1) +c_Y^2c_{\Delta Y}(-512 c_{\Delta Y}^4+320 c_{\Delta Y}^2+122 )\\ 
        & + c_Y^3(-448 c_{\Delta Y}^4+304 c_{\Delta Y}^2+16)+16  c_Y^4c_{\Delta Y} (32 c_{\Delta Y}^4-12 c_{\Delta Y}^2-11)\Big) 
        + \textcolor{cyan}{c_{\phi }^6}\Big((-32 c_{\Delta Y}^4+44 c_{\Delta Y}^2-14)\\ 
        &+8c_Y c_{\Delta Y} (-6 c_{\Delta Y}^2+5) +4c_Y^2 (64 c_{\Delta Y}^4-70 c_{\Delta Y}^2+11) +16 c_Y^3 c_{\Delta Y} (4 c_{\Delta Y}^2-3)-32c_Y^4 (8 c_{\Delta Y}^4-8 c_{\Delta Y}^2+1) \Big)  \bigg].
    \end{aligned}
\end{equation}
\begin{equation}
    \begin{aligned}
        C_4&=3 (c_Y+c_{\Delta Y})^4 \bigg[\Big((68 c_{\Delta Y}^4-149 c_{\Delta Y}^2+26)-4 c_Y c_{\Delta Y} (16 c_{\Delta Y}^4+18 c_{\Delta Y}^2-57)+2 c_Y^2 (76 c_{\Delta Y}^4+27 c_{\Delta Y}^2-78)\\ 
        &+4 c_Y^3 c_{\Delta Y} (16 c_{\Delta Y}^4-28 c_{\Delta Y}^2-39)-24 c_Y^4 (8 c_{\Delta Y}^4-9 c_{\Delta Y}^2-6)+48 c_Y^5 c_{\Delta Y} (2 c_{\Delta Y}^2-3)\Big)\\ 
        &+\textcolor{cyan}{c_{\phi }}\Big(-2 c_{\Delta Y} (32 c_{\Delta Y}^4-182 c_{\Delta Y}^2+5)+c_Y(468 c_{\Delta Y}^4-1045 c_{\Delta Y}^2-130) -8 c_Y^2c_{\Delta Y} (44 c_{\Delta Y}^4+100 c_{\Delta Y}^2-153)\\ 
        & +c_Y^3(280 c_{\Delta Y}^4+882 c_{\Delta Y}^2+36) +16  c_Y^4c_{\Delta Y} (24 c_{\Delta Y}^4-10 c_{\Delta Y}^2-93)+24c_Y^5 (-20 c_{\Delta Y}^4+31 c_{\Delta Y}^2+6) \Big)\\ 
        &+\textcolor{cyan}{ c_{\phi }^2}\Big((-520 c_{\Delta Y}^4-28 c_{\Delta Y}^2-21)+2 c_Yc_{\Delta Y} (-64 c_{\Delta Y}^4+734 c_{\Delta Y}^2+307) +6 c_Y^2(288 c_{\Delta Y}^4-444 c_{\Delta Y}^2-37)\\ 
        & -8c_Y^3 c_{\Delta Y} (72 c_{\Delta Y}^4+247 c_{\Delta Y}^2+9)+8 c_Y^4(-40 c_{\Delta Y}^4+477 c_{\Delta Y}^2+39) +64c_Y^5 c_{\Delta Y} (8 c_{\Delta Y}^4-15 c_{\Delta Y}^2-15) \Big)\\ 
        & +\textcolor{cyan}{c_{\phi }^3}\Big(8 c_{\Delta Y} (36 c_{\Delta Y}^4+19 c_{\Delta Y}^2+9)+ c_Y(-1736 c_{\Delta Y}^4-872 c_{\Delta Y}^2+23)+4 c_Y^2c_{\Delta Y} (-48 c_{\Delta Y}^4+668 c_{\Delta Y}^2+85)\\ 
        & +c_Y^3(2208 c_{\Delta Y}^4+500 c_{\Delta Y}^2-106)-16 c_Y^4c_{\Delta Y} (32 c_{\Delta Y}^4+288 c_{\Delta Y}^2+47) +8  c_Y^5(112 c_{\Delta Y}^4+188 c_{\Delta Y}^2+15)\Big) \\ 
        &+\textcolor{cyan}{c_{\phi }^4}\Big(2 (8 c_{\Delta Y}^4-92 c_{\Delta Y}^2-13)+4c_Y c_{\Delta Y} (160 c_{\Delta Y}^4+310 c_{\Delta Y}^2-37) +32c_Y^2 (-103 c_{\Delta Y}^4+3 c_{\Delta Y}^2+7)\\ 
        & -32c_Y^3 c_{\Delta Y} (28 c_{\Delta Y}^2+3)+8c_Y^4 (544 c_{\Delta Y}^4+92 c_{\Delta Y}^2-25) -32 c_Y^5 c_{\Delta Y} (32 c_{\Delta Y}^4+48 c_{\Delta Y}^2-3)\Big) \\ 
        &+\textcolor{cyan}{ c_{\phi }^5}\Big(8 c_{\Delta Y} (-16 c_{\Delta Y}^4+6 c_{\Delta Y}^2+11)+c_Y(-480 c_{\Delta Y}^4-68 c_{\Delta Y}^2+6) +8 c_Y^2c_{\Delta Y} (128 c_{\Delta Y}^4+144 c_{\Delta Y}^2-89) \\ 
        &+c_Y^3(-640 c_{\Delta Y}^4+400 c_{\Delta Y}^2+148)-32 c_Y^4c_{\Delta Y} (32 c_{\Delta Y}^4+52 c_{\Delta Y}^2-19) +32  c_Y^5(48 c_{\Delta Y}^4-4 c_{\Delta Y}^2-5)\Big) \\ 
        &+\textcolor{cyan}{c_{\phi }^6}\Big(8 c_{\Delta Y}^2 (8 c_{\Delta Y}^2-7)+8c_Y c_{\Delta Y} (4 c_{\Delta Y}^2+3)+c_Y^2(-512 c_{\Delta Y}^4+208 c_{\Delta Y}^2+32) +16 c_Y^3c_{\Delta Y} (24 c_{\Delta Y}^2-17)\\ 
        &  +32c_Y^4 (16 c_{\Delta Y}^4-4 c_{\Delta Y}^2-1)+256c_Y^5 c_{\Delta Y} (1-2 c_{\Delta Y}^2) \Big) \bigg].
    \end{aligned}
\end{equation}
\begin{equation}
    \begin{aligned}
        C_5&=3 (c_Y+c_{\Delta Y})^4\bigg[\Big(c_{\Delta Y}(-76 c_{\Delta Y}^4+207 c_{\Delta Y}^2-110 )+4 c_Y (20 c_{\Delta Y}^6-7 c_{\Delta Y}^4-47 c_{\Delta Y}^2+21)\\ 
        &-2 c_Y^2 c_{\Delta Y} (60 c_{\Delta Y}^4+25 c_{\Delta Y}^2-114)-16 c_Y^3 (4 c_{\Delta Y}^6-20 c_{\Delta Y}^4+9 c_{\Delta Y}^2+9)+8 c_Y^4 c_{\Delta Y} (4 c_{\Delta Y}^4-21 c_{\Delta Y}^2+18)\Big)\\ 
        &
        +\textcolor{cyan}{c_{\phi }}\Big((80 c_{\Delta Y}^6-432 c_{\Delta Y}^4+188 c_{\Delta Y}^2+68) -2 c_Yc_{\Delta Y} (204 c_{\Delta Y}^4-619 c_{\Delta Y}^2+252) +12 c_Y^2 (24 c_{\Delta Y}^6+48 c_{\Delta Y}^4\\ 
        &-100 c_{\Delta Y}^2-9)-16 c_Y^3 c_{\Delta Y}(38 c_{\Delta Y}^4+37 c_{\Delta Y}^2-93)-16 c_Y^4 (8 c_{\Delta Y}^6-54 c_{\Delta Y}^4+42 c_{\Delta Y}^2+9)\Big) \\ 
        &+\textcolor{cyan}{c_{\phi }^2}\Big(c_{\Delta Y} (556 c_{\Delta Y}^4-301 c_{\Delta Y}^2-87)+c_Y(-1520 c_{\Delta Y}^4+678 c_{\Delta Y}^2+32) +c_Y^2c_{\Delta Y}(-1304 c_{\Delta Y}^4+2258 c_{\Delta Y}^2\\
        &+438 )  +24 c_Y^3 (32 c_{\Delta Y}^6+82 c_{\Delta Y}^4-149 c_{\Delta Y}^2-10)+8 c_Y^4c_{\Delta Y} (-140 c_{\Delta Y}^4+71 c_{\Delta Y}^2+111) \Big) \\ 
        &
        +\textcolor{cyan}{c_{\phi }^3}\Big((-352 c_{\Delta Y}^6+196 c_{\Delta Y}^4+98 c_{\Delta Y}^2-74)+2 c_Y c_{\Delta Y}(1080 c_{\Delta Y}^4-758 c_{\Delta Y}^2+185) -4 c_Y^2 (16 c_{\Delta Y}^6+402 c_{\Delta Y}^4\\ 
        &+183 c_{\Delta Y}^2-25)+8 c_Y^3c_{\Delta Y}^3 (-320 c_{\Delta Y}^2+593) +16 c_Y^4 (64 c_{\Delta Y}^6-40 c_{\Delta Y}^4-69 c_{\Delta Y}^2-3)\Big) \\ 
        &+\textcolor{cyan}{ c_{\phi }^4}\Big(c_{\Delta Y}(-184 c_{\Delta Y}^4+24 c_{\Delta Y}^2+193 )-2 c_Y (352 c_{\Delta Y}^6-280 c_{\Delta Y}^4+160 c_{\Delta Y}^2+95)+c_Y^2  c_{\Delta Y}(2592 c_{\Delta Y}^4\\ 
        &-212 c_{\Delta Y}^2-226)+c_Y^3 (-3776 c_{\Delta Y}^4+832 c_{\Delta Y}^2+376)+8 c_Y^4 c_{\Delta Y}(16 c_{\Delta Y}^4+188 c_{\Delta Y}^2-75) \Big)\\ 
        &+\textcolor{cyan}{c_{\phi }^5}\Big(4 (64 c_{\Delta Y}^6+18 c_{\Delta Y}^4-80 c_{\Delta Y}^2+1)+c_Yc_{\Delta Y}(-992 c_{\Delta Y}^4+888 c_{\Delta Y}^2+284 ) -4c_Y^2 (128 c_{\Delta Y}^6+368 c_{\Delta Y}^4\\ 
        &-238 c_{\Delta Y}^2+21) +8c_Y^3 c_{\Delta Y} (512 c_{\Delta Y}^4-156 c_{\Delta Y}^2-137) -16c_Y^4 (64 c_{\Delta Y}^6+48 c_{\Delta Y}^4-46 c_{\Delta Y}^2-15) \Big) \\ 
        &+\textcolor{cyan}{c_{\phi }^6}\Big(c_{\Delta Y}(-224 c_{\Delta Y}^4+196 c_{\Delta Y}^2+22 )+c_Y (512 c_{\Delta Y}^6+64 c_{\Delta Y}^4-632 c_{\Delta Y}^2+60)\\ 
        &+4 c_Y^2c_{\Delta Y} (32 c_{\Delta Y}^4+68 c_{\Delta Y}^2-29) -32 c_Y^3 (32 c_{\Delta Y}^6+36 c_{\Delta Y}^4-49 c_{\Delta Y}^2+1)+32 c_Y^4 c_{\Delta Y} (48 c_{\Delta Y}^4-28 c_{\Delta Y}^2-9)\Big)\\ 
        & +\textcolor{cyan}{ c_{\phi }^7}\Big(32 c_{\Delta Y}^2 (c_{\Delta Y}^2-1)-8 c_Y c_{\Delta Y} (32 c_{\Delta Y}^4-34 c_{\Delta Y}^2+3)+8c_Y^2 (16 c_{\Delta Y}^4-26 c_{\Delta Y}^2+7)\\ 
        & +32 c_Y^3c_{\Delta Y} (16 c_{\Delta Y}^4-12 c_{\Delta Y}^2-1) -64 c_Y^4(8 c_{\Delta Y}^4-8 c_{\Delta Y}^2+1) \Big)\bigg].
    \end{aligned}
\end{equation}
\begin{equation}
    \begin{aligned}
        A_0&=3 \left((y_a-y_b)^2+4y_a\,  y_b\, \zeta\right) \left((y_a-y_b)^2-4 \zeta(1-\zeta-y_a\, y_b)\right) \bigg[\textcolor{cyan}{32\zeta^5} \Big(-y_a^8 (11 y_b^8+40 y_b^6+188 y_b^4-8 y_b^2+14)\\
        &+8y_a^6 (42 y_b^6-2 y_b^4+60 y_b^2-9) -4y_a^4 (183 y_b^4-68 y_b^2+35) +24 y_a^2(6 y_b^2-1) -3\Big)\\
        &+\textcolor{cyan}{16\zeta^4} \Big(y_a^9 y_b (17 y_b^8-76 y_b^6-304 y_b^4-388 y_b^2-34) + y_a^8(53 y_b^8+716 y_b^4-128 y_b^2+74) \\
        & +4 y_a^7y_b(8 y_b^6+394 y_b^4+328 y_b^2+49) -16y_a^6 (53 y_b^6+20 y_b^4+74 y_b^2-8) -4 y_a^5y_b (609 y_b^4+74 y_b^2+178)\\
        & +4 y_a^4(597 y_b^4-272 y_b^2+131) +4 y_a^3y_b (320 y_b^2-29) -336y_a^2 y_b^2 -51 y_a y_b +21\Big)\\
        &+\textcolor{cyan}{4\zeta^3} \Big(- y_a^{10}(15 y_b^{10}-190 y_b^8-576 y_b^6+1256 y_b^4+466 y_b^2+38) -64y_a^9 y_b (5 y_b^8-17 y_b^6+14 y_b^4-41 y_b^2+2)\\
        & -y_a^8 (619 y_b^8+1232 y_b^6-2856 y_b^4-2626 y_b^2+130) +64y_a^7 y_b (12 y_b^6-82 y_b^4-72 y_b^2-15)\\
        & +4y_a^6 (743 y_b^6-962 y_b^4-1062 y_b^2+190) +128 y_a^5y_b (48 y_b^4+31 y_b^2+17) -20y_a^4 (125 y_b^4-372 y_b^2+112)\\
        & -192y_a^3 y_b (28 y_b^2-3) -15y_a^2 (109 y_b^2-66) +192 y_a y_b -183\Big)\\
        &+\textcolor{cyan}{2\zeta^2} \Big(y_a^{11} y_b (3 y_b^{10}+66 y_b^8-16 y_b^6+1864 y_b^4-774 y_b^2-122) + y_a^{10}(147 y_b^{10}-1770 y_b^8-2088 y_b^6+24 y_b^4\\
        &+434 y_b^2+34) + y_a^9 y_b (839 y_b^8-3376 y_b^6+1480 y_b^4-2562 y_b^2+1690)+ y_a^8(3899 y_b^8+7568 y_b^6-1776 y_b^4\\
        &+106 y_b^2-662) -4y_a^7 y_b (675 y_b^6-1394 y_b^4+654 y_b^2+78) -4 y_a^6(3919 y_b^6-2166 y_b^4-1104 y_b^2+386) \\
        &+4y_a^5 y_b (1709 y_b^4-2860 y_b^2+796) +4 y_a^4(765 y_b^4-2372 y_b^2+650) + y_a^3 y_b (3799 y_b^2-1470)\\
        &+3 y_a^2(1269 y_b^2-686) +51y_a y_b +303\Big)\\
        &+\textcolor{cyan}{2\zeta} \Big(-y_a^{12} (9 y_b^{10}-56 y_b^8-68 y_b^6-598 y_b^4-85 y_b^2+30) - y_a^{11}y_b (15 y_b^{10}+38 y_b^8+80 y_b^6+2976 y_b^4-222 y_b^2\\
        &+170) - y_a^{10} (156 y_b^{10}-2039 y_b^8-272 y_b^6-45 y_b^4+664 y_b^2-165)-y_a^9 y_b (527 y_b^8-2712 y_b^6-4680 y_b^4-3422 y_b^2\\
        &+506) - y_a^8(4608 y_b^8+3940 y_b^6-230 y_b^4+3267 y_b^2-1042) +4 y_a^7y_b (351 y_b^6-2726 y_b^4-98 y_b^2-72) \\
        &+4y_a^6 (4192 y_b^6-4361 y_b^4+2748 y_b^2-495) -4 y_a^5y_b (477 y_b^4-2534 y_b^2+816) +y_a^4(5778 y_b^4-10713 y_b^2+2672) \\
        &-y_a^3y_b (2931 y_b^2-1690) +3 y_a^2(1084 y_b^2-467) -267 y_a y_b +150\Big)\\
        &+\Big(y_a^{13}y_b (y_b^2+1) (3 y_b^6-5 y_b^4+65 y_b^2+65)+y_a^{12}(22 y_b^{10}-127 y_b^8  -26 y_b^6-924 y_b^4-236 y_b^2+11)\\
        &+y_a^{11}y_b (27 y_b^{10}-86 y_b^8+248 y_b^6+2032 y_b^4-638 y_b^2-74)+y_a^{10}(59 y_b^{10} -1196 y_b^8+624 y_b^6+1570 y_b^4+1610 y_b^2\\
        &-188) +y_a^9y_b (446 y_b^8-2386 y_b^6-4907 y_b^4-1288 y_b^2+348) +y_a^8(3176 y_b^8+870 y_b^6-4225 y_b^4+1510 y_b^2-880)\\
        & +2y_a^7 y_b (234 y_b^6+4865 y_b^4+192 y_b^2+153)-2 y_a^6 (6158 y_b^6-12709 y_b^4+7564 y_b^2-1829)\\
        & -y_a^5y_b (3448 y_b^4+3474 y_b^2-1151) +y_a^4(-15456 y_b^4+22180 y_b^2-5683) +y_a^3y_b (1559 y_b^2-792) \\
        &-3y_a^2 (2539 y_b^2-1254) +138 y_a y_b -468\Big)\bigg]
    \end{aligned}
\end{equation}

\begin{equation}
    \begin{aligned}
        A_1&=3(y_a-y_b) (1-y_a)^4  (1-y_b)^4 \left((y_a-y_b)^2-4 \zeta(1-\zeta-y_a\, y_b)\right) \bigg[
        \textcolor{cyan}{256\zeta^6} \Big(y_a^4 y_b (11 y_b^3+24 y_b^2+22 y_b+4)\\
        & - y_a^3 y_b (7 y_b^2-4 y_b+6)+15 y_a^2y_b^2 -3 y_a y_b \Big)\\
        &+\textcolor{cyan}{16\zeta^5} \Big(-y_a^5 (83 y_b^5+424 y_b^4-116 y_b^3-56 y_b^2-146 y_b-16) - y_a^4 (739 y_b^4+184 y_b^3+892 y_b^2-64 y_b+30)\\
        &+4y_a^3 (247 y_b^3-28 y_b^2+133 y_b-6)-20y_a^2 (29 y_b^2-6 y_b +3) +3y_a (31 y_b-8) -3\Big)\\
        &+\textcolor{cyan}{16\zeta^4}\Big(-y_a^6 (7 y_b^6-12 y_b^5+315 y_b^4+564 y_b^3+68 y_b^2+56 y_b-13) +4 y_a^5(91 y_b^5+71 y_b^4-282 y_b^3+29 y_b^2-96 y_b\\
        &+5) +y_a^4 (76 y_b^4-1548 y_b^3+1085 y_b^2-416 y_b+118) -4y_a^3 (396 y_b^3-29 y_b^2+198 y_b-27)\\
        & +y_a^2 (499 y_b^2-168 y_b+129) -12y_a (5 y_b-4) +6\Big)\\
        &+\textcolor{cyan}{2\zeta^3} \Big(- y_a^7(9 y_b^7+24 y_b^6+230 y_b^5-328 y_b^4+1590 y_b^3+2016 y_b^2+210 y_b+80) + y_a^6(163 y_b^6-376 y_b^5\\
        &+8618 y_b^4+4576 y_b^3-2206 y_b^2+944 y_b-338)-y_a^5 (6349 y_b^5+600 y_b^4+562 y_b^3+7264 y_b^2-2226 y_b+720)\\
        & +y_a^4 (4535 y_b^4+5664 y_b^3-7090 y_b^2+2800 y_b-1126) + y_a^3(3695 y_b^3-2344 y_b^2+2954 y_b-984)\\
        & -3 y_a^2(583 y_b^2-120 y_b+258) +3y_a (25 y_b-88) -33\Big)\\
        &+\textcolor{cyan}{2\zeta^2} \Big( y_a^8(4 y_b^7-9 y_b^6-151 y_b^4+192 y_b^3-555 y_b^2-420 y_b-21) +4y_a^7 (8 y_b^7+18 y_b^6+135 y_b^5-193 y_b^4+1018 y_b^3\\
        &+284 y_b^2-105 y_b+26) -  y_a^6(222 y_b^6-856 y_b^5+11182 y_b^4+712 y_b^3+1564 y_b^2+1592 y_b-175)\\
        &+4y_a^5 (1987 y_b^5+21 y_b^4+2063 y_b^3+448 y_b^2-244 y_b+93)- y_a^4(6556 y_b^4+1028 y_b^3+1186 y_b^2+1132 y_b-383) \\
        & +4 y_a^3(310 y_b^3+132 y_b^2-61 y_b+87) +3 y_a^2 (62 y_b^2+32 y_b+71)+36y_a (y_b+2) +9\Big)\\
        &+\textcolor{cyan}{2\zeta} \Big(y_a^9(3 y_b^6+y_b^5+10 y_b^4-32 y_b^3+53 y_b^2-97 y_b-34) -y_a^8(3 y_b^7-28 y_b^6-20 y_b^5-233 y_b^4+289 y_b^3-904 y_b^2\\
        &-16 y_b+45) -y_a^7(29 y_b^7+30 y_b^6+404 y_b^5-758 y_b^4+3322 y_b^3-458 y_b^2+376 y_b+95) +y_a^6(203 y_b^6-522 y_b^5\\
        &+7216 y_b^4-926 y_b^3+2446 y_b^2-125 y_b-48) - y_a^5(4701 y_b^5-486 y_b^4+5648 y_b^3-710 y_b^2+483 y_b+54) \\
        &+ y_a^4(3623 y_b^4-490 y_b^3+1692 y_b^2-60 y_b-3) -3 y_a^3 (387 y_b^3-38 y_b^2+44 y_b+3)+9y_a^2 y_b (15 y_b+1) \Big)\\
        &+\Big((y_b-1) y_a^{10} (y_b^4+2 y_b^3+6 y_b^2+2 y_b+13)+y_a^9(-3 y_b^6+4 y_b^5-5 y_b^4+52 y_b^3-73 y_b^2+152 y_b-31) \\
        &+ y_a^8(2 y_b^7-25 y_b^6-9 y_b^5-162 y_b^4+258 y_b^3-699 y_b^2+229 y_b-74)+2 y_a^7(10 y_b^7+5 y_b^6+122 y_b^5-223 y_b^4\\
        &+968 y_b^3-317 y_b^2+234 y_b-42) +y_a^6(-130 y_b^6+250 y_b^5-3560 y_b^4+838 y_b^3-1348 y_b^2+345 y_b-87)\\
        & +3 y_a^5(728 y_b^5-134 y_b^4+788 y_b^3-177 y_b^2+116 y_b-15) -3 y_a^4(470 y_b^4-90 y_b^3+203 y_b^2-45 y_b+6)\\
        & +6 y_a^3 y_b (58 y_b^2-15 y_b+12)-54y_a^2 y_b^2 \Big)\bigg]
    \end{aligned}
\end{equation}
\begin{equation}
    \begin{aligned}
        A_2&=6 (1-y_a)^4 (1-y_b)^4 \sqrt{4 \zeta(1-\zeta-y_a\, y_b)-(y_a-y_b)^2} \bigg[
        \textcolor{cyan}{256 \zeta^7} \Big(y_a^5 y_b (11 y_b^4+44 y_b^3+84 y_b^2+44 y_b+14)\\
        & +12y_a^4 y_b (6 y_b^2+1) +36y_a^3 y_b (3 y_b^2+2 y_b+1) +12  y_a^2 y_b+3 y_a y_b \Big)
        +\textcolor{cyan}{64\zeta^6} \Big(-y_a^6 (18 y_b^6+127 y_b^5-10 y_b^4\\
        &+70 y_b^3-140 y_b^2-37 y_b-14) - y_a^5 (352 y_b^5+467 y_b^4+1368 y_b^3+205 y_b^2+184 y_b-15)+ y_a^4(512 y_b^4-726 y_b^3\\
        &+662 y_b^2-103 y_b+48) -6y_a^3 (216 y_b^3+59 y_b^2+60 y_b-5)+3y_a^2 (50 y_b^2-27 y_b+6) -3y_a (8 y_b-1) \Big)\\
        &+\textcolor{cyan}{32\zeta^5} \Big(- y_a^7 (3 y_b^7+284 y_b^5+570 y_b^4+102 y_b^3+212 y_b^2-56 y_b+2)+2  y_a^6(157 y_b^6+279 y_b^5-403 y_b^4+354 y_b^3\\
        &-624 y_b^2+23 y_b-39)+ y_a^5 (683 y_b^5-748 y_b^4+4344 y_b^3-970 y_b^2+798 y_b-104)-2  y_a^4(1486 y_b^4-691 y_b^3\\
        &+1727 y_b^2-192 y_b+114)+y_a^3 (2905 y_b^3-400 y_b^2+780 y_b-138) -6 y_a^2(111 y_b^2-13 y_b+13) +3y_a (13 y_b-4) \Big)\\
        &
        +\textcolor{cyan}{8\zeta^4} \Big(-  y_a^8(2 y_b^8+5 y_b^7+94 y_b^6-139 y_b^5+1254 y_b^4+1867 y_b^3+398 y_b^2+315 y_b-22)+ y_a^7(100 y_b^7-36 y_b^6\\
        &+6224 y_b^5+4662 y_b^4-856 y_b^3+2276 y_b^2-1120 y_b+111) -2y_a^6 (2246 y_b^6+898 y_b^5+704 y_b^4+4622 y_b^3-2832 y_b^2\\
        &+1134 y_b-219) + y_a^5 (3324 y_b^5+6358 y_b^4-14976 y_b^3+6340 y_b^2-4200 y_b+575)+2 y_a^4(4218 y_b^4-2659 y_b^3\\
        &+5344 y_b^2-1203 y_b+399) - y_a^3(6732 y_b^3-2428 y_b^2+2160 y_b-465)+6 y_a^2(306 y_b^2-22 y_b+41) \\
        &-3 y_a (28 y_b-11)\Big)
        +\textcolor{cyan}{8\zeta^3} \Big(y_a^9 (y_b^8-9 y_b^7-8 y_b^6-119 y_b^5+160 y_b^4-647 y_b^3-676 y_b^2-105 y_b-37) \\
        &+y_a^8 (15 y_b^8+26 y_b^7+415 y_b^6-554 y_b^5+4825 y_b^4+2554 y_b^3-47 y_b^2+533 y_b-103) -y_a^7(258 y_b^7-448 y_b^6\\
        &+12998 y_b^5+2884 y_b^4+2564 y_b^3+3520 y_b^2-919 y_b+278) +y_a^6(8850 y_b^6+952 y_b^5+12618 y_b^4+6672 y_b^3\\
        &-2364 y_b^2+1984 y_b-401)-y_a^5 (10064 y_b^5+3864 y_b^4-1294 y_b^3+3680 y_b^2-1967 y_b+438) +y_a^4(144 y_b^4\\
        &+2124 y_b^3-2578 y_b^2+1328 y_b-369) +  y_a^3(1318 y_b^3-720 y_b^2+753 y_b-198)-3y_a^2 (146 y_b^2-28 y_b+29) \\
        &+3y_a (11 y_b-3) \Big)
        +\textcolor{cyan}{2\zeta^2} \Big(  y_a^{10}(9 y_b^7-8 y_b^6+17 y_b^5-210 y_b^4+291 y_b^3-740 y_b^2-477 y_b-34)- y_a^9(9 y_b^8-132 y_b^7\\
        &-102 y_b^6-1268 y_b^5+1700 y_b^4-6972 y_b^3-1882 y_b^2+148 y_b-141) - y_a^8(124 y_b^8+119 y_b^7+2482 y_b^6-3919 y_b^5\\
        &+26632 y_b^4+1925 y_b^3+3674 y_b^2+1866 y_b-204) +y_a^7(1424 y_b^7-2510 y_b^6+56728 y_b^5-614 y_b^4+23200 y_b^3\\
        &+3749 y_b^2-476 y_b+509) -2 y_a^6(18164 y_b^6-567 y_b^5+28940 y_b^4+1477 y_b^3+2552 y_b^2+862 y_b-233) \\
        &+y_a^5 (38536 y_b^5+930 y_b^4+18536 y_b^3+1761 y_b^2-196 y_b+495) -2y_a^4 (6580 y_b^4+273 y_b^3+1007 y_b^2+315 y_b\\
        &-144)+y_a^3 (1744 y_b^3+135 y_b^2-228 y_b+135) -3y_a^2 (20 y_b^2+45 y_b-12) -36y_a y_b \Big)+\textcolor{cyan}{4\zeta} \Big( y_a^{11}(2 y_b^6+y_b^5+7 y_b^4\\
        &-18 y_b^3+32 y_b^2-55 y_b-17) -y_a^{10} (6 y_b^7-15 y_b^6+3 y_b^5-174 y_b^4+236 y_b^3-627 y_b^2-21 y_b+16) \\
        &+  y_a^9(4 y_b^8-69 y_b^7-33 y_b^6-534 y_b^5+826 y_b^4-2989 y_b^3+347 y_b^2-256 y_b-32)+ y_a^8 (53 y_b^8+29 y_b^7+858 y_b^6\\
        &-1382 y_b^5+8431 y_b^4-1359 y_b^3+2160 y_b^2-124 y_b+13)- y_a^7 (480 y_b^7-760 y_b^6+15480 y_b^5-2040 y_b^4+7100 y_b^3\\
        &-922 y_b^2+471 y_b-15)+y_a^6(9466 y_b^6-1032 y_b^5+13456 y_b^4-1702 y_b^3+2153 y_b^2-171 y_b+42) \\
        &-3 y_a^5(2748 y_b^5-312 y_b^4+1527 y_b^3-141 y_b^2+82 y_b-6) +3y_a^4 (962 y_b^4-89 y_b^3+202 y_b^2-18 y_b+3)\\
        & -6y_a^3 y_b (67 y_b^2-6 y_b+6) +27 y_a^2 y_b^2\Big)+\Big( y_a^{12}(y_b-1) (y_b^4+2 y_b^3+6 y_b^2+2 y_b+13)+y_a^{11}(-5 y_b^6+2 y_b^5-13 y_b^4\\
        &+60 y_b^3-95 y_b^2+178 y_b-31) +y_a^{10}(9 y_b^7-32 y_b^6+5 y_b^5-270 y_b^4+415 y_b^3-1016 y_b^2+291 y_b-74) \\
        &+y_a^9(-5 y_b^8+94 y_b^7+23 y_b^6+620 y_b^5-1035 y_b^4+3486 y_b^3-1123 y_b^2+616 y_b-84) +y_a^8(-65 y_b^8-19 y_b^7\\
        &-910 y_b^6+1400 y_b^5-8131 y_b^4+2335 y_b^3-2358 y_b^2+513 y_b-87) +y_a^7(504 y_b^7-696 y_b^6+13424 y_b^5-2712 y_b^4\\
        &+5528 y_b^3-1305 y_b^2+522 y_b-45) +y_a^6(-7928 y_b^6+1240 y_b^5-8896 y_b^4+1677 y_b^3-1392 y_b^2+225 y_b-18)\\
        &+3y_a^5 y_b (1728 y_b^4-267 y_b^3+754 y_b^2-135 y_b+36)-45y_a^4 y_b^2 (29 y_b^2-5 y_b+6) +180  y_a^3y_b^3\Big)\bigg]
    \end{aligned}
\end{equation}
\begin{equation}
    \begin{aligned}
        A_3&=\zeta (1-y_a^2) (1-y_b^2) \left(\left(y_a-y_b\right)^2+4 \zeta y_b y_a\right) \left(\left(y_a-y_b\right)^2-4 \zeta(1-\zeta-y_a\, y_b)\right) \bigg[
        \textcolor{cyan}{2\zeta^4} \Big(3 y_a^{10} (1-y_b^2)^3\\
        & +49  y_a^8(1-y_b^2)^3+y_a^6 (1346 y_b^6-1096 y_b^4+5847 y_b^2-1247) - y_a^4 (8826 y_b^4-3399 y_b^2+3239)\\
        &+27y_a^2 (201 y_b^2-56) -99\Big)
        +\textcolor{cyan}{\zeta^3} \Big(3y_a^{11} (y_b^2-1)^3 y_b  +y_a^{10}(y_b^2-1)^3 +65y_a^9 (y_b^2-1)^3 y_b  -69y_a^8 (y_b^2-1)^3\\
        &  -y_a^7y_b (1986 y_b^6-9384 y_b^4-2857 y_b^2-3951) +y_a^6(-3030 y_b^6+5720 y_b^4-15261 y_b^2+3381) -3  y_a^5y_b (3426 y_b^4\\
        &+5389 y_b^2+3107)+y_a^4(10110 y_b^4+8403 y_b^2+2317) +5y_a^3 y_b (5289 y_b^2-584)-3 y_a^2(6171 y_b^2-2408) \\
        & -1965y_a y_b -351\Big)
        +\textcolor{cyan}{2\zeta^2} \Big( y_a^{11}(y_b^2-1)^3 y_b  -y_a^{10}(y_b^2-1)^3  (3 y_b^2-1)+87 y_a^9(y_b^2-1)^3 y_b  +y_a^8(232 y_b^8-2140 y_b^6\\
        &-398 y_b^4+2111 y_b^2-34)+ y_a^7y_b (2170 y_b^6-7056 y_b^4+5731 y_b^2-3103) +2y_a^6 (2066 y_b^6-1242 y_b^4-1625 y_b^2\\
        &-200) +3y_a^5 y_b (546 y_b^4+233 y_b^2+1115) +2 y_a^4(1410 y_b^4-3392 y_b^2+2019) - y_a^3 y_b (2337 y_b^2+2020)\\
        &+3 y_a^2(2154 y_b^2-1669) +933y_a y_b +702\Big)
        +\textcolor{cyan}{3\zeta} \Big( y_a^{11} (y_b^2-1)^3 y_b - y_a^{10} (y_b^2-1)^3 (8 y_b^2-3)-y_a^9 y_b (17 y_b^8\\
        &+133 y_b^6-219 y_b^4+1503 y_b^2-427) - y_a^8 (287 y_b^8-2937 y_b^6-89 y_b^4-707 y_b^2+79)-3 y_a^7y_b (370 y_b^6-1224 y_b^4\\
        &+735 y_b^2-295) - y_a^6(4626 y_b^6+2616 y_b^4-2161 y_b^2+1449) +3 y_a^5y_b (222 y_b^4+401 y_b^2-93) +y_a^4 (6414 y_b^4\\
        &-5327 y_b^2+1203) - y_a^3y_b (4369 y_b^2-3048) +3y_a^2 (551 y_b^2-288) -504  y_a y_b+84\Big)\\
        &+6\Big( y_a^{11}(y_b^2-1)^3 y_b  +y_a^{10}(11 y_b^8-45 y_b^6+17 y_b^4-103 y_b^2-8) + y_a^9y_b (14 y_b^8-17 y_b^6+43 y_b^4+493 y_b^2-35)\\
        & + y_a^8(39 y_b^8-601 y_b^6+157 y_b^4-126 y_b^2+97) + y_a^7 y_b (226 y_b^6-848 y_b^4-365 y_b^2-247)+y_a^6 (1030 y_b^6+844 y_b^4\\
        &-951 y_b^2+485) + y_a^5y_b (54 y_b^4+635 y_b^2+65) -y_a^4(y_b^2-1)  (2810 y_b^2-1409) +3  y_a^3y_b (109 y_b^2-132)\\
        &-9y_a^2 (227 y_b^2-159) +51  y_a y_b-234\Big)\bigg]
    \end{aligned}
\end{equation}
\bibliography{EEC_forward}{}
\bibliographystyle{JHEP}

\end{document}